\documentstyle[amssymb,floats,10pt,epsf,prl,aps]{revtex}
\begin{document}
\draft
\title{Vortices in a trapped dilute Bose-Einstein condensate}
\author{Alexander L.~Fetter and Anatoly A.~Svidzinsky}
\address{Department of Physics, Stanford University, Stanford, CA 94305-4060}
\date{\today }
\maketitle

\begin{abstract}
We review the theory of vortices in trapped dilute Bose-Einstein condensates
and compare theoretical predictions with existing experiments. Mean-field
theory based on the time-dependent Gross-Pitaevskii equation describes the
main features of the vortex states, and its predictions agree well with
available experimental results. We discuss various properties of a single
vortex, including its structure, energy, dynamics, normal modes and
stability, as well as vortex arrays. When the nonuniform condensate contains
a vortex, the excitation spectrum includes unstable (``anomalous'') mode(s)
with negative frequency. Trap rotation shifts the normal-mode frequencies
and can stabilize the vortex. We consider the effect of thermal
quasiparticles on vortex normal modes as well as possible mechanisms for
vortex dissipation. Vortex states in mixtures and spinor condensates are
also discussed.
\end{abstract}


\tableofcontents

\section{Introduction}

The recent dramatic achievement of Bose-Einstein condensation in trapped
alkali-metal gases at ultra-low temperatures~\cite{Ande95,Brad95,Davi95} has
stimulated intense experimental and theoretical activity. The atomic
Bose-Einstein condensates (BECs) differ fundamentally from the helium BEC in
several ways. First, BECs in helium are uniform. In contrast, the trapping
potential that confines an alkali-metal-atom vapor BEC yields a
significantly nonuniform density. Another difference is that in bulk
superfluid $^4$He, measurements of the momentum distribution have shown that
the low-temperature condensate fraction is $\sim 0.1$, with the remainder of
the particles in finite momentum states~\cite{Grif93,Soko95}, whereas the
low-temperature atomic condensates can be prepared with essentially all
atoms in the Bose condensate. Finally, the condensates of alkali vapors are
pure and dilute (with mean particle density $\bar n$ and $\bar n|a|^3\ll 1$%
), so that the interactions can be accurately parametrized in terms of a
scattering length $a$ (in current experiments, alkali-metal-atom BECs are
much less dense than air at normal pressure). This situation differs from
superfluid $^4$He, where the relatively high density and strong repulsive
interactions greatly complicate the analytical treatments. As a result, a
relatively simple nonlinear Schr\"odinger equation (the Gross-Pitaevskii
equation) gives a precise description of the atomic condensates and their
dynamics (at least at low temperatures). One should mention, however, that
unlike the spinless $^4$He atoms, alkali atoms have nonzero hyperfine spins,
and various forms of spin-gauge effects can be important~\cite{Ho96}.

Bulk superfluids are distinguished from normal fluids by their ability to
support dissipationless flow. Such persistent currents are intimately
related to the existence of quantized vortices, which are localized phase
singularities with an integer topological charge. The superfluid vortex is
an example of topological defects that are well known in liquid helium~\cite
{Lifs80aa,Till86} and in superconductors~\cite{Vine69}. The occurrence of
quantized vortices in superfluids has been the object of fundamental
theoretical and experimental work~\cite{Onsa49,Feyn55a,Vine61,Yarm79,Donn91}%
. Vortex-like excitations exist in the earth's atmosphere \cite{Dolz90}, in
superfluid hadronic matter (neutron stars) \cite{Sedr91}, and even in
rotating nuclei \cite{Fowl85}. Examples of other topological defects that
could exist in dilute gas condensates are ``textures'' found in Fermi
superfluid $^3$He \cite{Voll90a}, skyrmions \cite{Makh92,Ho98} and spin
monopoles \cite{Garc00b}. Vortices in the $A$ and $B$ phases of $^3$He are
discussed in detail in the review articles \cite{Krus84,Volo84}. In
superfluid $^3$He the Cooper pairs have both orbital and spin angular
momentum. These internal quantum numbers imply a rich phase diagram of
allowed vortex structures, including nonquantized vortices with continuous
vorticity (see also Refs.~\cite{Fett86,Salo87}).

In the framework of hydrodynamics, the vortices obtained from the
Gross-Pitaevskii (GP) equation are analogous to vortices in classical fluids~%
\cite{Saff97}. Also the GP equation provides an approximate description of
some aspects of superfluid behavior of helium, such as the annihilation of
vortex rings~\cite{Jone82}, the nucleation of vortices~\cite{Fris92}, and
vortex-line reconnection~\cite{Kopl93,Kopl96}.

The initial studies of trapped Bose condensates concentrated on measuring
the energy and condensate fraction, along with the lowest-lying collective
modes and quantum-mechanical interference effects (see, for example, Ref.~%
\cite{Dalf99}). Although the possibility of trapped quantized vortices was
quickly recognized~\cite{Baym96}, successful experimental verification has
taken several years \cite{Matt99,Madi99,Madi00a,Chev00,Ande00}. This review
focuses on the behavior of quantized vortices in trapped dilute Bose
condensates, emphasizing the qualitative features along with the
quantitative comparison between theory and experiment.

The plan of the paper is the following. In Sec.~II we discuss the basic
formalism of mean-field theory (the time-dependent Gross-Pitaevskii
equation) that describes dilute Bose-Einstein condensates in the
low-temperature limit. We summarize properties of vortices in a uniform
condensate and also introduce relevant length and energy scales of a
condensate in a harmonic trap. In Sec.~III we discuss the structure of
stationary vortex states in trapped condensates. We analyze the energy of a
straight vortex as a function of displacement from the trap center and
consider conditions of vortex stability when the trap rotates. Also we
discuss the recent experimental creation of a single vortex and vortex
arrays. In Sec.~IV we introduce the concept of elementary excitations (the
Bogoliubov equations) and analyze the lowest (unstable) mode of the vortex
for different values of the interaction parameter. We also consider the
splitting of the condensate normal modes due to presence of a vortex line.

In Sec.~V we investigate the general dynamical behavior of a vortex, based
on a time-dependent variational analysis and on the method of matched
asymptotic expansions. The latter method allows us take into account effects
of both nonuniform condensate density and vortex curvature. We consider
normal modes of a vortex in two- and three-dimensional condensates. Also we
discuss the energy of a curved vortex line and a nonlinear tilting of a
vortex in slightly anisotropic condensates. In Sec.~VI we analyze the effect
of thermal quasiparticles on the vortex normal modes and discuss possible
mechanisms of vortex dissipation. Also we discuss the influence of vortex
generation on energy dissipation in superfluids. In Sec.~VII we consider
vortices in multicomponent condensates and analyze various spin-gauge
effects. In particular, we focus on the successful method of vortex
generation in a two-component system that was recently used by the JILA
group to create a vortex. In Sec.~VIII we draw our conclusions and discuss
perspectives in the field.

\section{Time-dependent Gross-Pitaevskii equation}

Bogoliubov's seminal treatment \cite{Bogo47} of a uniform Bose gas at zero
temperature emphasized the crucial role of (repulsive) interactions both for
the structure of the ground state and for the existence of superfluidity.
Subsequently, Gross \cite{Gros61,Gros63} and Pitaevskii \cite{Pita61}
independently considered an {\em inhomogeneous\/} dilute Bose gas,
generalizing Bogoliubov's approach to include the possibility of nonuniform
states, especially quantized vortices.

An essential feature of a dilute Bose gas at zero temperature is the
existence of a macroscopic wave function (an ``order parameter'') $\Psi $
that characterizes the Bose condensate. For a uniform system with $N$
particles in a stationary box of volume $V$, the order parameter $\Psi =%
\sqrt{N_0/V}$ reflects the presence of a macroscopic number $N_0$ of
particles in the zero-momentum state, with the remaining $N^{\prime }=N-N_0$
particles distributed among the various excited states with ${\bf k}\neq 0$.
The single-particle states for periodic boundary conditions are plane waves $%
V^{-1/2}e^{i{\bf k\cdot r}}$ labeled with the wave vector ${\bf k}$, and the
corresponding creation and annihilation operators $a_{{\bf k}}^{\dagger }$
and $a_{{\bf k}}$ obey the usual Bose-Einstein commutation relations $[a_{%
{\bf k}},a_{{\bf k^{\prime }}}^{\dagger }]=\delta _{{\bf k,k^{\prime }}}$.
In the presence of a uniform Bose condensate with ${\bf k}=0$, the
ground-state expectation value $\langle a_0^{\dagger }a_0\rangle _0=N_0$ is
macroscopic, whereas the ground-state expectation value of the commutator of
these zero-mode operators $\langle [a_0,a_0^{\dagger }]\rangle _0$
necessarily equals 1. Hence the commutator is of order $1/\sqrt{N_0}$
relative to each separate operator, and they can be approximated by
classical numbers $a_0\approx a_0^{\dagger }\approx \sqrt{N_0}$. This
``Bogoliubov'' approximation identifies these classical fields as the order
parameter for the stationary uniform condensate. In contrast, the
ground-state expectation value for all the other normal modes $\langle a_{%
{\bf k}}^{\dagger }a_{{\bf k}}\rangle _0$ is of order unity, and the
associated operators $a_{{\bf k}}^{\dagger }$ and $a_{{\bf k}}$ require a
full quantum-mechanical treatment.

The existence of nonuniform states of a dilute Bose gas can be understood by
considering a second-quantized Hamiltonian

\begin{equation}
\label{1}\hat H=\int dV\left[ \hat \psi ^{\dagger }\left( T+V_{{\rm tr}%
}\right) \hat \psi +\case{1}{2}g\hat \psi ^{\dagger }\hat \psi ^{\dagger }%
\hat \psi \hat \psi \right] ,
\end{equation}
expressed in terms of Bose field operators $\hat \psi ({\bf r})$ and $\hat
\psi ^{\dagger }({\bf r})$ that obey Bose-Einstein commutation relations
\begin{equation}
\label{2}[\hat \psi ({\bf r}),\hat \psi ^{\dagger }({\bf r}^{\prime
})]=\delta ({\bf r-r}^{\prime }),\quad [\hat \psi ({\bf r}),\hat \psi ({\bf r%
}^{\prime })]=[\hat \psi ^{\dagger }({\bf r}),\hat \psi ^{\dagger }({\bf r}%
^{\prime })]=0.
\end{equation}
Here $T=-\hbar ^2\nabla ^2/2M$ is the kinetic energy operator for the
particles of mass $M$, $V_{{\rm tr}}({\bf r})$ is an external (trap)
potential, and the interparticle potential has been approximated by a
short-range interaction $\approx g\,\delta ({\bf r-r}^{\prime })$, where $g$
is a coupling constant with the dimensions of energy $\times $ volume. For a
dilute cold gas, only binary collisions at low energy are relevant, and
these collisions are characterized by a single parameter, the $s$-wave
scattering length $a$, independent of the details of the two-body potential.
An analysis of the scattering by such a potential (see, for example~\cite
{Fett71,Fett99}) shows that $g\approx 4\pi a\hbar ^2/M$. Determinations of
the scattering length for the atomic species used in the experiments on Bose
condensation give: $a=2.75 $ nm for $^{23}$Na \cite{Ties96}, $a=5.77$ nm for
$^{87}$Rb~\cite{Boes97}, and $a=-1.45$ nm for $^7$Li~\cite{Abra95}. In a
uniform bulk system, $a$ must be positive to prevent an instability leading
to a collapse, but a Bose condensate in an external confining trap can
remain stable for $a<0$ as long as the number of condensed atoms $N_0$
remains below a critical value $N_{cr}\sim d/|a|$, where $d$ is the
oscillator length~\cite{Dalf99,Fett99}. If the interparticle potential is
attractive ($a<0$), the gas tends to increase its density in the trap center
to lower the interaction energy. The kinetic energy opposes this tendency,
and the resulting balance can stabilize inhomogeneous gas. A vortex line
located along the trap axis reduces the peak central density in the cloud of
atoms. Thus a vortex can help stabilize a larger trapped condensate with
attractive interactions in the sense it can contain a larger number of atoms~%
\cite{Dalf96}.

The time-dependent Heisenberg operator $\hat\psi({\bf r},t)=\exp(i\hat H
t/\hbar) \,\hat\psi({\bf r})\,\exp(-i\hat H t/\hbar) $ obeys the equation of
motion $i\hbar\, \partial\hat\psi({\bf r},t)/\partial t = [\hat\psi({\bf r}%
,t),\hat H]$, which yields a nonlinear operator equation

\begin{equation}
\label{3}i\hbar\, \frac{\partial\hat\psi({\bf r},t)}{\partial t} =
\left(T+V_{{\rm tr}}\right)\hat\psi({\bf r},t) + g\,\hat\psi^\dagger({\bf r}%
,t)\,\hat\psi({\bf r},t) \,\hat\psi({\bf r},t).
\end{equation}
The macroscopic occupation of the condensate makes it natural to write the
field operator as a sum $\hat\psi({\bf r},t) = \Psi({\bf r},t) + \hat \phi(%
{\bf r},t)$ of a classical field $\Psi({\bf r},t)$ that characterizes the
macroscopic condensate and a quantum field $\hat \phi({\bf r},t)$ referring
to the remaining noncondensed particles. To leading order, the Bogoliubov
approximation omits the quantum fluctuations entirely, giving the
time-dependent Gross-Pitaevskii (GP) equation \cite{Gros61,Pita61}

\begin{equation}
\label{4}i\hbar\, \frac{\partial\Psi({\bf r},t)}{\partial t} = \left[T+V_{%
{\rm tr}}+g\,|\Psi({\bf r},t)|^2\right] \Psi({\bf r},t)
\end{equation}
for the condensate wave function $\Psi({\bf r},t)$. Since $\hat\psi({\bf r}%
,t) $ reduces the number of particles by one, its off-diagonal matrix
element $\langle N-1|\hat\psi({\bf r},t)|N\rangle$ oscillates at a frequency
corresponding to the chemical potential $\mu \approx E_0(N)-E_0(N-1)$
associated with removing one particle from the ground state. Thus the
stationary solutions take the form $\Psi({\bf r},t)=\Psi({\bf r})\,e^{-i\mu
t/\hbar}$, where $\Psi({\bf r})$ obeys the stationary GP equation
(frequently identified as a nonlinear Schr\"odinger equation, although the
eigenvalue $\mu$ is not the energy per particle)

\begin{equation}
\label{5}(T+V_{{\rm tr}} +g|\Psi|^2)\Psi = \mu\Psi.
\end{equation}

Apart from very recent work on $^{85}$Rb using a Feshbach resonance to tune $%
a$ to large positive values~\cite{Corn00a}, essentially all studies of
trapped atomic gases involve the dilute limit ($\bar n|a|^3\ll 1$, where $%
\bar n$ is the average density of the gas), so that depletion of the
condensate is small with $N^{\prime }=N-N_0\propto \sqrt{\bar n|a|^3}N\ll N$%
. Typically $\bar n|a|^3$ is always less than $10^{-3}$. Hence most of the
particles remain in the condensate, and the difference between the
condensate number $N_0$ and the total number $N$ can usually be neglected.
In this case, the stationary GP equation (\ref{5}) for the condensate wave
function follows by minimizing the Hamiltonian functional

\begin{equation}
\label{6}H = \int dV\left[\Psi^*\left(T+V_{{\rm tr}}\right)\Psi + \case{1}{2}
g|\Psi|^4\right],
\end{equation}
subject to a constraint of fixed condensate number $N_0=\int dV\,|\Psi|^2
\approx N$ (readily included with a Lagrange multiplier that is simply the
chemical potential~$\mu$).

\subsection{Unbounded Condensate}

The nonlinear Schr\"odinger equation (\ref{5}) contains a local
self-consistent Hartree potential energy $V_H({\bf r} ) = g|\Psi({\bf r})|^2$
arising from the interaction with the other particles at the same point. In
an unbounded condensate with $V_{{\rm tr}}=0$, the left-hand side of Eq.~(%
\ref{5}) involves both the kinetic energy $T$ and this repulsive Hartree
potential $g|\Psi|^2= gn$ for a uniform medium with bulk density $n$. On
dimensional grounds, the balance between these two terms implies a
``correlation'' or ``healing'' length
\begin{equation}
\label{7}\xi = \frac{\hbar}{\sqrt{2Mng}} = \frac{1}{\sqrt{8\pi na}}.
\end{equation}
This length characterizes the distance over which the condensate wave
function heals back to its bulk value when perturbed locally (for example,
at a vortex core, where the density vanishes).

For a uniform system in a box of volume $V$, the condensate wave function is
$\Psi =\sqrt{N_0/V}\approx \sqrt{N/V}$, and Eq.~(\ref{6}) shows that the
ground-state energy $E_0$ arises solely from the repulsive interparticle
energy of the condensate $E_{{\rm int}}\approx \frac{1}{2}\,gN^2/V$. The
bulk chemical potential is then given by

\begin{equation}
\label{8}\mu = \left(\frac{\partial E_0}{\partial N}\right)_{\!\! V} = gn =
\frac{4\pi a\hbar^2n}{M}.
\end{equation}
The corresponding pressure follows from the thermodynamic relation

\begin{equation}
\label{9}p=-\left(\frac{\partial E_0}{\partial V}\right)_{\!\! N}=
\case{1}{2}gn^2=\frac{E_{{\rm int}}}{V}.
\end{equation}
Finally, the compressibility determines the bulk speed of sound $s$:

\begin{equation}
\label{10}s^2 = \frac{1}{M}\,\left(\frac{\partial p}{\partial n}\right)=%
\frac{gn}{M}=\frac{\mu}{M}= \frac{4\pi a\hbar^2n}{M^2},\>\>%
\hbox{or, equivalently,}\>\> s = \frac{\hbar}{\sqrt 2M\xi}\,;
\end{equation}
Equations (\ref{7}) and (\ref{10}) both indicate that a bulk uniform Bose
condensate requires a repulsive interaction ($a>0$), since otherwise the
healing length and the speed of sound become imaginary.

\subsection{Quantum-Hydrodynamic Description of the Condensate}

It is often instructive to represent the condensate wave function in an
equivalent ``quantum-hydrodynamic'' form

\begin{equation}
\label{11}\Psi({\bf r},t)= |\Psi({\bf r},t)|\,e^{iS({\bf r},t)},
\end{equation}
with the condensate density

\begin{equation}
\label{12}n({\bf r},t)=|\Psi({\bf r},t)|^2.
\end{equation}
The corresponding current density ${\bf j} = (\hbar/2Mi)[\Psi^*\bbox{\nabla}%
\Psi-(\bbox{\nabla}\Psi^*)\Psi]$ automatically assumes a hydrodynamic form

\begin{equation}
\label{13}{\bf j(r},t) = n({\bf r},t)\,{\bf v(r},t),
\end{equation}
with an irrotational flow velocity

\begin{equation}
\label{14}{\bf v(r},t)=\bbox{\nabla}\Phi({\bf r},t)
\end{equation}
expressed in terms of a velocity potential
\begin{equation}
\label{15}\Phi({\bf r},t) = \frac{\hbar S({\bf r},t)}{M}.
\end{equation}

Substitute Eq.~(\ref{11}) into the time-dependent GP equation (\ref{4}). The
imaginary part yields the familiar continuity equation for compressible flow

\begin{equation}
\label{16}\frac{\partial n}{\partial t} +\bbox{\nabla}\cdot \left(n{\bf v}%
\right)=0.
\end{equation}
Correspondingly, the real part constitutes the analog of the Bernoulli
equation for this condensate fluid

\begin{equation}
\label{17}\case{1}{2} Mv^2 + V_{{\rm tr}} +\frac{1}{\sqrt n}\,T\sqrt n + gn+M%
\frac{\partial \Phi}{\partial t}= 0.
\end{equation}
To interpret this equation, note that the assumption of a zero-temperature
condensate implies vanishing entropy; furthermore, the conventional
Bernoulli equation for irrotational compressible {\it isentropic\/} flow can
be rewritten as~\cite{Land59,Fett80}

\begin{equation}
\label{18}\case{1}{2} Mv^2 + U +\frac{e+p}{n} +M\frac{\partial \Phi}{%
\partial t}= 0,
\end{equation}
where $U$ is the external potential energy, $e$ is the energy density and $%
e+p$ is the enthalpy density. Comparison with Eqs.~(\ref{6}) and (\ref{9})
shows that Eq.~(\ref{17}) for the condensate dynamics indeed incorporates
the appropriate constitutive relations for the enthalpy per particle $%
(e+p)/n = (\sqrt n)^{-1}T\sqrt n + g n$.

As a result, the hydrodynamic form of the time-dependent Gross-Pitaevskii
equation in Eqs.~(\ref{16}) and (\ref{17}) necessarily reproduces all the
standard hydrodynamic behavior found for classical irrotational compressible
isentropic flow. In particular, the dynamics of vortex lines at zero
temperature follows from the Kelvin circulation theorem \cite{Land59,Fett80}%
, namely that each element of the vortex core moves with the local
translational velocity induced by all the sources in the fluid (self-induced
motion for a curved vortex, other vortices, and net applied flow). The only
explicitly quantum-mechanical feature in Eq.~(\ref{17}) is the ``quantum
kinetic pressure " $(\sqrt n)^{-1}T\sqrt n\,$; as seen from Eq.~(\ref{7}),
this contribution determines the healing length $\xi$ that will fix the size
and structure of the vortex core.

In classical hydrodynamics, the flow can be considered incompressible when
the velocity $|v|$ is small compared to the speed of sound. More generally,
classical compressible flow becomes irreversible when the flow becomes
supersonic because of the emission of sound waves (which are still part of
the hydrodynamic formalism). In a dilute Bose gas, however, Eqs.~(\ref{16})
and (\ref{17}) neglect the normal component entirely. As discussed below in
Sec.~IV.B, the system becomes unstable with respect to the emission of
quasiparticles once the flow speed exceeds the Landau critical velocity
(which here is simply the speed of sound). The normal component then plays
an essential role and must be included in addition to the condensate. In
this sense, a dilute Bose gas is intrinsically more complicated than a
classical compressible fluid.

\subsection{Vortex Dynamics in Two Dimensions}

Vinen's experiment~\cite{Vine61} on the dynamics of a long fine wire in
rotating superfluid $^4$He strikingly confirmed Onsager's and Feynman's
theoretical prediction of quantized circulation \cite{Onsa49,Feyn55a}. These
remarkable observations stimulated the study of the nonlinear stationary GP
equation (\ref{5}) in the absence of a confining potential, building on an
earlier analysis by Ginzburg and Pitaevskii of vortex-like solutions for
superfluid $^4$He near $T_\lambda$ \cite{Ginz58}. Gross and Pitaevskii
independently investigated stationary two-dimensional solutions of the form $%
\Psi({\bf r}) = \sqrt n\,\chi({\bf r})$, where $n$ is the bulk density far
from the origin. Specifically, they considered axisymmetric solutions

\begin{equation}
\label{19}\chi({\bf r}) = e^{i\phi}f\!\left(\frac{r_\perp}{\xi}\right),
\end{equation}
where ($r_\perp,\phi$) are two-dimensional cylindrical polar coordinates,
and $f\to 1$ for $r_\perp\gg \xi$. Equations (\ref{14}) and (\ref{15})
immediately give the local circulating flow velocity

\begin{equation}
\label{20}{\bf v} = \frac{\hbar}{Mr_\perp}\,\hat\phi,
\end{equation}
which represents circular streamlines with an amplitude that becomes large
as $r_\perp\to 0$. Comparison of Eqs.~(\ref{10}) and (\ref{20}) shows that
the circulating flow becomes supersonic ($v\approx s$) when $r_\perp\approx
\xi$.

The particular condensate wave function (\ref{19}) describes an infinite
straight vortex line with quantized circulation

\begin{equation}
\label{22}\kappa =\oint d{\bf l}\cdot {\bf v}=\frac hM,
\end{equation}
precisely as suggested by Onsager and Feynman~\cite{Onsa49,Feyn55a}.
Stokes's theorem then yields $h/M=\int d{\bf S}\cdot {\bbox
\nabla }\times {\bf v}$, with the corresponding localized vorticity
\begin{equation}
\label{21}\bbox{\nabla}\times {\bf v}=\frac hM\,\delta ^{(2)}({\bf r}_{\perp
})\,\,\hat z.
\end{equation}
Hence the velocity field around a vortex in a dilute Bose condensate is
irrotational except for a singularity at the origin.

\begin{figure}
\bigskip
\centerline{\epsfxsize=0.41\textwidth\epsfysize=0.41\textwidth
\epsfbox{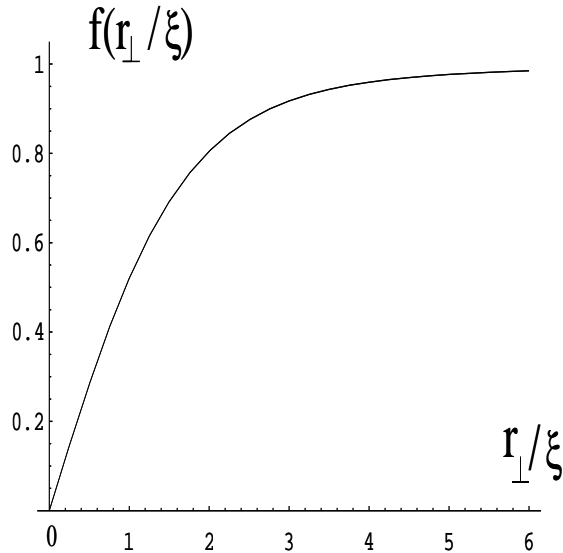}}

\vspace{0.6cm}

\caption{Radial wave function $ f(r_\perp/\xi)$ obtained by numerical
solution of the stationary GP equation for a straight vortex line.}
\label{fig1}
\end{figure}

The kinetic energy per unit length is given by
\begin{equation}
\label{23}\int d^2r_{\perp }\,\Psi ^{*}\left( -\frac{\hbar ^2\nabla ^2}{2M}%
\right) \Psi =\frac{\hbar ^2}{2M}\int d^2r_{\perp }\,|\bbox{\nabla}\Psi |^2=%
\frac{\hbar ^2n}{2M}\int d^2r_{\perp }\,\left[ \left( \frac{df}{dr_{\perp }}%
\right) ^{\!\!2}+\frac{f^2}{r_{\perp }^2}\right] ,
\end{equation}
and the centrifugal barrier in the second term forces the amplitude to
vanish linearly within a core of radius $\approx \xi $ (see Fig.~\ref{fig1}%
). This core structure ensures that the particle current density ${\bf j}=n%
{\bf v}$ vanishes and the total kinetic-energy density remains finite as $%
r_{\perp }\to 0$. The presence of the vortex produces an additional energy $%
E_v$ per unit length, both from the kinetic energy of circulating flow and
from the local compression of the fluid. Numerical analysis with the GP
equation ~\cite{Ginz58} yields $E_v\approx (\pi \hbar ^2n/M)\ln \left(
1.46R/\xi \,\right) $, where $R$ is an outer cutoff; apart from the additive
numerical constant, this value is simply the integral of $\case{1}{2}Mv^2n$.

To illustrate that the time-dependent GP equation indeed incorporates the
correct classical vortex dynamics, consider a state of the form

\begin{equation}
\label{24}\Psi({\bf r},t) = \sqrt n\,e^{i{\bf q\cdot r}}\,\chi({\bf r-r}%
_0)\,e^{-i\mu t/\hbar},
\end{equation}
where $\chi$ is the previous stationary solution (\ref{19}) of the GP
equation for a quantized vortex, now shifted to the instantaneous position $%
{\bf r}_0(t)$, and $\mu$ is now a modified chemical potential. The total
flow velocity is the sum of a uniform velocity ${\bf v}_0 = \hbar{\bf q}/M$
and the circulating flow around the vortex. Substitute this wave function
into the time-dependent GP equation (\ref{4}). Since $\chi$ itself obeys the
stationary GP equation (\ref{5}) with chemical potential $\mu=gn$, a
straightforward analysis shows that $\mu = \frac{1}{2}Mv_0^2 +gn$, where the
first term arises from the center of mass motion of the condensate. The
remaining terms yield

\begin{equation}
\label{25}i\hbar\frac{\partial \chi({\bf r-r}_0)}{\partial t}\equiv -i\hbar
\frac{d{\bf r}_0}{dt}\cdot \bbox{\nabla}\chi({\bf r-r}_0) = -i\hbar {\bf v}%
_0\cdot \bbox{\nabla}\chi({\bf r-r}_0).
\end{equation}
This equation shows that $d{\bf r}_0(t)/dt = {\bf v}_0$, so that the vortex
wave function moves rigidly with the applied flow velocity ${\bf v}_0$,
correctly reproducing classical irrotational hydrodynamics.

A similar method applies to the self-induced motion of two well-separated
vortices at ${\bf r}_1$ and ${\bf r}_2$ with $|{\bf r}_1-{\bf r}_2|\gg \xi $%
; in this case,

\begin{equation}
\label{26}\Psi({\bf r},t) = \sqrt n\,\chi({\bf r-r}_1)\,\chi({\bf r-r}%
_2)\,e^{-i \mu t/\hbar}
\end{equation}
represents an approximate solution with $\mu = ng$ because there is no net
flow velocity at infinity. The density $n\,|f({\bf r-r}_1)|^2|f({\bf r-r}%
_2)|^2$ is essentially constant except near the two vortex cores, and the
phase is the sum $S({\bf r-r}_1)+S({\bf r-r}_2)$ of the two azimuthal angles
for the variable ${\bf r}$ measured from the local vortex cores.
Substitution into the time-dependent GP equation readily shows that each
vortex moves with the velocity induced by the other, for example

\begin{equation}
\label{27}\frac{d{\bf r}_1}{dt} \approx \frac{\hbar}{M}\bbox{\nabla}S({\bf %
r-r}_2)\big|_{{\bf r=r}_1}.
\end{equation}
This method also describes the two-dimensional motion of many well-separated
line vortices \cite{Fett65,Fett66}. The dynamics of the many-vortex case in
2D was also studied in \cite{Lund91,Rica90,Rica92}.

\subsection{Trapped Condensate}

The usual condition for a uniform dilute gas requires that the interparticle
spacing $\sim n^{-1/3}$ be large compared to the scattering length ($%
n^{-1/3}\gg a$ or $na^3\ll 1$). The situation is more complicated in the
case of a dilute trapped gas, because of the three-dimensional harmonic
trapping potential $V_{{\rm tr}}=\case{1}{2}M\left( \omega _x^2x^2+\omega
_y^2y^2+\omega _z^2z^2\right) $. The stationary GP equation (\ref{5})
provides a convenient approach to study the structure of the condensate in
such a harmonic confining potential.

For an ideal noninteracting gas (with $g=0$), the states are the familiar
harmonic-oscillator wave functions with the characteristic spatial scale set
by the oscillator lengths $d_j = \sqrt{\hbar/M\omega_j}$ ($j = x$, $y$, and $%
z$). In particular, the ground-state wave function can be obtained by
optimizing the competition between the kinetic energy $E_{{\rm kin}} =
\langle T\rangle$ and the confining energy $E_{{\rm tr}}=\langle V_{{\rm tr}
}\rangle$, where $\langle\cdots\rangle=N^{-1}\int dV\,\Psi^*\cdots\Psi $
denotes the expectation value for the state with the condensate wave
function $\Psi$. The situation is more complicated for an interacting
system, however, because the additional interaction energy $E_{{\rm int}} =
\langle \frac{1}{2} g|\Psi|^2 \rangle$ provides a new dimensionless
parameter. The ratio $E_{{\rm int} }/N\hbar\omega_0$ serves to quantify the
effect of the interactions, where $\omega_0=
\left(\omega_x\omega_y\omega_z\right)^{1/3}$ is the mean oscillator
frequency. It is not difficult to show that this ratio is of order $Na/d_0$
for $Na/d_0\lesssim 1$ where $d_0 = \sqrt{\hbar/M\omega_0}$ is the mean
oscillator length~\cite{Baym96,Dalf99,Fett99}, and of order $(Na/d_0)^{2/5}$
for $Na/d_0\gg 1$. Thus the presence of the confining trap significantly
alters the physics of the problem, for the additional characteristic length $%
d_0$ and energy $\hbar\omega_0$ now imply the existence of two distinct
regimes of dilute trapped gases:

\subsubsection{Near-ideal regime}

In the limit $Na/d_0\ll1$, the condensate states are qualitatively similar
to those of an ideal gas in a three-dimensional harmonic trap, with
ground-state wave function $\Psi({\bf r}) \propto \exp\left[-\case{1}{2}%
\left(x^2/d_x^2+y^2/d_y^2+z^2/d_z^2\right)\right]$. The repulsive
interactions play only a small role, and the condensate dimensions are
comparable with the oscillator lengths $d_j$.

\subsubsection{Thomas-Fermi regime}

In the opposite limit $Na/d_0\gg 1$, which is relevant to current
experiments on trapped Bose condensates, the repulsive interactions
significantly expand the condensate, so that the kinetic energy associated
with the density variation becomes negligible compared to the trap energy
and interaction energy. As a result, the kinetic-energy operator $T$ can be
omitted in the stationary GP equation (\ref{5}), which yields the
Thomas-Fermi (TF) parabolic profile for the ground-state density~\cite
{Baym96}

\begin{equation}
\label{28}n({\bf r}) \approx |\Psi_{TF}({\bf r})|^2 = \frac{1}{g}\,
\left[\,\mu - V_{{\rm tr}}({\bf r})\,\right]\Theta\left[\,\mu - V_{{\rm tr}}(%
{\bf r})\,\right]= n(0)\left(1-\sum_{j=x,y,z}\frac{x_j^2}{R_j^2}\right)
\Theta\left(1-\sum_{j=x,y,z}\frac{x_j^2}{R_j^2}\right),
\end{equation}
where $n(0)= \mu/g$ is the central density and $\Theta(x)$ denotes the unit
positive step function. The resulting ellipsoidal three-dimensional density
is characterized by two physically different types of parameters: (a) the
central density $n(0)$ fixed by the chemical potential [note that $n(0)$
plays essentially the same role as the bulk density $n$ does for the uniform
condensate, where $\mu = gn$], and (b) the three condensate radii

\begin{equation}
\label{29}R_j^2 = \frac{2\mu}{M\omega_j^2}.
\end{equation}

The normalization integral $\int dV n({\bf r}) = N$ yields the important TF
relation~\cite{Baym96}

\begin{equation}
\label{30}N= \frac{8\pi}{15}n(0)\,R_0^3 =\frac{R_0^5}{15 \,a \,d_0^4},\>\>%
\hbox{or,
equivalently,}\>\> \frac{R_0^5}{d_0^5} = 15 \,\frac{Na}{d_0}\gg 1,
\end{equation}
where $R_0=\left(R_xR_y R_z\right)^{1/3}$ is the mean condensate radius.
This last equality shows that the repulsive interactions expand the mean TF
condensate radius $R_0$ proportional to $N^{1/5}$. The TF chemical potential
becomes

\begin{equation}
\label{31}\mu =\case{1}{2} M\omega_0^2R_0^2= \case{1}{2} \hbar\omega_0 \,%
\frac{R_0^2}{d_0^2},
\end{equation}
so that $\mu\gg \hbar\omega_0$ in this limit. The corresponding ground-state
energy $E_0 = \case{5}{14}\hbar\omega_0 (R_0^2/d_0^2) N = \case{5}{7} \mu
\,N $ follows immediately from the thermodynamic relation $\mu = \partial
E_0/\partial N$.

The TF limit leads to several important simplifications. For a trapped
condensate, it is natural to define the healing length (\ref{7}) in terms of
the central density, with $\xi~=~[8\pi n(0)\,a]^{-1/2}$. In the TF limit,
this choice implies that

\begin{equation}
\label{32}\xi \,R_0 = d_0^2,\quad\hbox{or, equivalently,}\quad \frac{\xi}{d_0%
}=\frac{d_0}{R_0}\ll 1.
\end{equation}
Thus the TF limit provides a clear separation of length scales $\xi\ll
d_0\ll R_0$, and the (small) healing length $\xi$ characterizes the small
vortex core. In contrast, the healing length (and vortex-core radius) in the
near-ideal limit are comparable with $d_0$ and hence with the size of the
condensate.

The quantum-hydrodynamic equations also simplify in the TF limit, because
the quantum kinetic pressure in Eq.~(\ref{17}) becomes negligible. For the
static TF ground-state density given in Eq.~(\ref{28}), the small
perturbations $n^{\prime}$ in the density and $\Phi^{\prime}$ in the
velocity potential can be combined to yield the generalized wave equation~%
\cite{Stri96}

\begin{equation}
\label{33}M\frac{\partial^2n^{\prime}}{\partial t^2}= \bbox{\nabla}\cdot
\left[\left(\mu - V_{{\rm tr}}\right)\bbox{\nabla}n^{\prime}\right]\quad%
\hbox{or,
equivalently,}\quad \frac{\partial^2n^{\prime}}{\partial t^2}= \bbox{\nabla}%
\cdot \left[s^2({\bf r})\bbox{\nabla}n^{\prime}\right],
\end{equation}
where $s^2({\bf r})= \left[\mu-V_{{\rm tr}}({\bf r})\right]/M$ defines a
spatially varying local sound speed. Stringari has used this equation to
analyze the low-lying normal modes of the TF condensate, and several
experimental studies have verified these predictions in considerable
detail~(see, for example, Ref.\cite{Dalf99}).

\section{Static vortex states}

In the context of rotating superfluid $^4$He, Feynman~\cite{Feyn55a} noted
that solid-body rotation with ${\bf v}_{{\rm sb}} = \bbox{\Omega}\times {\bf %
r}$ has constant vorticity $\bbox{\nabla}\times {\bf v}_{{\rm sb}} = 2%
\bbox{\Omega}$. Since each quantized vortex line in rotating superfluid $^4$%
He has an identical localized vorticity associated with the singular
circulating flow (\ref{21}), he argued that a uniform array of vortices can
``mimic'' solid-body rotation on average, even though the flow is strictly
irrotational away from the cores. He then considered the circulation $%
\Gamma=\oint_C d{\bf l}\cdot {\bf v}$ along a closed contour $C$ enclosing a
large number ${\cal N}_v$ of vortices. The quantization of circulation
ensures that $\Gamma = {\cal N}_v\cdot \kappa$, where $\kappa= h/M$ is the
quantum of circulation. If the vortex array mimics solid-body rotation,
however, the circulation should also be $\Gamma = 2\Omega\cdot {\cal A}_v$,
where ${\cal A}_v$ is the area enclosed by the contour $C$. In this way, the
areal vortex density in a rotating superfluid becomes

\begin{equation}
\label{34}n_v =\frac{{\cal N}_v}{{\cal A}_v} = \frac{2\Omega}{\kappa}.
\end{equation}
Equivalently, the area per vortex is simply $1/n_v=\kappa/2\Omega$, which
decreases with increasing rotation speed. Note that Eq.~(\ref{34}) is
directly analogous to the density of vortices (flux lines) $n_v=B/\Phi_0$ in
a type-II superconductor, where $B$ is the magnetic flux density and $%
\Phi_0= h/2e$ is the quantum of magnetic flux in SI units (see, for example,
Ref.~\cite{Tink85})

\subsection{Structure of Single Trapped Vortex}

\subsubsection{Axisymmetric trap}

Consider an axisymmetric trap with oscillator frequencies $\omega_z$ and $%
\omega_{\perp }$ and axial asymmetry parameter $\lambda \equiv \omega
_z/\omega _{\perp }$ (note that $\lambda \lesssim 1$ yields an elongated
cigar-shape condensate, and $\lambda \gtrsim 1$ yields a flattened
disk-shape condensate). The conservation of angular momentum $L_z$ allows a
simple classification of the states of the condensate. For example, the
macroscopic wave function for a singly quantized vortex located along the $z$%
-axis takes the form

\begin{equation}
\label{35}\Psi ({\bf r})=e^{i\phi }\,|\Psi (r_{\perp },z)|.
\end{equation}
The circulating velocity is identical to Eq.~(\ref{20}), and the centrifugal
energy [compare Eq.~(\ref{23})] gives rise to an additional term $\case{1}{2}%
Mv^2=\hbar ^2/2Mr_{\perp }^2$ in the GP equation (\ref{5}). In principle, a $%
q$-fold vortex with $\Psi \propto e^{iq\phi }$ also satisfies the GP
equation, but the corresponding energy increases like $q^2$ [compare the
discussion below Eq.~(\ref{23})]; consequently, a multiply quantized vortex
is expected to be unstable with respect to the formation of $q$ singly
quantized vortices.

For a noninteracting gas in an axisymmetric trap, the condensate wave
function for a singly quantized vortex on the symmetry axis involves the
first excited radial harmonic-oscillator state with the noninteracting
condensate vortex wave function

\begin{equation}
\label{36}\Psi({\bf r}) \propto e^{i\phi}\, r_\perp\,\exp\left[-\case{1}{2}%
\left(\frac{r_\perp^2}{d_\perp^2} +\frac{z^2}{d_z^2}\right)\right],
\end{equation}
of the anticipated form (\ref{35}). The inclusion of interactions for a
singly quantized vortex in small to medium axisymmetric condensates with $%
Na/d_0\lesssim 1$ requires numerical analysis~\cite{Dalf96,Kono00}. Some
phases of rotating BEC in a spherically symmetric harmonic well in the
near-ideal-gas limit ($\xi \gtrsim d_0$) were considered by Wilkin and Gunn
\cite{Wilk00}. By exact calculation of wave functions and energies for small
number of particles, they show that the ground state in a rotating trap is
reminiscent of those found in the fractional quantum Hall effect. These
states include ``condensates" of composite bosons of the atoms attached to
an integer number of quanta of angular momenta, as well as the Laughlin and
Pfaffian~\cite{Moor91} states.

\begin{figure}
\bigskip
\centerline{\epsfxsize=0.31\textwidth\epsfysize=0.26\textwidth
\epsfbox{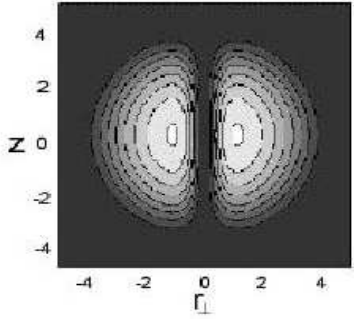}}

\vspace{0.4cm}

\caption{Contour plot in the $xz$ plane for a condensate with $10^4$ $^{87}$Rb
atoms containing a vortex along the
$z$ axis.   The trap is spherical and distances are in units of the
oscillator length $d= 0.791\ \mu$m. The interaction  parameter is
$Na/d=72.3$.  Luminosity is proportional to density, the white area being the
most dense.}
\noindent (Taken from Ref.~\cite{Dalf99}).
\label{fig2}
\end{figure}

In general, the density for a central vortex vanishes along the symmetry
axis, and the core radius increases away from the center of the trap,
yielding a toroidal condensate density (see Fig.~\ref{fig2}). This behavior
is particularly evident for a vortex in the TF limit $Na/d_0 \gg 1$, when

\begin{equation}
\label{37}n(r_{\perp },z)\approx n(0)\left( 1-\frac{\xi ^2}{r_{\perp }^2}-%
\frac{r_{\perp }^2}{R_{\perp }^2}-\frac{z^2}{R_z^2}\right) \Theta \left( 1-%
\frac{\xi ^2}{r_{\perp }^2}-\frac{r_{\perp }^2}{R_{\perp }^2}-\frac{z^2}{%
R_z^2}\right) .
\end{equation}
Here, the density differs from Eq.~(\ref{28}) for an axisymmetric
vortex-free TF condensate only because of the dimensionless centrifugal
barrier $\xi ^2/r_{\perp }^2$. This term forces the density to vanish within
a core whose characteristic radius is $\xi $ in the equatorial region $%
|z|\ll R_z$ and then flares out with increasing $|z|$. The TF separation of
length scales ensures that the vortex affects the density only the immediate
vicinity of the core~\cite{Dalf96,Sinh97,Rokh97}; this behavior can usually
be approximated with a short-distance cutoff. For such a quantized TF
vortex, the chemical potential $\mu _1$ differs from $\mu _0$ for a
vortex-free TF condensate by small fractional corrections of order $%
(d_0/R_0)^4\ln \,(R_0/d_0)$.

\subsubsection{Nonaxisymmetric trap}

If a singly quantized vortex is oriented along the $z$ axis of a
nonaxisymmetric trap ($R_x\neq R_y$) the condensate wave function is no
longer an eigenfunction of the angular momentum operator $L_z$. In the TF
limit near the trap center the phase $S$ of the condensate wave function has
the form \cite{Svid00B}:
\begin{equation}
S\approx \phi -\frac 14\left( \frac 1{R_x^2}-\frac 1{R_y^2}\right) r_{\perp
}^2\ln \left( \frac{r_{\perp }}{R_{\perp }}\right) \sin (2\phi ),
\end{equation}
and the condensate velocity is
\begin{equation}
{\bf v}\approx \frac \hbar M\left\{ \frac{\hat \phi }{r_{\perp }}-\frac 12%
\left( \frac 1{R_x^2}-\frac 1{R_y^2}\right) r_{\perp }\ln \left( \frac{%
r_{\perp }}{R_{\perp }}\right) \left[ \cos (2\phi )\hat \phi +\sin (2\phi )%
\hat r_{\perp }\right] \right\} ,
\end{equation}
where $R_{\perp }^2=2R_x^2R_y^2/(R_x^2+R_y^2)$. Near the vortex core the
condensate wave function and the condensate velocity possess cylindrical
symmetry, while far from the vortex core the condensate velocity adjusts to
the anisotropy of the trap and becomes asymmetric.

\subsection{Thermodynamic Critical Angular Velocity for Vortex Stability}

If the condensate is in rotational equilibrium at an angular velocity $%
\Omega $ around the $\hat z$ axis, the integrand of the GP Hamiltonian~(\ref
{6}) acquires an additional term $-\Psi^*\Omega L_z\Psi$~\cite{Lifs80},
where $L_z = xp_y-yp_x= -i\hbar\left(x\partial_y-y\partial_x\right)$ is the $%
z$ component of the angular-momentum operator. Thus the Hamiltonian $%
H^{\prime}$ in the rotating frame becomes

\begin{equation}
\label{38}H^{\prime}= H-\Omega L_z = \int dV\left[\Psi^*\left(T+V_{{\rm tr}%
}-\Omega L_z\right)\Psi + \case{1}{2} g|\Psi|^4\right],
\end{equation}
where the variables in the integrand are now those in the rotating frame.
Similarly, the GP equations (\ref{4}) and (\ref{5}) acquire an additional
term $-\Omega L_z\Psi$.

\subsubsection{Axisymmetric trap}

The situation is especially simple for an axisymmetric trap, where the
states can be labeled by the eigenvalues of $L_z$. For example, the energy
of a vortex-free condensate $E_0^{\prime }(\Omega )$ in the rotating frame
is numerically equal to the energy $E_0$ in the laboratory frame because the
corresponding angular momentum vanishes. A singly quantized vortex along the
trap axis has the total angular momentum $N\hbar $, so that the
corresponding energy of the system in the rotating frame is $E_1^{\prime
}(\Omega )=E_1-N\hbar \Omega $. The difference between these two energies is
the increased energy

\begin{equation}
\label{39}\Delta E^{\prime}(\Omega) =
E_1^{\prime}(\Omega)-E_0^{\prime}(\Omega) = E_1-E_0-N\hbar\Omega
\end{equation}
associated with the formation of the vortex at an angular velocity $\Omega$.
In the laboratory frame ($\Omega=0$), it is clear that $E_1>E_0$ because of
the added kinetic energy of the circulating flow. If the condensate is in
equilibrium in the rotating frame, however, $E_1^{\prime}(\Omega)$ decreases
linearly with increasing $\Omega$, and the relative energy of the vortex
vanishes at a ``thermodynamic'' critical angular velocity $\Omega_c$
determined by $\Delta E^{\prime}(\Omega_c)=0$. Equation (\ref{39})
immediately yields

\begin{equation}
\label{40}\Omega _c=\frac{E_1-E_0}{N\hbar },
\end{equation}
expressed solely in terms of energy of a condensate with and without the
vortex evaluated in the laboratory frame.

For a noninteracting trapped gas, the difference $E_1-E_0=
N\hbar\omega_\perp $ follows immediately from the excitation energy for the
singly quantized vortex in Eq.~(\ref{36}) relative to the stationary ground
state. In this noninteracting case, Eq.~(\ref{40}) gives $%
\Omega_c=\omega_\perp$, so that the noninteracting thermodynamic critical
angular velocity is just the radial trap frequency. Indeed, the same
critical angular velocity value also applies to a $q$-fold vortex in a
noninteracting condensate, because of the special form of the noninteracting
excitation energy $E_q-E_0 = N q\hbar\omega_\perp$ and the corresponding
angular momentum $N q\hbar$. Thus the noninteracting condensate becomes
massively degenerate as $\Omega\to\omega_\perp$\cite{Wilk98,Butt99}.
Physically, this degeneracy reflects the cancellation between the
centrifugal potential $-\frac{1}{2}M\Omega^2 r_\perp^2$ and the radial trap
potential $\case{1}{2}M\omega_\perp^2r_\perp^2$ as $\Omega\to\omega_\perp$.

\begin{figure}
\bigskip
\centerline{\epsfxsize=0.45\textwidth\epsfysize=0.40\textwidth
\epsfbox{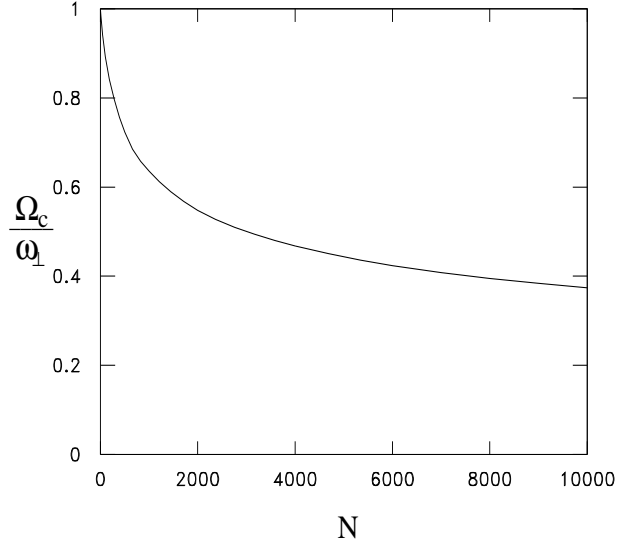}}

\vspace{0.3cm}

\caption{Thermodynamic critical angular velocity $\Omega_c$ for the formation
of a singly quantized vortex in a spherical trap with $d_0=0.791 \ \mu$m and
$N$ atoms of $^{87}$Rb.}
\noindent (Taken from Ref.~\cite{Dalf99}).
\label{fig3}
\end{figure}

Numerical analysis~\cite{Dalf96} for small and medium values of $Na/d_0$
shows that $\Omega_c/\omega_\perp$ decreases with increasing $N$, and a
perturbation analysis~\cite{Butt99,Linn99} confirms this behavior for a
weakly interacting system, with the analytical result $\Omega_c/\omega_\perp%
\approx 1-1/(2\sqrt {2\pi})\,(Na/d_z)$ for small values of the interaction
parameter $Na/d_z$. Figure~\ref{fig3} shows the behavior of $\Omega_c(N)$ in
a spherical trap, based on numerical analysis of the GP equation with
parameters relevant for $^{87}$Rb~\cite{Dalf96}.

In the strongly interacting (TF) limit, the chemical potential $\mu_1(N)$
for a condensate containing a singly quantized vortex can be evaluated with
Eq.~(\ref{37}), and the thermodynamic identity $\mu_1=\partial E_1/\partial
N $ then yields $E_1(N)$. Use of the corresponding expressions for the
vortex-free condensate gives the approximate expression~\cite
{Sinh97,Lund97,Cast99}

\begin{equation}
\label{41}\Omega _c\approx \frac 52\,\frac{\hbar ^2}{MR_{\perp }^2}\ln
\left( \frac{0.67R_{\perp }}\xi \right) \>\>\hbox {for a TF condensate}.
\end{equation}
This expression exceeds the usual estimate~\cite{Donn91} $\Omega _c\approx
(\hbar /MR_{\perp }^2)\ln (1.46R_{\perp }/\xi )$ for uniform superfluid in a
rotating cylinder of radius $R_{\perp }$ because the nonuniform density in
the trapped gas reduces the total angular momentum relative to that for a
uniform fluid. Equation (\ref{41}) has the equivalent form

\begin{equation}
\label{42}\frac{\Omega _c}{\omega _{\perp }}\approx \frac 52\,\frac{d_{\perp
}^2}{R_{\perp }^2}\ln \left( \frac{0.67R_{\perp }}\xi \right) .
\end{equation}
This ratio is small in the TF limit, because $d_{\perp }^2/R_{\perp }^2\sim
\xi /R_{\perp }\ll 1$. For an axisymmetric condensate with axial asymmetry $%
\lambda \equiv \omega _z/\omega _{\perp }$, the TF relation $d_{\perp
}^2/R_{\perp }^2=(d_{\perp }/15Na\lambda )^{2/5}$ shows how this ratio
scales with $N$ and $\lambda $.

In contrast to the case for repulsive interactions, the thermodynamic
critical angular velocity $\Omega_c$ for the vortex state with attractive
interactions {\em increases} as the number of atoms grows~\cite{Shi97,Dalf96}%
. Since $\Omega_c=\omega_\perp$ for a noninteracting condensate, $\Omega_c$
for a vortex in a condensate with attractive interactions necessarily
exceeds $\omega_\perp$. The stability or metastability of such a vortex is
unclear because $\Omega=\omega_\perp$ is also the limit of mechanical
stability for a noninteracting condensate.

\begin{figure}
\bigskip
\centerline{\epsfxsize=0.45\textwidth\epsfysize=0.40\textwidth
\epsfbox{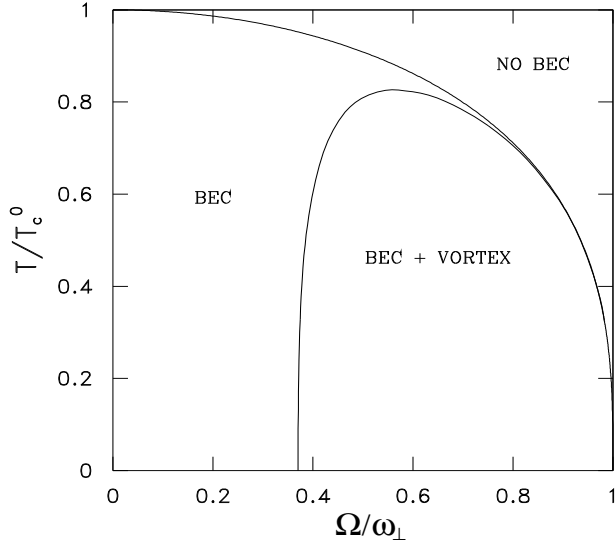}}

\vspace{0.3cm}

\caption{Phase diagram for vortices in a harmonically trapped Bose gas,
$N=10^4$, $a/d_{\perp}=7.36\times 10^{-3}$ and $\lambda=1$.}
\noindent (Taken from Ref.~\cite{Stri99}).
\label{fig4s}
\end{figure}

Approximately the same functional relationship holds between the
thermodynamic critical frequency $\Omega _c$ and the number of atoms in the
condensate $N_0$ \cite{Stri99} for nonzero temperatures. A new feature,
however, is that the number of atoms in the condensate becomes
temperature-dependent:
\begin{equation}
\label{t1}\frac{N_0}N=1-\left( \frac T{T_c}\right) ^3,
\end{equation}
where $T_c$ is the critical temperature of Bose condensation. If the trap
rotates at an angular velocity $\Omega $, the distribution function of the
thermal atoms changes due to the centrifugal force. As a result the critical
temperature decreases according to \cite{Stri99}
\begin{equation}
\label{t2}\frac{T_c(\Omega )}{T_c^0}=\left( 1-\frac{\Omega ^2}{\omega
_{\perp }^2}\right) ^{1/3},
\end{equation}
where $T_c^0$ is the critical temperature in the absence of rotation.
Equations (\ref{42})-(\ref{t2}) allows one to calculate the critical
temperature $T_v(\Omega )$, below which the vortex corresponds to a stable
configuration in a trap rotating with frequency $\Omega $. In Fig.~\ref
{fig4s} we show the critical curves $T_c(\Omega )$ and $T_v(\Omega )$. For
temperatures below $T_c(\Omega )$ the gas exhibits Bose-Einstein
condensation. Only for temperatures below $T_v(\Omega )$ does the vortex
state become thermodynamically stable. From Fig.~\ref{fig4s} one can see
that the critical temperature for the creation of stable vortices exhibits a
maximum as a function of $\Omega $.

\subsubsection{Nonaxisymmetric trap}

A rotating nonaxisymmetric trap introduces significant new physics, because
the moving walls induce an irrotational flow velocity even in the absence of
a vortex~\cite{Land59,Lamb45,Fett74,Fede99,Fede99a}. In the simplest case of
a classical uniform fluid in a rotating elliptical cylinder, the
instantaneous induced velocity potential in the laboratory frame is~\cite
{Land59,Lamb45}

\begin{equation}
\label{43}\Phi _{{\rm cl}}=\Omega \,xy\,\frac{A^2-B^2}{A^2+B^2},
\end{equation}
where $A$ and $B$ are the semi-axes of the elliptical cylinder. The induced
angular momentum and kinetic energy are reduced from the usual solid-body
values by the factor $I_0/I_{{\rm sb}}=\left[ (A^2-B^2)/(A^2+B^2)\right] ^2$%
. In the extreme case $B\ll A$, the moment of inertia can approach the
solid-body value, even though the flow is everywhere irrotational.

The thermodynamic critical angular velocity $\Omega _c$ for vortex creation
in the same uniform classical fluid depends on the asymmetry ratio $B/A$~%
\cite{Fett74}, and experiments on superfluid $^4$He confirm the theoretical
predictions in considerable detail~\cite{Deco75}. In the limit $B\ll A$, a
detailed calculation shows that $\Omega _c\approx (\hbar /2MB^2)\,\ln (B/\xi
)$; the appearance of $B$ here is readily understood from Feynman's picture
of a vortex occupying an area $\approx h/2M\Omega $ [compare Eq.~(\ref{34})]
and hence having to fit the area $\pi B^2$ fixed by the smaller lateral
dimension $B$.

The preceding analysis for an axisymmetric dilute trapped Bose gas can be
generalized to treat the TF limit in a totally anisotropic disk-shape
harmonic trap with $\omega _x^2+\omega _y^2\ll \omega _z^2$, starting from
Eq.~(\ref{38}) for the Hamiltonian in the rotating frame~\cite{Svid00}. The
presence of a vortex leaves the TF condensate density essentially unchanged,
and this Hamiltonian can serve as an energy functional to determine the
phase $S$ and hence the superfluid motion of the condensate. Since $R_x$, $%
R_y\gg R_z$, the curvature of the vortex is negligible. Hence we consider a
singly quantized straight vortex displaced laterally from the center of the
rotating trap to a transverse position ${\bf r}_0=(x_0,y_0)$ that serves as
a new origin of coordinates. The condensate wave function then has the form

\begin{equation}
\label{44}\Psi =|\Psi |\,e^{i\phi +iS_0},
\end{equation}
where $\phi $ in the first term is the polar angle around the vortex axis
and $S_0$ is a periodic function of $\phi $. Varying the Hamiltonian gives
an Euler-Lagrange equation for $S_0$, and it can be well approximated by the
solution for a vortex-free condensate, which is $M/\hbar $ times the
classical expression (\ref{43}) with $A$ and $B$ replaced by the TF radii $%
R_x$ and $R_y$ given in Eq.~(\ref{29}), and with $x$ and $y$ shifted to the
new origin.

As in Eq.~(\ref{39}) for an axisymmetric trap, $\Delta E^{\prime
}(x_0,y_0,\Omega )$ gives the increased energy in the rotating frame
associated with the presence of the straight vortex. A detailed calculation
with logarithmic accuracy yields~\cite{Svid00}

\begin{equation}
\label{45}\Delta E^{\prime}(x_0, y_0,\Omega)= \frac{8\pi}{3}\mu
R_z\,\xi^2n(0)\left(1-\zeta_0^2\right)^{3/2} \left[\ln\left(\frac{R_\perp}{%
\xi}\right)- \frac{8}{5}\,\frac{\mu\Omega}{\hbar\left(\omega_x^2+\omega_y^2%
\right)} \left(1-\zeta_0^2\right)\right],
\end{equation}
where $\zeta_0^2 \equiv x_0^2/R_x^2+y_0^2/R_y^2\le 1$ is a dimensionless
displacement of the vortex from the trap center. Here, the mean transverse
condensate radius $R_\perp$ is given by the arithmetic mean of the inverse
squared radii

\begin{equation}
\label{46}\frac{1}{R_\perp^2} = \frac{1}{2}\left(\frac{1}{R_x^2}+\frac{1}{%
R_y^2}\right)= \frac{M\left(\omega_x^2+\omega_y^2\right)}{4\mu}.
\end{equation}

\begin{figure}
\bigskip
\centerline{\epsfxsize=0.45\textwidth\epsfysize=0.45\textwidth
\epsfbox{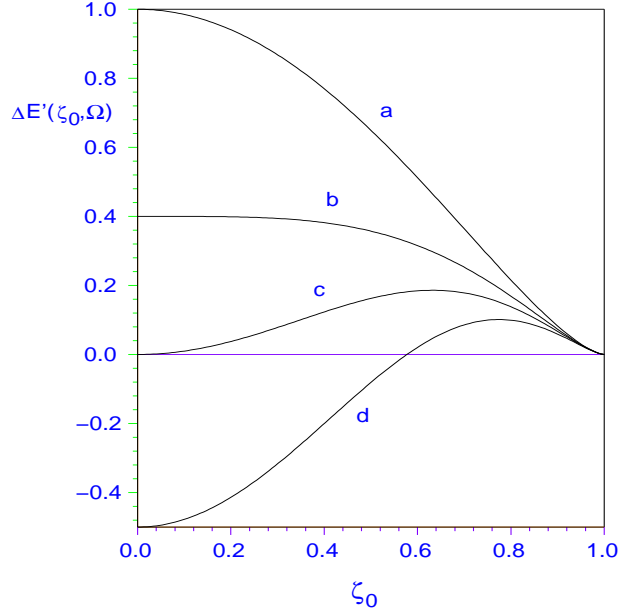}}

\vspace{0.3cm}

\caption{Energy (\ref{45}) [in units of $\Delta E'(0,0)$]
associated with a singly quantized straight vortex in a rotating asymmetric
trap in the TF limit as a function of a fractional vortex displacement
$\zeta_0$ from the symmetry axis. Different curves represent
different fixed values of the external angular velocity $\Omega$:  (a)
$\Omega=0$ (unstable); (b) $\Omega=\Omega_m$ [given in Eq.~(\ref{47})]
(onset of metastability at the origin); (c) $\Omega=\Omega_c$
[given in Eq.~(\ref{48})] (onset of stability at origin); (d)
$\Omega=\case{3}{2}\Omega_c$, where the thin barrier hinders vortex
tunneling from the surface.}
\label{fig4}
\end{figure}

Figure \ref{fig4} shows the behavior of $\Delta E^{\prime}(\zeta_0,\Omega)$
as a function of $\zeta_0$ for various fixed values of $\Omega$. Curve (a)
for $\Omega=0$ shows that the corresponding energy $\Delta
E^{\prime}(\zeta_0,\Omega=0)$ decreases monotonically with increasing $%
\zeta_0$, with negative curvature at $\zeta_0=0$. In the absence of
dissipation, energy is conserved and the vortex follows an elliptical
trajectory at fixed $\zeta_0$ around the center of the trap along a line $V_{%
{\rm tr}} = {\rm const}$. At low but finite temperature, however, the vortex
experiences weak dissipation; thus it slowly reduces its energy by moving
outward along curve (a), executing a spiral trajectory in the $xy$ plane.

With increasing fixed rotation speed $\Omega$, the function $\Delta
E^{\prime}(\zeta_0,\Omega)$ flattens. Curve (b) shows the special case of
zero curvature at $\zeta_0=0$. It corresponds to the rotation speed

\begin{equation}
\label{47}\Omega_m= \frac{3}{2}\frac{\hbar}{M R_\perp^2}\ln\left(\frac{%
R_\perp}{\xi}\right)\quad\hbox{for a disk-shape
condensate},
\end{equation}
at which angular velocity a central vortex first becomes metastable in a
large disk-shape condensate. For $\Omega<\Omega_m$, the negative local
curvature at $\zeta_0=0$ means that weak dissipation impels the vortex away
from the center. For $\Omega>\Omega_m$, however, the positive local
curvature means that weak dissipation now impels the vortex back toward the
center of the trap. In this regime, the central position is {\it locally\/}
stable; it is not {\em globally\/} stable, however, because $\Delta
E^{\prime}(0,\Omega)$ is positive for $\Omega\approx \Omega_m$.

Curve (c) shows that $\Delta E^{\prime}(0,\Omega_c)$ vanishes at the
thermodynamic critical angular velocity

\begin{equation}
\label{48}\Omega_c= \frac{5}{2}\frac{\hbar}{M R_\perp^2}\ln\left(\frac{%
R_\perp}{\xi}\right)= \frac{5}{3}\Omega_m\quad%
\hbox{for a disk-shape condensate}.
\end{equation}
As expected, this expression (\ref{48}) reduces to Eq.~(\ref{41}) in the
limit of an axisymmetric disk-shape condensate. For $\Omega>\Omega_c$, the
central vortex is both locally and globally stable relative to the
vortex-free state, and the energy barrier near the outer surface of the
condensate becomes progressively narrower. Curve (d) illustrates this
behavior for $\Omega=\case{3}{2}\Omega_c$. Eventually, the barrier thickness
becomes comparable with the thickness of the boundary layer within which the
TF approximation fails\cite{Dalf96a}, and it has been suggested that a
vortex might then nucleate spontaneously through a surface instability~\cite
{Dalf97,Fede99a,Isos99b}. For a two-dimensional condensate, a phase diagram
for different critical velocities of trap rotation {\it vs.} the system
parameter $an_z$ ($n_z$ is the area density) is given in~\cite{Isos99b}.

\subsection{Experimental Creation of a single vortex}

The first experimental detection of a vortex involved a nearly spherical $%
^{87}$Rb TF condensate containing two different internal (hyperfine)
components~\cite{Matt99} that tend to separate into immiscible phases. The
JILA group in Boulder created the vortex through a somewhat intricate
coherent process that controlled the interconversion between the two
components (discussed below in Sec.~\ref{mixt}). In essence, the coupled
two-component system acts like an $SU(2)$ spin-$\case{1}{2}$ system whose
topology differs from the usual $U(1)$ complex one-component order parameter
$\Psi$ familiar from superfluid $^4$He (and conventional BCS
superconductivity). Apart from the magnitude $|\Psi|$ that is fixed by the
temperature in a uniform system, a one-component order parameter has only
the phase that varies between 0 and $2\pi$. This topology is that of a
circle and yields quantized vorticity to ensure that the order parameter is
single-valued~\cite{Onsa49,Feyn55a}. In contrast, a two-component system has
two degrees of freedom in addition to the overall magnitude; its topology is
that of a sphere and does not require quantized vorticity. The qualitative
difference between the two cases can be understood as follows: the single
degree of freedom of the one-component order parameter is like a rubber band
wrapped around a cylinder, while the corresponding two degrees of freedom
for the two-component order parameter is like a rubber band around the
equator of a sphere. The former has a given winding number that can be
removed only be cutting it (ensuring the quantization of circulation),
whereas the latter can be removed simply by pulling it to one of the poles
(so that there is no quantization).

\begin{figure}
\bigskip
\centerline{\epsfxsize=0.55\textwidth\epsfysize=0.45\textwidth
\epsfbox{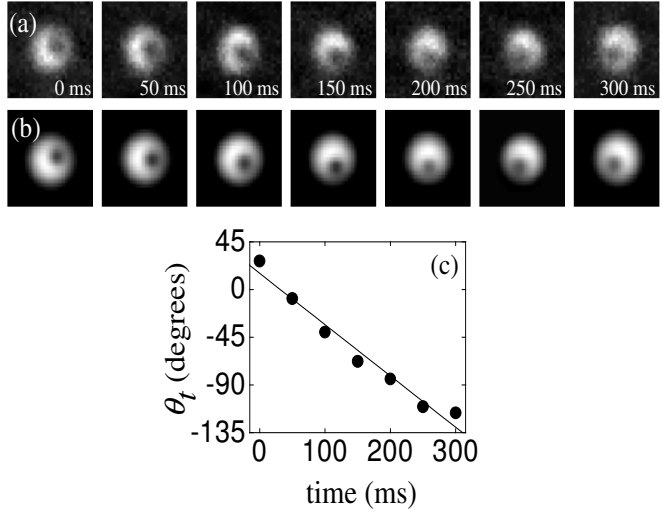}}

\vspace{-0.3cm}

\caption{(a) Successive images of a condensate with a vortex. The recorded
profile of each trapped condensate is fit with a smooth TF distribution (b).
The vortex core is the dark region within the bright condensate image. (c) The
azimuthal angle of the core is determined for each image, and plotted
{\it vs.} time held in the trap. A linear fit to the data gives a precession
frequency 1.3(1) Hz.}
\noindent (Taken from Ref.~\cite{Ande00}).
\label{figJILA1}
\end{figure}

\begin{figure}
\bigskip
\centerline{\epsfxsize=0.45\textwidth\epsfysize=0.35\textwidth
\epsfbox{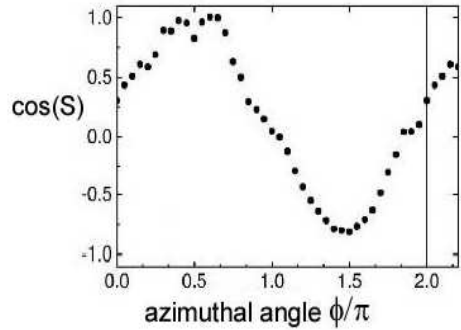}}


\caption{Cosine of the phase  around the vortex, showing the
sinusoidal variation expected for the azimuthal angle.} \noindent (Taken from
Ref.~\cite{Matt99}).
\label{figJILA2}
\end{figure}

The JILA group was able to spin up the condensate by coupling the two
components. They then turned off the coupling, leaving the system with a
residual trapped quantized vortex consisting of one circulating component
surrounding a nonrotating core of the other component, whose size is
determined by the relative fraction of the two components. By selective
tuning, they can image either component nondestructively~\cite{Ande00}; Fig.~%
\ref{figJILA1} shows the precession of the filled vortex core around the
trap center. In addition, an interference procedure allowed them to map the
variation of the cosine of the phase around the vortex, clearly showing the
expected sinusoidal variation (Fig.~\ref{figJILA2}).

The JILA group has also been able to remove the component filling the core,
in which case they obtain a single-component vortex~\cite{Ande00}. This
one-component vortex has a small core size and can only be imaged by
expanding both the condensate and the core, which becomes visible through
its reduced density~\cite{Lund98,Cast99}. They first make an image of the
two-component vortex, next remove the component filling the core, and then
make an image of the one-component vortex after a variable time delay. In
this way, they can measure the precession rate of the one-component
empty-core vortex and compare it with theoretical predictions~\cite{Fede00a}%
. The data show no tendency for the core to spiral outward, suggesting that
the thermal damping is negligible on the time scale of $\sim 1$ s.

\begin{figure}
\bigskip
\centerline{\epsfxsize=0.45\textwidth\epsfysize=0.17\textwidth
\epsfbox{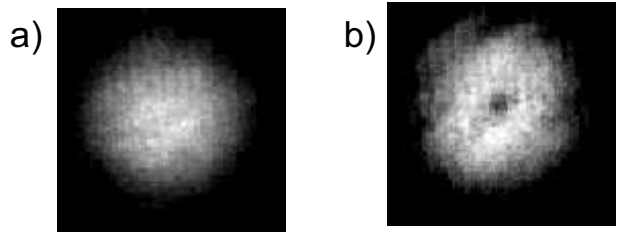}}

\vspace{0.3cm}

\caption{Optical thickness of the expanded clouds in the transverse
direction showing the difference between the states (a) without and (b) with a
vortex.}
\noindent (Taken from Ref.~\cite{Madi99}).
\label{figENS1}
\end{figure}

Separately, the ENS group in Paris observed the formation of one and more
vortices in a single-component $^{87}$Rb elongated cigar-shape TF condensate
with a weak nonaxisymmetric deformation that rotates about its long axis~%
\cite{Madi99,Madi00a,Chev00}. In essence, a static cylindrically symmetric
magnetic trap is augmented by a nonaxisymmetric attractive dipole potential
created by a stirring laser beam. The combined potential produces a
cigar-shape harmonic trap with a slightly anisotropic transverse profile.
The transverse anisotropy rotates slowly at a rate $\Omega \lesssim 200$ Hz.
In the first experiments~\cite{Madi99}, the trap was rotated in the normal
state and then cooled, with the clear signal of the vortex shown in Fig.~\ref
{figENS1} (the trap was turned off, allowing the atomic cloud to expand so
that the vortex core becomes visible). This order was reversed (cool first,
then rotate) in a later series of runs~\cite{Chev00}. In both cases, the
observed critical angular velocity $\sim 0.7\omega _{\perp }$ for creating
the first (central) vortex was roughly 70\% higher than the predicted
thermodynamic value $\Omega _c$ in Eq.~(\ref{41}). These observations agree
qualitatively with the suggestion that a surface instability might nucleate
a vortex~\cite{Dalf97,Fede99a,Isos99b}. Alternative explanations of this
discrepancy involve the bending modes of the vortex (discussed below in
Sec.~IV.D.4 and V.D.2).

\subsection{Vortex Arrays}

Under appropriate stabilization conditions, such as steady applied rotation,
vortices can form a regular array. In a rotating uniform superfluid, the
quantized vortex lines parallel to the axis of rotation form a lattice. This
lattice rotates as a whole around the axis of rotation, thus simulating
rigid rotation~\cite{Andr61}. At nonzero temperature, dissipative mutual
friction from the normal component ensures that the array rotates with the
same angular velocity as the container. Early experiments on rotating
superfluid $^4$He~\cite{Will74,Yarm79,Yarm82} provided memorable
``photographs'' of vortex lines and arrays with relatively small numbers of
vortices, in qualitative agreement with analytical~\cite{Tkac66,Hess67} and
numerical~\cite{Stau68,Camp79} predictions. A triangular array is favored
for vortices near the rotation axis of rapidly rotating vessels of
superfluid helium \cite{Tkac66}. Vortex lattices also occur in the neutron
superfluid in rotating neutron stars \cite{Sedr91}.

\begin{figure}
\bigskip
\centerline{\epsfxsize=0.45\textwidth\epsfysize=0.42\textwidth
\epsfbox{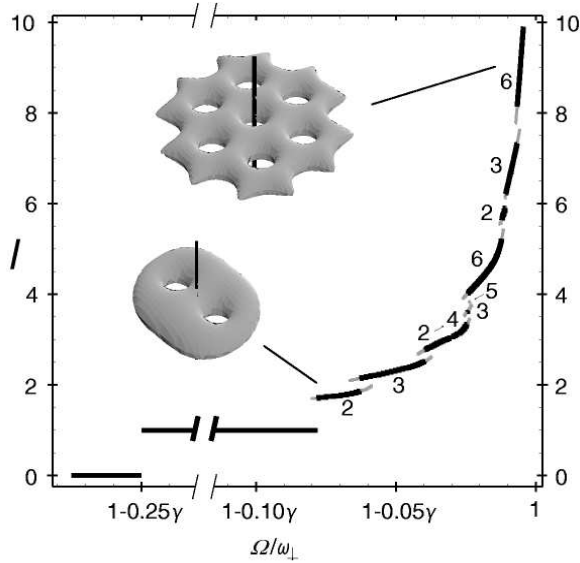}}

\vspace{0.3cm}

\caption{Dimensionless angular momentum $l$ per particle  {\it vs.}
dimensionless angular velocity $\Omega/\omega_\perp$. In the figure
$\gamma=(2/\pi)^{1/2}aN/d_z$.
Black lines show stable states and gray lines show metastable states.  There
are no stable or metastable states in the forbidden ranges $l=0$-1 and
$l=1$-1.70. The rotational symmetry of each branch is indicated. The total
angular momentum diverges as $\Omega$ approaches the maximum angular velocity
$\omega_\perp$.
Three-dimensional plots of constant density show states with
two-fold and six-fold symmetry. Reprinted by permission from Nature
{\bf 397}, 327, (1999), \copyright 1999
Macmillan Magazines Ltd.}
\label{fig8}
\end{figure}

Even before the recent observation of vortex arrays in an elongated rotating
trapped condensate~\cite{Madi99,Madi00a}, several theoretical groups had
analyzed many of the expected properties. In a weakly interacting
(near-ideal) axisymmetric condensate, the thermodynamic critical angular
velocity $\Omega _c$ for the appearance of the first vortex is already close
to the radial trap frequency $\omega _{\perp }$, so that the creation of
additional vortices involves many states $\phi _m({\bf r}_{\perp })\propto
e^{im\phi }\,r_{\perp }^m\,\exp (-\frac 12r_{\perp }^2/d_{\perp }^2)$ with
low energy $m\hbar (\omega _{\perp }-\Omega )$ per particle in the rotating
frame. Butts and Rokhsar~\cite{Butt99} used a linear combination of these
nearly degenerate states as a variational condensate wave function,
minimizing the total energy in the laboratory frame $E_{{\rm lab}}$ subject
to the condition of fixed number $N$ of particles and fixed angular momentum
$l$ per particle. As expected from the theoretical and experimental results
for liquid helium, the system undergoes a sequence of transitions between
states that break rotational symmetry. Several of these have $p$-fold
symmetry where $p$ is a small integer. Each vortex represents a node in the
condensate wave function, and their positions can vary with the specified
angular momentum. Indeed, as $l$ increases from 0 to 1, the first vortex
moves continuously from the edge of the condensate to the center. For larger
number of vortices, the centrifugal forces tend to flatten and expand the
condensate in the radial direction. In this approach of keeping $l$ fixed,
the angular velocity follows from the relation $\hbar \Omega =\partial E_{%
{\rm lab}}/\partial l$. Figure~\ref{fig8} shows the angular momentum versus
the angular velocity for the first several states. Reference~\cite{Kavo00}
has carried out more detailed studies of the states for relatively small
values of the angular momentum per particle $l\lesssim 2$.

These analyses work at fixed angular momentum $Nl$, in which case the
angular velocity $\Omega $ must be determined from the resulting $E_{{\rm lab%
}}(l)$. In contrast, the ENS experiments fix $\Omega $ (as do experiments on
superfluid helium) and then measure $L_z$ from the splitting of the
quadrupole modes~\cite{Chev00} (see Sec.~IV.D.3). The JILA group~\cite
{Halj00} also uses this technique to detect the presence of a vortex in a
nonrotating condensate. The transition from fixed $L_z$ to fixed $\Omega $
can be considered a Legendre transformation to the Hamiltonian (\ref{38}) in
the rotating frame. Even though it is easier to work at fixed $\Omega $
(because there is no constraint of fixed $L_z/N=l$), no such analysis has
yet been carried out in the weak-coupling limit.

In the strong-coupling (TF) limit, Castin and Dum~\cite{Cast99} have
performed extensive numerical studies of equilibrium vortex arrays in two
and three dimensions, based on the Hamiltonian in the rotating frame (thus
working at fixed $\Omega$). They also propose an intuitive variational
calculation based on a factorization approximation that is very similar to
Eq.~(\ref{26}), apart from a different analytic form of the radial function~%
\cite{Fett65,Fett66}.

The nucleation of vortices and the resulting structures of vortex arrays in
zero temperature BECs are also investigated numerically by Feder, Clark and
Schneider~\cite{Fede99a}. In their simulations, vortices are generated by
rotating a three-dimensional, nonaxisymmetric harmonic trap. Vortices first
appear at a rotation frequency significantly larger than the critical
frequency for vortex stabilization. At higher frequencies, the trap geometry
strongly influences the structure of the vortex arrays, but the lattices
approach triangular arrays at large vortex densities.

\begin{figure}
\bigskip
\centerline{\epsfxsize=0.45\textwidth\epsfysize=0.15\textwidth
\epsfbox{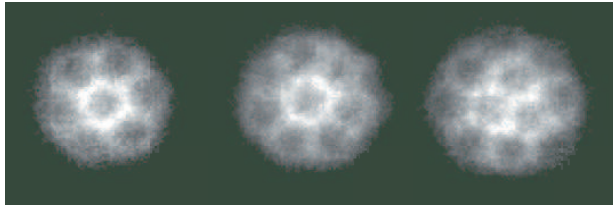}}

\vspace{0.3cm}

\caption{Arrays of vortices in a  Bose-Einstein condensate stirred  with a
laser beam.}
\noindent (Taken from Ref.~\cite{Madi00a}).
\label{figENS2}
\end{figure}

The ENS experiments~\cite{Madi99,Madi00a} have produced remarkable images of
vortex arrays. Figure~\ref{figENS2} shows three different arrays with up to
11 vortices (obtained after an expansion of 27 ms). The initial condensate
is very elongated (along with the vortices), so that the radial expansion
predominates once the trap is turned off. As a result, the expanded
condensate acquires a pancake shape similar to that in Fig.~\ref{fig8}.

\section{Bogoliubov equations: stability of small-amplitude perturbations}

This section considers only the behavior of a dilute one-component Bose gas,
for which the analysis of the eigenfrequencies is particularly direct. In
the more general case of two interpenetrating species, even a uniform system
can have imaginary frequencies for sufficiently strong interspecies
repulsion~\cite{Nepo74,Cols78}; this dynamical instability signals the onset
of phase separation.

\subsection{General features for nonuniform condensate}

The special character of an elementary excitation in a dilute Bose gas
largely arises from the role of the Bose condensate that acts as a particle
reservoir. This situation is especially familiar in the uniform system,
where an elementary excitation with wave vector ${\bf k}$ can arise from the
interacting ground state $\Psi_0$ either through the creation operator $a_{%
{\bf k}}^\dagger$ or through the annihilation operator $a_{-{\bf k}}$ (in
the thermodynamic limit, these two states $a_{{\bf k}}^\dagger\Psi_0$ and $%
a_{-{\bf k}}\Psi_0$ differ only by a normalization factor). The true excited
eigenstates are linear combinations of the two states, and the corresponding
operator for the Bogoliubov quasiparticle is a weighted linear combination~%
\cite{Bogo47,Fett71,Fett99}

\begin{equation}
\label{49}\alpha_{{\bf k}}^\dagger = u_k\,a_{{\bf k}}^\dagger +v_k \,a_{-%
{\bf k}},
\end{equation}
where $u_k$ and $v_k$ are the (real) Bogoliubov coherence factors. This
linear transformation (\ref{49}) is canonical if the quasiparticle operators
also obey Bose-Einstein commutation relations, which readily yields the
condition

\begin{equation}
\label{50}u_k^2-v_k^2 =1,\quad \hbox{for all ${\bf k}\neq 0$}.
\end{equation}

More generally, the second-quantized Bose field operator $\hat\psi $ in Eq.~(%
\ref{2}) can be written as $\hat\psi({\bf r})\approx \Psi({\bf r})+\hat\phi(%
{\bf r})$, where $\hat\phi$ is a small deviation operator from the
macroscopic condensate wave function $\Psi$. These deviation operators obey
the approximate Bose-Einstein commutation relations

\begin{equation}
\label{51}\left[\hat\phi({\bf r}),\hat\phi^\dagger({\bf r}^{\prime})\right]
\approx \delta\!\left({\bf r-r}^{\prime}\right),\quad\left[\hat\phi({\bf r}),%
\hat\phi({\bf r}^{\prime})\right]= \left[\hat\phi^\dagger({\bf r}),\hat\phi%
^\dagger({\bf r}^{\prime})\right]\approx 0.
\end{equation}

Since $\hat\psi({\bf r})$ does not conserve particle number, it is
convenient to use a grand canonical ensemble, with the new Hamiltonian
operator $\hat K=\hat H-\mu\hat N$ instead of the Hamiltonian (\ref{1}). To
leading (second) order in the small deviations, the perturbation in $\hat K$
contains not only the usual ``diagonal'' terms involving $\hat\phi^\dagger%
\hat\phi$, but also ``off-diagonal'' terms proportional to $\hat\phi\hat\phi$
and $\hat\phi^\dagger\hat\phi^\dagger$. Consequently, the resulting
Heisenberg operators $\hat\phi $ and $\hat\phi^\dagger$ obey coupled linear
equations of motion (it is here that the role of the condensate is evident,
for this coupling vanishes if $\Psi$ vanishes). Pitaevskii~\cite{Pita61}
developed this approach for the particular case of a vortex line in
unbounded condensate, and the formalism was subsequently extended to include
a general nonuniform condensate~\cite{Fett72,Fett96}.

In direct analogy to the Bogoliubov transformation for the uniform system,
assume the existence of a linear transformation to quasiparticle operators $%
\alpha_j$ and $\alpha_j^\dagger$ for a set of normal modes labeled by $j$

\begin{mathletters}
\label{52}
\begin{equation}
\hat\phi({\bf r},t) = {\sum_j}'\left[u_j({\bf r})\alpha_j(t)-v_j^*({\bf
r})\alpha_j^\dagger(t)\right],
\end{equation}
\begin{equation}\hat\phi^\dagger ({\bf r},t) =
{\sum_j}'\left[u_j^*({\bf r})\alpha_j^\dagger (t)-v_j({\bf
r})\alpha_j(t)\right],
\end{equation}
\end{mathletters}
where the primed sum means to omit the condensate mode. Here, the
quasiparticle operators $\alpha_j$ and $\alpha_k^\dagger$ obey Bose-Einstein
commutation relations $\left[\alpha_j,\alpha_k^\dagger\right] = \delta_{jk}$
and have simple harmonic time dependences $\alpha_j(t) =
\alpha_j\,\exp\left(-i E_j t/\hbar\right)$ and $\alpha_j^\dagger(t) =
\alpha_j^\dagger \,\exp\left(i E_j t/\hbar\right)$. Comparison with the
equations of motion for $\hat\phi$ and $\hat\phi^\dagger$ shows that the
corresponding spatial amplitudes obey a set of coupled linear ``Bogoliubov
equations''

\begin{mathletters}
\label{53}
\begin{equation}
L u_j-g\left(\Psi\right)^2 v_j = E_ju_j,
\end{equation}
\begin{equation}
L v_j-g\left(\Psi^*\right)^2 u_j = -E_j v_j,
\end{equation}
\end{mathletters}
where

\begin{equation}
\label{54}L = T+V_{{\rm tr}}-\mu + 2g|\Psi|^2
\end{equation}
is a Hermitian operator.

Straightforward manipulations with the Bogoliubov equations show that $E_j
\int dV\left(|u_j|^2-|v_j|^2\right)$ is real. If the integral $\int
dV\left(|u_j|^2-|v_j|^2\right)$ is nonzero, then $E_j$ itself is real. Like
Eq.~(\ref{50}) for a uniform condensate, the Bose-Einstein commutation
relations (\ref{51}) for the deviations from the nonuniform condensate can
be shown to imply the following {\em positive\/} normalization~\cite{Fett72}

\begin{equation}
\label{55}\int dV\left( |u_j|^2-|v_j|^2\right) =1.
\end{equation}
For each solution $u_j,v_j$ with eigenvalue $E_j$ and positive
normalization, the Bogoliubov equations always have a second solution $%
v_j^{*},u_j^{*}$ with eigenvalue $-E_j$ and {\em negative} normalization.
The only exception to the requirement of real eigenvalues arises for
zero-norm solutions with $\int dV\left( |u_j|^2-|v_j|^2\right) =0$. In this
case the character of the eigenvalue requires additional analysis. Numerical
investigations~\cite{Pu99} of vortices in nonuniform trapped condensates
have reported imaginary and/or complex eigenfrequencies for doubly quantized
vortices but only real eigenfrequencies for singly quantized vortices.
Specifically, for a repulsive interparticle interaction, Pu {\it et al.}~%
\cite{Pu99} found that singly quantized vortices are always intrinsically
stable; in contrast, multiply quantized vortices have alternating stable and
unstable regions with complex excitation energy as the interaction parameter
$Na/d$ increases. The most unstable vortex state decays after several
periods of the harmonic trapping potential. In the case of multiply
quantized vortices ($q>1$), the vortex core contains localized quasiparticle
bound states with small exponential tails; these modes have complex
frequencies and are responsible for splitting the multicharged core~\cite
{Aran96}. For an attractive interaction, stable vortices exist only for the
singly quantized case in the weak-interaction regime; a multiply quantized
vortex state is always unstable. Similar imaginary and complex solutions
have been found for dark solitons~\cite{Mury99,Fedi99,Fede00}. For
additional results on complex eigenfrequencies, see Ref.~\cite{Garc99} and
the Appendix of Ref.~\cite{Gara00}.

In terms of the quasiparticle operators, the approximate perturbation
Hamiltonian operator takes the simple intuitive form

\begin{equation}
\label{56}\hat K^{\prime}\approx {\sum_j}^{\prime}\,
E_j\,\alpha_j^\dagger\alpha_j,
\end{equation}
apart from a constant ground-state contribution of all the normal modes.
Here, the sum is over all the states with positive normalization, and it is
clear that the sign of the energy eigenvalues $E_j$ is crucial for the
stability. If one or more of the eigenvalues is negative, the Hamiltonian is
no longer positive definite, and the system can lower its energy by creating
quasiparticles in the unstable modes.

The present derivation of the Bogoliubov equations and their properties
emphasizes the quantum-mechanical basis for the positive normalization
condition (\ref{55}) and the sign of the eigenvalues. It is worth noting an
alternative purely ``classical'' treatment~\cite{Edwa96,Dalf99} based
directly on small perturbations of the time-dependent GP equation (\ref{4})
around the static condensate $\Psi({\bf r})$. The solution is assumed to
have the form

\begin{equation}
\label{57}\Psi({\bf r},t) = e^{-i\mu t/\hbar}\left[\Psi({\bf r})+ u({\bf r})
e^{-i\omega t}-v^*({\bf r}) e^{i\omega t}\right],
\end{equation}
and the appropriate eigenvalue equations then reproduce Eqs.~(\ref{53}).

\subsection{Uniform condensate}

For a uniform condensate, the solutions of Eq.~(\ref{53}) are plane waves,
and the corresponding energy is the celebrated Bogoliubov spectrum~\cite
{Bogo47}

\begin{equation}
\label{58}E_k=
\sqrt{gn\hbar ^2k^2/M+\left( \hbar ^2k^2/2M\right) ^2},
\end{equation}
where ${\bf k}$ is the wave vector of the excitation and $n$ is the
condensate density. For long wavelengths $k\xi \ll 1$, Eq.~(\ref{58})
reduces to a linear phonon spectrum $E_k\approx \hbar sk$ with the speed of
compressional sound $s=\sqrt{gn/M}$ given by Eq.~(\ref{10}). In the opposite
limit $k\xi \gg 1$, the spectrum reduces to the free-particle form plus a
mean-field Hartree shift from the interaction with the background condensate
$E_k\approx (\hbar ^2k^2/2M)+gn$.

To understand the importance of the sign of the eigenfrequency, it is
instructive to consider the case of a condensate that moves uniformly with
velocity ${\bf v}_0$. As noted in connection with Eq.~(\ref{24}), the
condensate wave function is $\Psi({\bf r}) = \sqrt n \,e^{i{\bf q\cdot r}}$,
where ${\bf q} = M{\bf v}_0/\hbar$ and the chemical potential becomes $\mu =
\frac{1}{2}Mv_0^2 + gn$. The Bogoliubov amplitudes for an excitation with
wave vector ${\bf k}$ relative to the moving condensate have the form

\begin{equation}
\label{59}\pmatrix{u_{\bf k}({\bf r}) \cr v_{\bf k}({\bf r}) }=
\pmatrix{e^{i\bf
q\cdot r}\,u_k e^{i\bf k\cdot r}\cr e^{-i\bf
q\cdot r}\, v_k e^{i\bf k\cdot r}},
\end{equation}
where the different signs $\pm i{\bf q\cdot r}$ arise from the different
phases $\pm i 2{\bf q\cdot r}$ in the off-diagonal coupling terms in the
Bogoliubov equations (\ref{53}). The solution with positive norm has the
eigenvalue

\begin{equation}
\label{60}E_{{\bf k}}({\bf v}_0) = \hbar {\bf k\cdot v}_0 +E_k,
\end{equation}
as expected from general considerations~\cite{Land41,Lifs80a}. In the
long-wavelength limit, this excitation energy reduces to $E_{{\bf k}}({\bf v}%
_0) \approx \hbar k (v_0\cos\theta + s)$, where $\theta$ is the angle
between ${\bf k}$ and ${\bf v}_0$. For $v_0<s$, the quasiparticle energy is
positive for all angles $\theta$, but for $v_0>s$, the quasiparticle energy
becomes negative for certain directions, indicating the onset of an
instability. This behavior simply reflects the well-known Landau critical
velocity for the onset of dissipation, associated with the emission of
quasiparticles. It has many analogies with supersonic flow in classical
compressible fluids~\cite{Land87} and Cherenkov radiation of photons in a
dielectric medium~\cite{Land84,Jack98}. For $v_0>s$, the GP description
becomes incomplete because the excitation of quasiparticles means that the
noncondensate is no longer negligible.

\subsection{Quantum-hydrodynamic description of small-amplitude normal modes}

The quantum-hydrodynamic forms (\ref{16}) and (\ref{17}) of the
time-dependent GP equation provide a convenient alternative basis for
studying the small-amplitude normal modes. The small perturbations in the
density $n^{\prime}e^{-i\omega t}$ and the velocity potential $%
\Phi^{\prime}e^{-i\omega t}$ obey coupled linear equations~\cite
{Fett96,Fett98a,Svid98} that reduce to Eq.~(\ref{33}) in the TF limit for a
static condensate~\cite{Stri96}. A comparison with Eqs.~(\ref{52}) shows
that the quantum-hydrodynamic amplitudes

\begin{mathletters}
\label{61}
\begin{equation}
n_j' = \Psi^*\,u_j-\Psi\,v_j= |\Psi|\left(e^{-i S}u_j-e^{i S}v_j\right),
\end{equation}
\begin{equation}
\Phi_j' = \frac{\hbar}{2 M i \,|\Psi|^2}\left(\Psi^*\,u_j+\Psi\,v_j\right)=
\frac{\hbar}{2 M i \,|\Psi|}\left(e^{-i S}u_j+e^{i S}v_j\right)
\end{equation}
\end{mathletters}
are simply linear combinations of the Bogoliubov amplitudes $u_j$ and $v_j$
in the presence of the given condensate solution $\Psi = e^{iS}|\Psi|$. The
positive normalization condition (\ref{55}) yields the equivalent
quantum-hydrodynamic form

\begin{equation}
\label{62}\int dV\,i\left({n_j^{\prime}}^*\Phi_j^{\prime}-{\Phi_j^{\prime}}%
^*n_j^{\prime}\right) = \frac{\hbar}{M}.
\end{equation}
For many purposes, the quantum-hydrodynamic modes provide a clearer picture
of the dynamical motion.

\subsection{Singly quantized vortex in axisymmetric trap}

Early numerical studies for small and medium values of the interaction
parameter $Na/d_0\lesssim ~1$ examined the small-amplitude excitations of a
condensate with a singly quantized vortex~\cite{Dodd97}. In particular, the
spectrum contained an ``anomalous'' mode with a {\it negative\/} excitation
frequency and {\it positive\/} normalization associated with a large
Bogoliubov amplitude $u$ localized in the vortex core (see also relevant
comments in Ref.~\cite{Rokh97} concerning the relationship between the sign
of the normalization and the sign of the eigenfrequency). The anomalous mode
corresponds to a precession of the vortex line around $z$ axis. As seen from
the general discussion of the Bogoliubov equations, this anomalous mode
indicates the presence of an instability.

Since the condensate wave function has an explicit phase $\Psi({\bf r}%
)=e^{i\phi}\,|\Psi(r_\perp,z)|$, the Bogoliubov amplitudes for an excitation
with angular momentum $m\hbar$ relative to the vortex condensate take the
form

\begin{equation}
\label{63}\pmatrix{u_m({\bf r})  \cr v_m({\bf r}) }=
\pmatrix{e^{i\phi}
\,e^{im\phi}\tilde
u_m(r_\perp,z)\cr e^{-i\phi}
\,e^{im\phi}\tilde
v_m(r_\perp,z)}.
\end{equation}
analogous to those in Eq.~(\ref{59}) for a condensate in uniform motion.
Here, the azimuthal quantum number $m$ characterizes the associated density
and velocity deformations of the vortex proportional to $e^{im\phi}$ [for
example, $n_m^{\prime}= |\Psi|\left(\tilde u_m-\tilde v_m\right)e^{im\phi}$,
as is clear from Eqs.~(\ref{61})]. The numerical studies~\cite{Dodd97} found
that the anomalous mode has an azimuthal quantum number $m_a=-1$. Its
frequency $\omega_a$ is negative throughout the relevant range of $%
Na/d_0\lesssim 1$; in the noninteracting limit, $\omega_a$ approaches $%
-\omega_\perp$, and $\omega_a$ increases toward $0$ from below with
increasing $Na/d_0$.

To understand the particular value $m_a=-1$, it is helpful to recall the
noninteracting limit, when the negative anomalous mode for the vortex
condensate signals the instability associated with Bose condensation in the
first excited harmonic-oscillator state with excitation energy $\hbar
\omega_\perp$ and unit angular momentum. A particle in the condensate can
make a transition from the vortex state back to the true harmonic-oscillator
ground state, with a change in frequency $-\omega_\perp$ and a change in
angular momentum quantum number $-1$. More generally, the density
perturbation $n_a^{\prime}$ for the anomalous mode with negative frequency $%
-|\omega_a|$ is proportional to $\exp\left[i\left(|\omega_a|t-\phi\right)%
\right]$ and hence precesses in the {\it positive\/} sense (namely
counterclockwise) at the frequency $|\omega_a|$. Thus the anomalous mode
describes the JILA observations of the precession frequency of a
one-component vortex~\cite{Ande00,Fede00a}.

\subsubsection{Near-ideal regime}

An explicit perturbation analysis~\cite{Fett98,Linn99} of the GP equation
for the condensate wave function in the weakly interacting limit found the
thermodynamic critical angular velocity

\begin{equation}
\label{64}\frac{\Omega_c}{\omega_\perp} = 1-\frac{1}{\sqrt{8\pi}}\frac{Na}{%
d_z}+ \Omega_c^{(2)}(\lambda)\left(\frac{Na}{d_z}\right)^{\!\!2}+\cdots\ ,
\end{equation}
where the second-order correction depends explicitly on the axial asymmetry $%
\lambda = \omega_z/\omega_\perp$. Similarly, a perturbation expansion of the
Bogoliubov equations in the weak-coupling limit verified the numerical
analysis and found the explicit expression for the frequency of the
anomalous mode

\begin{equation}
\label{65}\frac{\omega_a}{\omega_\perp} = -1+\frac{1}{\sqrt{8\pi}}\frac{Na}{%
d_z}+ \omega_a^{(2)}(\lambda)\left(\frac{Na}{d_z}\right)^{\!\!2}+\cdots\ .
\end{equation}
It is evident that $\Omega_c+\omega_a$ vanishes through first order, and the
detailed analysis shows that the second-order contribution to the sum is
positive for all values of the axial asymmetry parameter $\lambda$.

The physics of the anomalous mode can be clarified by considering an
axisymmetric condensate in rotational equilibrium at an angular velocity $%
\Omega$ around the $\hat z$ axis. In the rotating frame, the Hamiltonian
becomes $H-\Omega L_z$, and the Bogoliubov amplitudes have frequencies $%
\omega_j(\Omega) = \omega_j -m_j\Omega$, where $\omega_j$ is the frequency
in the nonrotating frame and $m_j$ is the azimuthal quantum number [see Eq.~(%
\ref{63})]. For the anomalous mode with $m_a=-1$, the resulting frequency in
the rotating frame is

\begin{equation}
\label{66}\omega _a(\Omega )=\omega _a+\Omega ,
\end{equation}
which is directly analogous to Eq.~(\ref{60}) for uniform translation. Since
$\omega _a$ is negative, the anomalous frequency in a rotating frame
increases linearly toward zero with increasing $\Omega $; in particular, $%
\omega _a(\Omega )$ vanishes at a characteristic rotation frequency

\begin{equation}
\label{67}\Omega^*= -\omega_a = |\omega_a|
\end{equation}
that signals the onset of the regime $\Omega\ge \Omega^*$ for which the
singly quantized vortex becomes locally stable. Equation~(\ref{65}) gives an
explicit expression for $\Omega^*$ in the weak-coupling limit, and detailed
comparison with Eq.~(\ref{64}) indicates that $\Omega^*<\Omega_c$ for any
axial asymmetry $\lambda$ (but only because of the second-order
contributions). It is natural to identify $\Omega^*$ with the angular
velocity for the onset of local stability with respect to small
perturbations; this quantity was denoted $\Omega_m$ in connection with the
equilibrium energy in the TF limit (see Fig.~\ref{fig4}).

\subsubsection{Thomas-Fermi regime for disk-shape trap}

The anomalous negative-frequency mode exists only because the condensate
contains a vortex. Hence it cannot be analyzed by treating the vortex itself
as a perturbation. In the TF limit, however, it is possible to use Gross's
and Pitaevskii's~\cite{Gros61,Pita61} solution (\ref{19}) for a vortex in a
laterally unbounded fluid as the basis for a perturbation expansion. A
detailed analysis of the Bogoliubov equations for an axisymmetric rotating
flattened trap in the TF limit yields the explicit expression for the
anomalous mode~\cite{Svid98a}

\begin{equation}
\label{68}\omega_a(\Omega) =\Omega-\frac{3\hbar\omega_\perp^2}{4\mu}\ln\left(%
\frac{R_\perp}{\xi}\right)= \Omega-\frac{3}{2}\frac{\hbar}{MR_\perp^2}%
\ln\left(\frac{R_\perp}{\xi}\right).
\end{equation}
As in Eq.~(\ref{67}) for the weak-coupling limit, Eq.~(\ref{68}) yields

\begin{equation}
\label{69}\Omega^* = \frac{3}{2}\frac{\hbar}{MR_\perp^2}\ln\left(\frac{
R_\perp}{\xi}\right) =\Omega_m=\frac{3}{5}\Omega_c,
\end{equation}
where the last two equalities follow from (\ref{47}) and (\ref{48}). This
relation further supports the identification of $\Omega^*$ with the
metastable rotation frequency $\Omega_m$ associated with local stability of
a vortex for small lateral displacements from the center of the trap. Note
that $\Omega_m<\Omega_c$ for a disk-shape condensate (in the TF limit) [see
Eqs.~(\ref{47}) and (\ref{48})], similar to the behavior for the
weak-coupling regime.

\subsubsection{Quantum-hydrodynamic analysis of condensate normal modes in
the Thomas-Fermi regime}

In addition to the anomalous mode described above, the condensate has a
sequence of normal modes that occur both with and without a vortex. Indeed,
one of the early triumphs of the quantum-hydrodynamic description~\cite
{Stri96} was the detailed agreement between the theoretical predictions and
the measured frequencies of the lowest few collective normal modes~\cite
{Dalf99}. For an axisymmetric condensate, the normal modes can be classified
by their azimuthal quantum number $m$, and modes with $\pm m$ are degenerate
for a stationary condensate.

When the condensate contains a vortex, however, the various collective modes
are perturbed. In particular, the vortex breaks time-reversal symmetry by
imposing a preferred sense of rotation, so that modes with $\pm m$ are split
(this behavior is analogous to the Zeeman effect in which an applied
magnetic field splits the magnetic sublevels). In fact, the splitting of
these degenerate hydrodynamic modes has been used to detect the presence of
a vortex~\cite{Chev00,Halj00} and to infer its circulation and angular
momentum.

\begin{figure}
\bigskip

\vspace{3cm}

\centerline{\epsfxsize=0.50\textwidth\epsfysize=0.17\textwidth
\epsfbox{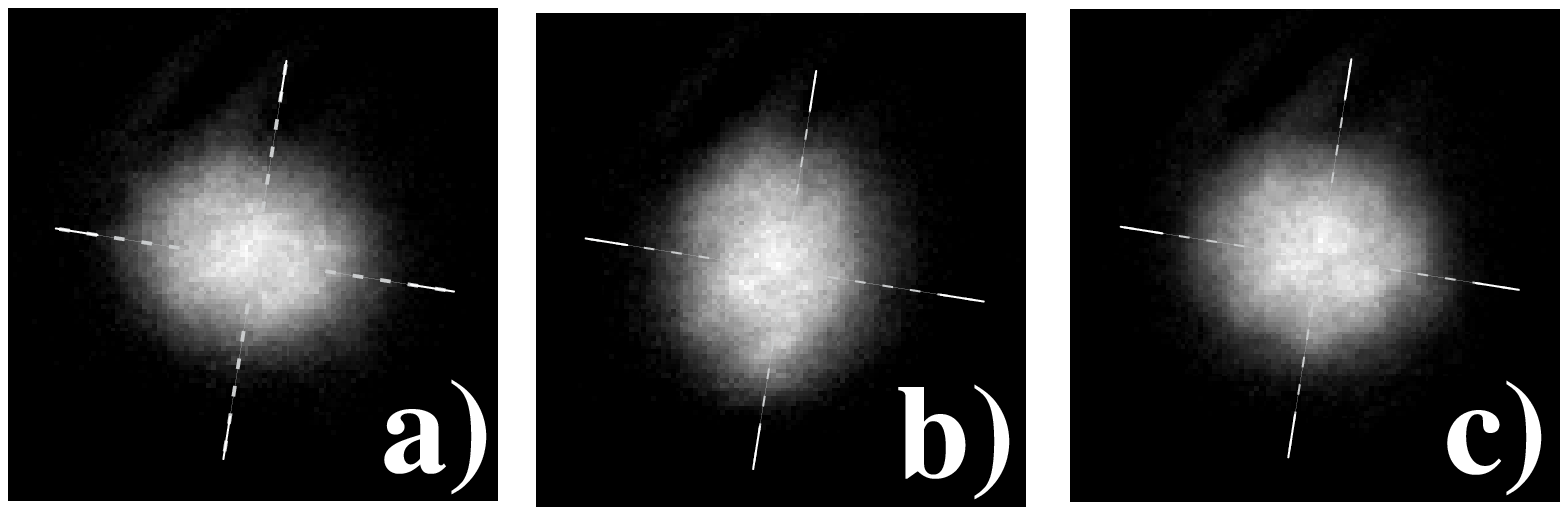}}

\vspace{0.9cm}

\centerline{\epsfxsize=0.50\textwidth\epsfysize=0.17\textwidth
\epsfbox{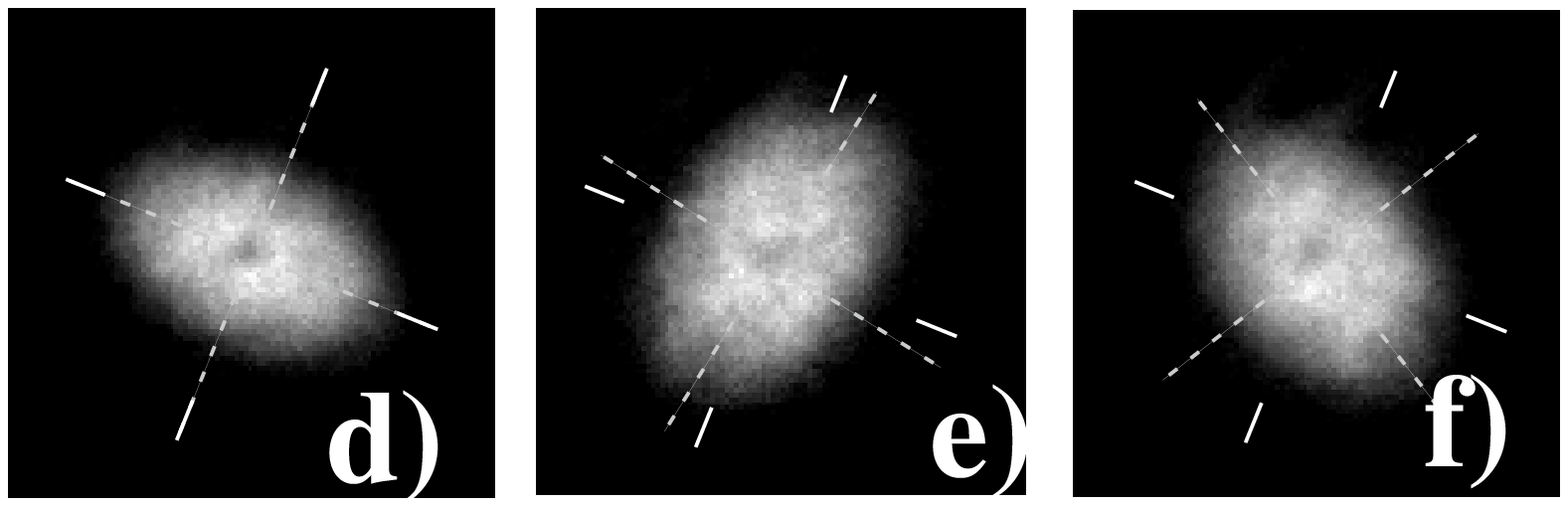}}

\vspace{-2cm}

\caption{Transverse oscillations of a stirred condensate with $3.7\times
10^5$ atoms in an elongated trap with $\omega_\perp/2\pi = 171 $ Hz.  For
(a)-(c), the stirring frequency $\Omega/2\pi = 114$ Hz is below the threshold
for vortex nucleation, whereas for (d)-(f), the stirring frequency
$\Omega/2\pi = 120$ Hz has nucleated a vortex (visible at the center of the
condensate). The sequences of pictures correspond to time delays $\tau
=1,3$ and
$5$ ms for which the ellipticity in the $xy$ plane is maximum. The fixed
axes indicate the excitation basis of the quadrupole mode and the rotating
ones indicate the condensate axes.}
\noindent (Taken from Ref.~\cite{Chev00}).
\label{figENS3}
\end{figure}

In the context of the quantum-hydrodynamic description, the principal effect
of the vortex arises through its circulating velocity field ${\bf v}$, which
shifts the time derivative $\partial _t\to \partial _t+\bbox{\nabla}\cdot
{\bf v}$. For a normal mode $\propto e^{im\phi }$ with azimuthal quantum
number $m$, the perturbation in the frequency has the form $m\hbar
/Mr_{\perp }^2$. A detailed analysis shows that the fractional splitting of
the modes is of order $(\omega _{+}-\omega _{-})/\omega _{+}\sim |m|d_{\perp
}^2/R_{\perp }^2$, with a numerical coefficient that depends on the
particular mode in question~\cite{Sinh97,Svid98}. Independently, Zambelli
and Stringari~\cite{Zamb98} used sum rules to calculate the vortex-induced
splitting of the lowest quadrupole mode with $m=\pm 2$; the two approaches
yield precisely the same expressions. In the absence of a vortex, the $|m|=2$
mode simply involves an oscillating quadrupole distortion, but the
vortex-induced splitting means that the quadrupole distortion precesses
slowly in a sense determined by the circulation around the vortex. The
angular frequency of precession of the eigenaxes of the quadrupole mode is
equal to $(\omega _{+}-\omega _{-})/2|m|=(\omega _{+}-\omega _{-})/4=\frac 74%
\omega _{\perp }{d_{\perp }^2}/{R_{\perp }^2}$. Figure~\ref{figENS3} shows
the difference between the two cases (with and without a vortex) for a
condensate with $\approx 3.7\times 10^5$ $^{87}$Rb atoms in an elongated
trap with $\omega _{\perp }/2\pi =171$ Hz. In the ENS experiment \cite
{Chev00}, when one vortex is nucleated at the center of the condensate, the
measured frequency splitting of the quadrupole mode ($\omega _{+}/2\pi =250$
Hz) is $(\omega _{+}-\omega _{-})/2\pi =66(\pm 7)$ Hz. For the experimental
parameters ($R_{\perp }=3.8\;\mu $m), theory predicts $(\omega _{+}-\omega
_{-})/2\pi =7\hbar /2\pi MR_{\perp }^2=56$ Hz. The result holds in the TF
limit and is valid with an accuracy of order $d_{\perp}^2\ln (R_{\perp }/\xi
)/R_{\perp }^2\sim 0.15$. With this uncertainty, the theoretical prediction $%
56(\pm 8)$ Hz agrees with the experimental value.

One should note that the vortex-induced splitting of the condensate modes is
maximum if the vortex is located at the trap center. If a straight vortex
line is displaced a distance $\zeta _0=r_0/R_{\perp }$ from the $z$ axis of
the TF condensate, then the splitting of the quadrupole mode ($m=\pm 2$) is
given by the expression
\begin{equation}
\omega _{+}-\omega _{-}=7\omega _{\perp }\frac{d_{\perp }^2}{R_{\perp }^2}%
\left( 1-\frac 54\zeta _0^2\left[ 1+\frac 12\zeta _0^4-\zeta _0^6+\frac 3{10}%
\zeta _0^8\right] \right)
\end{equation}
The splitting goes to zero if the vortex moves out of the condensate ($\zeta
_0=1$).

\subsubsection{Numerical analysis for general interaction parameter}

Garc\'\i a-Ripoll and P\'erez-Garc\'\i a~\cite{Garc99} have performed
extensive numerical analyses of the stability of vortices in axisymmetric
traps with an axial asymmetry parameter $\lambda =\omega _z/\omega _{\perp
}=1$ (a sphere) and $\lambda =\frac 12$ (one particular cigar-shape
condensate). They conclude that a doubly quantized vortex line has normal
modes with imaginary frequencies and that an external rotation cannot
stabilize it. For a singly quantized vortex in a spherical trap, however,
they confirm the presence of one negative-frequency (anomalous) mode with $%
|\omega _a|<\Omega _c$. For their cigar-shape condensate, they find
additional negative-frequency modes and suggest that such elongated
condensates are less stable than spherical or disk-shape ones. More recent
numerical work~\cite{Garc00a,Fede00a} confirms these findings for other
geometries, especially that for the ENS experiment~\cite{Madi99}, where the
axial asymmetry is large ($\omega _{\perp }/\omega _z\approx R_z/R_{\perp
}\approx 14$). It is expected that a vortex in an elongated condensate
becomes stable only for an external angular velocity $\Omega _m=\max |\omega
_a|$, where $\max |\omega _a|$ is the absolute value of the most negative of
these anomalous modes. For only modestly elongated traps, the metastable
frequency $\Omega _m$ exceeds the thermodynamic critical value $\Omega _c$;
these results provide an alternative explanation of the ENS observation that
the first vortex appears at an applied rotation $\approx 70\%$ higher than $%
\Omega _c$. Independently, an analysis of the bending modes of a trapped
vortex~\cite{Svid00a} in the TF limit finds that a vortex in a spherical or
disk-shape condensate has only one negative frequency (anomalous) mode, but
the number of such modes in an elongated condensate increases with the axial
asymmetry ratio $R_z/R_{\perp }$ (discussed below in Sec.~V.D.2).

\section{Vortex dynamics}

The preceding sections considered the equilibrium and stability of a vortex
in a trapped Bose condensate, using the stationary GP equation and the
Bogoliubov equations that characterize the small perturbations of the
stationary vortex. These approaches are somewhat indirect, for they do not
consider the dynamical motion of the vortex core. The present section treats
two different methods that address such questions directly.

\subsection{Time-dependent Variational Analysis}

Consider a variational problem for the action $\int dt\,{\cal L}(t)$
obtained from the Lagrangian

\begin{equation}
\label{70}{\cal L}(t) = \int dV\,\left[\frac{i\hbar}{2}\left(\Psi^*\,\frac{%
\partial\Psi}{\partial t} -\Psi\, \frac{\partial\Psi^*}{\partial t}%
\right)-\Psi^*\left(T + V_{{\rm tr}}-\Omega L_z\right)\Psi-\case{1}{2}%
g\,|\Psi|^4\right].
\end{equation}
It is easy to verify that the Euler-Lagrange equation for this action is
precisely the time-dependent GP equation in the rotating frame.

If, instead of $\Psi ({\bf r},t)$, we substitute a trial function that
contains different variational parameters (for example, the location of the
vortex core), the resulting time evolution of these parameters characterizes
the dynamics of the condensate. This method is not exact, but it provides an
appealing physical picture. For example, it determined the low-energy
excitations of a trapped vortex-free condensate at zero temperature~\cite
{Pere96,Pere97} for general values of the interaction parameter. In the TF
limit, this work reproduced the expressions derived by Stringari~\cite
{Stri96} based on Eq.~(\ref{33}).

\subsubsection{Near-ideal regime}

In the near-ideal limit, only the axisymmetric case has been studied, and it
is natural to start from the noninteracting vortex state (\ref{36}),
incorporating small lateral displacements of the vortex and the center of
mass of the condensate, along with a phase that characterizes the velocity
field induced by the motion of the condensate~\cite{Linn00}. In addition to
the rigid dipole mode (in which the condensate and the vortex oscillate
together at the transverse trap frequency $\omega_\perp$), an extra normal
mode arises at the anomalous (negative) frequency $\omega_a$ given in Eq.~(%
\ref{65}) omitting the second-order corrections that are beyond the present
approximation. In this weak-coupling limit, the resulting displacement of
the vortex is twice that of the center of mass, so that both must be
included to obtain the correct dynamical motion. Detailed analysis confirms
the positive normalization and relative displacements found from the
Bogoliubov equations for the same axisymmetric trap~\cite{Fett98}.

\subsubsection{Thomas-Fermi regime for straight vortex in disk-shape trap}

For a nonaxisymmetric trap in the TF regime, only the nonrotating case ($%
\Omega =0$) has been analyzed, using the fully anisotropic TF wave function
as an appropriate trial state, again with parameters describing the small
displacements of the straight vortex and the center of mass of the
condensate~\cite{Svid00}. The trial wave function was chosen in the form

\begin{equation}
\Psi \left( {\bf r},t\right) =B(t)f\left[ {\bf r}-{\bbox r}_0(t)\right] F[%
{\bf r}-{\bbox \eta }_0(t)] \prod_{j=x,y,z}\exp \left[ ix_j\alpha
_j(t)+ix_j^2\beta _j(t)\right] .
\end{equation}
Here the function $f\left( {\bf r}\right) $ characterizes the vortex line
inside the trap and far away from the vortex core has the approximate form $%
f\left( {\bf r}\right) =e^{i\phi }$; the function $F({\bf r})$ is the TF\
condensate density. The time-dependent vector ${\bbox \eta }_0(t)=(\eta
_{0x},\eta _{0y},\eta _{0z})$ describes the motion of the center of the
condensate, while ${\bbox
r}_0(t)=(x_0,y_0,0)$ describes the motion of the vortex line in the $xy$
plane. The other variational parameters are the amplitude $B(t)$ and the set
$\alpha _j(t)$ and $\beta _j(t)$. Substitution of the trial wave function
into~(\ref{70}) yields an effective Lagrangian as a function of the
variational parameters (and their first time derivatives). The resulting
Lagrangian equations have a solution that corresponds to the motion of the
vortex relative to the condensate. For this solution the vortex motion is
described by
\begin{equation}
x_0=\varepsilon _0R_x\sin \left( \omega _at+\phi _0\right) ,\quad
y_0=\varepsilon _0\,R_y\cos \left( \omega _at+\phi _0\right) ,
\end{equation}
while the displacement of the condensate is given by
\begin{equation}
\eta _{0x}=-\frac{15\varepsilon _0\xi ^2}{2R_y}\ln \left( \frac{R_{\perp }}%
\xi \right) \frac{R_x}{R_x+R_y}\sin \left( \omega _at+\phi _0\right) ,
\end{equation}
\begin{equation}
\eta _{0y}=-\frac{15\varepsilon _0\xi ^2\,}{2R_x}\ln \left( \frac{R_{\perp }}%
\xi \right) \frac{R_y}{R_x+R_y}\cos \left( \omega _at+\phi _0\right) ,
\end{equation}
where
\begin{equation}
\label{71}\omega _a=-\frac{3\hbar \omega _x\omega _y}{4\mu }\ln \left( \frac{%
R_{\perp }}\xi \right) =-\frac{3\hbar }{2MR_xR_y}\ln \left( \frac{R_{\perp }}%
\xi \right) ,
\end{equation}
in agreement with that found in Eq.~(\ref{68}). The quantity $%
x_0^2/R_x^2+y_0^2/R_y^2=\varepsilon _0^2$ remains constant as the vortex
line follows an elliptic trajectory around the center of a trap along the
line $V_{{\rm tr}}=const$, and the energy of the system is conserved [as
follows from Eq.~(\ref{45})]. The condensate also precesses with the
relative phase shift $\pi $ at the same frequency, but the amplitude of the
condensate motion is smaller than that of the vortex line by a factor $\sim
\xi ^2\ln \left( R_{\perp }/|q|\xi \right) /R_xR_y$.

For an axisymmetric TF condensate in rotational equilibrium at an angular
velocity $\Omega$, the Lagrangian (\ref{70}) provides a more general result
for the precession frequency. With the hydrodynamic variables $\Psi =
e^{iS}|\Psi|$, the first term of the Lagrangian becomes $-\hbar\int
dV\,|\Psi|^2\partial S/\partial t$. Since the TF condensate density vanishes
at the surface, the particle current also vanishes there, and it usually
suffices to assume a single straight vortex displaced laterally to ${\bf r}%
_0(t)$, with $S({\bf r,r}_0)=\arctan[(y-y_0)/(x-x_0)]$ and no image vortex.
The Lagrangian becomes

\begin{equation}
\label{Lagr}{\cal L}=\int dV\,M n({\bf r})\,\dot{{\bf r}}_0\cdot {\bf v}_0(%
{\bf r})-E(r_0)+\Omega L_z(r_0),
\end{equation}
where

\begin{equation}
\label{rot1}{\bf v}_0({\bf r})=\frac \hbar M\bbox{\nabla}S({\bf r,r}_0)=-%
\frac \hbar M\bbox{\nabla}_0S({\bf r,r}_0)=(\kappa /2\pi )\,\frac{\hat z%
\times ({\bf r-r}_0)}{|{\bf r-r}_0|^2}
\end{equation}
is the circulating velocity field about the vortex line. In the special case
of a two-dimensional condensate with the TF density $n(r)=n(0)(1-r_{\perp
}^2/R_{\perp }^2)$ per unit length, Eq.~(\ref{Lagr}) becomes

\begin{equation}
\label{rot2}{\cal L}_2=(\dot \phi _0+\Omega )L_{z2}(r_0)-\dot \phi
_0L_{z2}(0)-E_2(r_0),
\end{equation}
where $\phi _0$ is the azimuth angle describing position of the vortex line,
\begin{equation}
\label{rot3}L_{z2}(r_0)=\case{1}{2}n(0)\pi R_{\perp }^2\hbar (1-\zeta
_0^2)^2
\end{equation}
and

\begin{equation}
\label{rot4}E_2(r_0)=\frac{\kappa ^2Mn(0)}{8\pi }\left[ 2(1-\zeta _0^2)\ln
\left( \frac{R_{\perp }}\xi \right) +(1-\zeta _0^2)\ln (1-\zeta
_0^2)-1+2\zeta _0^2\right]
\end{equation}
with $\zeta _0=r_0/R_{\perp }$ [note that $\frac 12n(0)$ is the mean
particle density $\overline{n}$ per unit length]. These expressions differ
from the classical results for a uniform fluid in a rotating cylinder~\cite
{Hess67,Pack72} because of the parabolic TF density; in particular, the TF
angular momentum per unit length $L_{z2}$ here is proportional to $(1-\zeta
_0^2)^2$, whereas that for a uniform density is proportional to $1-\zeta
_0^2 $.

The Lagrangian dynamical equations show that the vortex precesses at fixed $%
r_0$ with the angular frequency

\begin{equation}
\label{prec2}\dot \phi _0=-\Omega +\frac{\partial E_2/\partial r_0}{\partial
L_{z2}/\partial r_0}=-\Omega -\frac{\partial E_2/\partial r_0}{\kappa
Mr_0n(r_0)}.
\end{equation}
This result is just that expected from the Magnus force on a straight vortex~%
\cite{Jack00,Lund00,McGe00}. For small displacements from the center, the
precession frequency in a nonrotating two-dimensional condensate reduces to $%
\dot \phi _0\approx (\kappa /2\pi R_{\perp }^2)\ln (R_{\perp }/\xi )\approx
\case{1}{2}\Omega _c$~\cite{Cast99}, but $\dot \phi _0$ increases with
increasing $r_0$ and eventually diverges near the edge of the condensate
where the density vanishes.

The corresponding results for a three-dimensional disk-shape TF condensate
follow from Eqs.~(\ref{45}) and (\ref{Lagr}). In particular, the integration
over $z$ means that the total angular momentum $L_{z3}=N\hbar (1-\zeta
_0^2)^{5/2}$ associated with the presence of the vortex differs from the
two-dimensional result proportional to $(1-\zeta _0^2)^2$. Apart from
numerical factors reflecting the three-dimensional geometry, Eq.~(\ref{prec2}%
) remains correct. For a straight vortex, it yields

\begin{equation}
\label{prec3}\dot \phi _0=-\Omega +\frac{\Omega _m}{1-r_0^2/R_{\perp }^2},
\end{equation}
where $\Omega _m=\frac 32(\hbar /MR_{\perp }^2)\ln (R_{\perp }/\xi )$ is the
metastable frequency (\ref{47}) for the appearance of a central vortex in a
disk-shape condensate. In the special case of a vortex near the center ($%
r_0\to 0$), this precession frequency reduces to (minus) the corresponding
anomalous frequency $\omega _a(\Omega )$ in Eq.~(\ref{68}) for a condensate
with a single central vortex line. To understand why the precession
frequency $\dot \phi _0$ is the negative of the anomalous frequency, recall
that the linearized perturbation in the density for the anomalous mode is
proportional to $\exp i[m_a\phi -\omega _a(\Omega )t]=\exp i[-\phi -\omega
_a(\Omega )t]$ because $m_a=-1$; this latter form shows clearly that the
normal mode propagates around the symmetry axis at an angular frequency $%
-\omega _a(\Omega )$, with the sense of rotation fixed by the sign of $%
-\omega _a(\Omega )$.

According to (\ref{prec3}), for a nonrotating trap the precession velocity
of a displaced vortex increases with the vortex displacement as $v=\Omega
_mr_0/(1-r_0^2/R_{\perp }^2)$. It is interesting to estimate at what
displacement the vortex velocity becomes supersonic \cite{FedePC}. Assuming
the speed of sound varies radially with the local density as $c=c_0\sqrt{%
1-r_0^2/R_{\perp }^2}$, where $c_0=\sqrt{\mu /M}=\omega _{\perp }R_{\perp }/%
\sqrt{2}$, we obtain $v/c=(\sqrt{2}\Omega _mr_0/\omega _{\perp }R_{\perp
})(1-r_0^2/R_{\perp }^2)^{-3/2}$. As a result, the vortex velocity becomes
supersonic if
\begin{equation}
\frac{r_0}{R_{\perp }}>{\frac{\omega _{\perp }}{\sqrt{2}\Omega _m}}\left( 1-%
\frac{r_0^2}{R_{\perp }^2}\right) ^{3/2}=\frac{\sqrt{2}R_{\perp }}{3\xi \ln
(R_{\perp }/\xi )}\left( 1-\frac{r_0^2}{R_{\perp }^2}\right) ^{3/2}.
\end{equation}
For parameters of JILA experiments \cite{Ande00} $R_{\perp }/\xi \approx 33$%
, this gives a critical displacement of $r_0/R_{\perp }\approx 0.82$ where
the precession vortex velocity becomes supersonic.

\subsection{Method of Matched Asymptotic Expansions}

At zero temperature, the dynamics of a condensate in a rotating
nonaxisymmetric trap follows from the appropriate time-dependent GP equation

\begin{equation}
\label{72}i\hbar\frac{\partial\Psi}{\partial t} =\left(-\frac{\hbar^2\nabla^2%
}{2M} +V_{{\rm tr}} +g|\Psi|^2 -\mu(\Omega) +i\hbar\bbox{\Omega}\cdot({\bf r}%
\times \bbox{\nabla})\right)\Psi.
\end{equation}
A vortex line in the condensate will, in general, move in response to the
effect of the nonuniform trap potential and the external rotation, as well
as self-induced effects caused by its own local curvature. This problem can
be solved in the case of a large condensate, where the TF separation of
length scales means that the vortex-core radius $\xi$ is much smaller than
the condensate radii $R_j$. The relevant mathematics involves the method of
matched asymptotic expansions~\cite{Pism91,Rubi94,Pism99}.

\subsubsection{Dynamics of straight vortex in Thomas-Fermi regime for
disk-shape trap}

As an introduction to these techniques, it is helpful first to concentrate
on the case of a straight singly quantized vortex line~\cite{Svid00}, which
is applicable to disk-shape condensates with $R_z\ll R_\perp$; this analysis
generalizes two-dimensional results found by Rubinstein and Pismen~\cite
{Rubi94}. Assume that the vortex is located near the center of the trap at a
transverse position ${\bf r}_{\perp 0}(t)$. In this region, the trap
potential does not change significantly on a length scale comparable with
the vortex core size $\xi$. The method of matched asymptotic expansions
compares the solution of Eq.~(\ref{72}) on two very different length scales:

First, consider the detailed structure of the vortex core. Assume that the
vortex moves with a transverse velocity ${\bf V}\perp \hat z$, and transform
to a co-moving frame centered at the vortex core. Away from the trap center,
the trap potential exerts a force proportional to $\bbox{\nabla}_\perp V_{%
{\rm tr}}$ evaluated at the position ${\bf r}_{\perp 0}(t)$. The resulting
steady solution includes the ``asymptotic'' region $|{\bf r}_\perp-{\bf r}%
_{\perp 0}|\gg \xi$.

Second, consider the region far from the vortex (on this scale, the vortex
core is effectively a singularity). The short-distance behavior of this
latter solution also includes the region $\xi \ll |{\bf r}_\perp-{\bf r}%
_{\perp 0}|$. The requirement that the two solutions match in the
overlapping region of validity determines the translational velocity ${\bf V}
$ of the vortex line.

Unfortunately, the details become rather intricate, but the final answer is
elegant and physical:

\begin{equation}
\label{73}{\bf V} =\frac{3\hbar}{4 M \mu}\left[\ln\left(\frac{R_\perp}{\xi}%
\right)-\frac{8 \mu \Omega}{3\hbar\left(\omega_x^2+\omega_y^2\right)}%
\right]\,\left(\hat z\times\bbox{\nabla}_\perp V_{{\rm tr}}\right)=\frac{%
3\hbar}{4 M \mu}\left[\ln\left(\frac{R_\perp}{\xi}\right)-\frac{2 M
R_\perp^2 \Omega}{3\hbar}\right]\,\left(\hat z\times\bbox{\nabla}_\perp V_{%
{\rm tr}}\right),
\end{equation}
where $R_\perp$ for an asymmetric trap is defined in Eq.~(\ref{46}). This
expression has several notable features.

(a) The motion is along the direction $\hat z\times\bbox{\nabla}_\perp V_{%
{\rm tr}} $ and hence follows an equipotential line of $V_{{\rm tr}}$. Thus
the trajectory conserves energy, which is expected because the GP equation
omits dissipative processes. In the present case of an anisotropic harmonic
trap, the trajectory is elliptical.

(b) For a nonrotating trap ($\Omega=0$), the motion is counterclockwise in
the positive sense at the frequency given by Eq.~(\ref{71}), proportional to
$\omega_x\omega_y$.

(c) With increasing applied rotation $\Omega$, the translational velocity $%
{\bf V}$ decreases and vanishes at the special value

\begin{equation}
\label{74}\Omega_m= \frac{3\hbar\left(\omega_x^2+\omega_y^2\right)}{8\mu}
\ln\left(\frac{R_\perp}{\xi}\right) = \frac{3\hbar}{2 M R_\perp^2} \ln\left(%
\frac{R_\perp}{\xi}\right),
\end{equation}
proportional to $\case{1}{2}\left(\omega_x^2+\omega_y^2\right)$. This value
precisely reproduces Eq.~(\ref{47}) associated with the onset of
metastability for small transverse displacements of the vortex from the trap
center.

(d) For $\Omega > \Omega_m$, the motion is clockwise as seen in the rotating
frame. A detailed analysis based on the normalization of the Bogoliubov
amplitudes shows that the positive-norm state has a frequency [compare Eq.~(%
\ref{71})]

\begin{equation}
\label{75}\omega _a(\Omega )=\frac{2\omega _x\omega _y}{\omega _x^2+\omega
_y^2}\left( \Omega -\Omega _m\right) .
\end{equation}
Note that this expression differs somewhat from Eq.~(\ref{66}) because the
trap here is anisotropic. The normal-mode frequency is negative and hence
unstable for $\Omega <\Omega _m$, but it becomes positive and hence stable
for $\Omega >\Omega _m$. \vskip.1cm \noindent This direct analysis of the
motion of a straight vortex reproduces the physics of the onset of (static)
metastability (\ref{47}) studied with the GP Hamiltonian and the (dynamic)
anomalous mode (\ref{69}) and (\ref{71}) studied with the Bogoliubov
equations and with the Lagrangian method.

\subsubsection{Dynamics of curved vortex in Thomas-Fermi regime}

Consider a nonaxisymmetric trap that rotates with an angular velocity $%
\bbox
\Omega $ (for convenience, $\bbox\Omega $ is often taken along the $z$
axis). At low temperature in a frame rotating with the same angular
velocity, the trap potential is time independent, and Eq.~(\ref{72})
describes the evolution of the condensate wave function. In the TF limit,
the method of matched asymptotic expansions again yields an approximate
solution for the motion of a singly quantized vortex line with instantaneous
configuration ${\bf r}_0(z,t)$. Let $\hat t$ be the local tangent to the
vortex (defined with the usual right-hand rule), $\hat n$ be the
corresponding normal, and $\hat b\equiv \hat t\times \hat n$ be the
binormal. A generalization of the work of Pismen and Rubinstein~\cite
{Pism91,Rubi94} eventually yields the explicit expression for the local
translational velocity~of the vortex \cite{Svid00a}

\begin{equation}
\label{76}{\bf V(r}_0)=-\frac{\hbar}{2M}\left(\frac{\hat t\times {\bbox %
\nabla}V_{{\rm tr}}({\bf r}_0)}{g|\Psi_{TF}|^2}+k\hat b\right)\ln\left(\xi\,%
\sqrt{\frac{1}{R_\perp^2}+\frac{k^2}{8}}\right)+ \frac{2\,{\bbox\nabla}V_{%
{\rm tr}}({\bf r}_0)\times{\bbox\Omega}}{\Delta_\perp V_{{\rm tr}}({\bf r}_0)%
},
\end{equation}
where $k$ is the local curvature (assumed small, with $k\xi\ll 1$) and $%
\Delta_\perp$ is the Laplacian operator in the plane perpendicular to ${\bf %
\Omega}$.

This vector expression holds for general orientation of the gradient of the
trap potential, the normal to the vortex line, and the angular velocity
vector. Near the TF boundary of the condensate, the denominator of the first
term becomes small, implying that the numerator $\hat t\times {\bbox \nabla}%
\,V_{{\rm tr}}({\bf r}_0)$ must also vanish near the boundary. As a result,
the axis of the vortex line $\hat t$ is parallel to ${\bbox \nabla}V_{{\rm tr%
}}$ at the surface and hence obeys the intuitive boundary condition that the
vortex must be perpendicular to the condensate surface.

\subsection{Normal modes of a vortex in a rotating two-dimensional TF
condensate}

This very general Eq.~(\ref{76}) applies in many different situations~\cite
{Svid00a}. The simplest case is an initially straight vortex in a
two-dimensional asymmetric TF condensate with ${\bf \Omega }=\Omega \hat z$
and $\omega _z=0$ (hence no confinement in the $z$ direction). For small
displacements, the $x$ and $y$ coordinates of the vortex core execute
harmonic motion $\propto \exp [i(\kappa z-\omega t)]$ that can vary between
helical and planar depending on the relative phase of the $x$ and $y$
motion. The dispersion relation $\omega _\kappa (\Omega )$ depends on the
continuous parameter $\kappa $ and the rotation frequency $\Omega $, along
with the TF radii $R_x$ and $R_y$ \cite{Svid00a}:
\begin{equation}
\omega _\kappa (\Omega )=\pm \frac \hbar {2MR_xR_y}\sqrt{\left( 2-\kappa
^2R_x^2-\tilde \Omega \right) \left( 2-\kappa ^2R_y^2-\tilde \Omega \right) }%
\ln \left( \xi \sqrt{\frac 1{R_{\perp }^2}+\frac{|\kappa |^2}8}\right) .
\end{equation}
where

\begin{equation}
\tilde \Omega =\frac{4MR_x^2R_y^2}{\hbar (R_x^2+R_y^2)\ln \left( \xi \sqrt{%
\frac 1{R_{\perp }^2}+\frac{|\kappa |^2}8}\right) ^{-1}}\Omega
\end{equation}
is a dimensionless rotation speed.

Of all the various normal modes, a straight vortex line ($\kappa=0$) has the
most negative (anomalous) frequency

\begin{equation}
\label{77}\omega_a(\Omega)= -\frac{\hbar}{2MR_xR_y}\,\left[ \ln\left(\frac{%
R_\perp}{\xi}\right) -\frac{4\mu\Omega}{\hbar\left(\omega_x^2+\omega_y^2%
\right)}\right],
\end{equation}
where an analysis similar to that for Eq.~(\ref{75}) shows that the minus
sign corresponds to the Bogoliubov solution with positive norm. For $%
\Omega=0 $, the vortex precesses counterclockwise about the $z$ axis in the
positive sense. With increasing rotation frequency $\Omega$, the precession
frequency decreases and vanishes at $\Omega=\Omega_m$, where the metastable
rotation frequency in two dimensions is

\begin{equation}
\label{78}\Omega_m = \frac{\hbar\left(\omega_x^2+\omega_y^2\right)}{4\mu}
\,\ln\left(\frac{R_\perp}{\xi}\right)= \frac{\hbar}{MR_\perp^2} \,\ln\left(%
\frac{R_\perp}{\xi}\right);
\end{equation}
as expected, this value is the precession frequency $\case{1}{2}\Omega_c$
discussed below Eq.~(\ref{prec2}) [compare Eq.~(\ref{47}) for $\Omega_m$ in
a three-dimensional disk-shape TF condensate; the different numerical
coefficient arises from the integration over the parabolic density in the $z$
direction].

More generally, for $\kappa ^2>0$ and a nonaxisymmetric trap ($R_x> R_y$),
the oscillation frequency can be imaginary (and hence unstable) within a
range of axial wave numbers determined by $\sqrt{(2-\tilde \Omega )}%
/R_x<|\kappa |<\sqrt{(2-\tilde \Omega )}/R_y$. For sufficiently fast
rotation, however, the frequencies become real, and the small oscillations
become stable at a rotation frequency $\tilde \Omega >\tilde \Omega _m=2$.
In the limit of a uniform unbounded condensate ($R_x,R_y\to \infty $), the
general dispersion relation reduces to the familiar one for helical waves on
a long straight vortex line~\cite{Lifs80aa}

\begin{equation}
\label{79}\omega=\pm\frac{\hbar}{2M}\,\kappa^2\ln\left(|\kappa|\xi\right).
\end{equation}

Using this dispersion relation, Barenghi~\cite{Bare96} estimated the
amplitude of the vortex waves due to thermal excitation (the cloud is
assumed to rotate at an angular velocity $\Omega >\Omega _c$, so that the
vortex is stable). He showed that finite-temperature effects in a Bose
condensate can distort the vortex state significantly, even at the very low
temperatures relevant to the experiments. For $T=10^{-7}$ K, $\bar n\approx
10^{12}-10^{13}$ cm$^{\!\!-3}$ and $R\approx 5\;\mu $m, the amplitude of
vortex oscillations can be 4-14 times the size of the vortex core. At the
same time, the thermal excitation of vortex waves in superfluid $^4$He is
negligible (much smaller than the corresponding vortex-core size).

\subsection{Normal modes of a vortex in a rotating three-dimensional TF
condensate}

Consider a three-dimensional TF condensate with $\omega_z> 0$, confined
within a TF region $z^2\le R_z^2=2\mu/M\omega_z^2$.

\subsubsection{General formalism}

For a vortex that initially lies along the $z$ axis, it is straightforward
to find the pair of coupled equations for the small transverse displacements
of the vortex $x(z,t)$ and $y(z,t)$. In particular, we seek solutions of the
form

\begin{equation}
\label{80}x=x(z)\sin (\omega t+\varphi _0),\qquad y=y(z)\cos (\omega
t+\varphi _0),
\end{equation}
in which case the amplitudes $x(z)$ and $y(z)$ describe the vortex shape and
obey coupled ordinary differential equations. Introducing dimensionless
scaled coordinates $x\to R_xx$, $y\to R_yy$, $z\to R_zz$, we find from Eq.~(%
\ref{76})

\begin{equation}
\label{81}\tilde \omega (1-z^2)x=-\frac d{dz}\left[ \beta (1-z^2)\frac{dy}{dz%
}\right] -y+\tilde \Omega (1-z^2)y,
\end{equation}

\begin{equation}
\label{82}\tilde \omega (1-z^2)y=-\frac d{dz}\left[ \alpha (1-z^2)\frac{dx}{%
dz}\right] -x+\tilde \Omega (1-z^2)x,
\end{equation}
where

\begin{equation}
\label{83}\alpha = \frac{R_x^2}{R_z^2},\quad \beta=\frac{R_y^2}{R_z^2}
\end{equation}
characterize the trap anisotropy and

\begin{equation}
\label{84}\tilde\omega =\frac{2MR_xR_y}{\hbar\,\ln(R_\perp/\xi)}%
\,\omega,\quad\tilde\Omega =\frac{4MR_x^2R_y^2}{\hbar(R_x^2+R_y^2)\ln(R_%
\perp/\xi)}\,\Omega
\end{equation}
are dimensionless angular velocities.

These equations (\ref{81}) and (\ref{82}) constitute a two-component
Sturm-Liouville system with natural boundary conditions~\cite{Fett80a}
because the factor $1-z^2$ vanishes at $z=\pm 1$. Consequently, the
eigenfunctions merely must remain bounded at the surface of the condensate.
A straightforward generalization of the usual analysis shows that the
eigenfunctions obey the orthogonality condition

\begin{equation}
\label{85}\int_{-1}^1 dz \,(1-z^2)x_my_n\propto \delta_{mn}\,.
\end{equation}

\subsubsection{Special solutions}

In the general case of a nonaxisymmetric trap, the resulting equations
remain coupled, but they separate in the particular case of stationary
solutions with $\omega =0$. For a nonrotating trap, such configurations
reflect a balance between the effects of curvature and the nonuniform trap
potential. For example, the small-amplitude stationary solutions $x_n(z)$
remain finite at the surface $z=\pm 1$ only for certain special values of
the trap anisotropy

\begin{equation}
\label{86}\alpha =\alpha _n=\frac 2{n(n+1)},
\end{equation}
where $n\ge 0$ is an integer. The corresponding solutions have the form $%
x_n(z)\propto P_n(z)$, where $P_n$ is the familiar Legendre polynomial. The
solutions have $n$ nodes and cross the $z$ axis $n$ times. If $\alpha $
differs from one of these special values (\ref{86}), there is no stationary
configuration. Similarly, the equation for the $y$ displacement has
stationary solutions only if $\beta \equiv R_y^2/R_z^2=2/[m(m+1)]$.

This classification of the solutions by the number of nodes remains more
generally valid. In the special case of an axisymmetric condensate ($%
\alpha=\beta$), we can consider the precession frequency $\omega_n$ of the
mode with $n$ nodes as a function of the axial trap anisotropy $\alpha$.
Evidently, the function $\omega_n$ changes sign at the special value $\alpha
=\alpha_n =2/[n(n+1)]$. This observation allows us to determine the number
of modes with negative frequencies at a fixed value of the anisotropy
parameter $\alpha$. For $\alpha\ge 1$ (a spherical or disk-shape
condensate), only one mode has a negative frequency. If $\case{1}{3}%
<\alpha<1 $, there are two such anomalous modes, and so on. If $%
\alpha_n<\alpha<\alpha_{n-1}$, a nonrotating axisymmetric TF condensate has $%
n$ anomalous modes with negative frequency.

The special case of a nearly disk-shape anisotropic rotating TF condensate
is particularly tractable because $\alpha^{-1}$ and $\beta^{-1}$ provide
small expansion parameters. There is only one relevant normal mode, with
frequency

\begin{equation}
\label{87}\omega_a(\Omega)=-\Omega_m+\Omega,
\end{equation}
where

\begin{equation}
\label{88}\Omega_m=\frac{\hbar\left(\omega_x^2+\omega_y^2\right)}{8\mu}
\left[3+\frac{1}{10}\left(\frac{1}{\alpha}+\frac{1}{\beta}\right)\right]
\ln\left(\frac{R_\perp}{\xi}\right)\quad%
\hbox{for nearly disk-shape TF
condensate}.
\end{equation}
If $\Omega<\Omega_m =|\omega_a(0)|$, the frequency is negative, and the mode
is therefore unstable. This value generalizes that found previously in Eqs.~(%
\ref{47}) and (\ref{74}) for the angular velocity at which a straight vortex
at the center of a thin disk-shape condensate becomes metastable, now
including the first corrections of order $\alpha^{-1}$ and $\beta^{-1}$.

This result (\ref{88}) remains approximately correct for a spherical
condensate ($\alpha =\beta =1$), which is the geometry used in recent JILA
experiments~\cite{Ande00}. Since $\Omega _m$ is numerically equal to the
frequency $|\omega _a|$ of the one anomalous mode in the nonrotating
condensate, Eq.~(\ref{88}) also yields the precession frequency of a nearly
straight vortex moving counterclockwise around the center of the condensate%
\cite{Svid00,Fede00a}. In particular, we find $|\omega _a|/\omega =%
\case{8}{5}(\xi/R)\ln (1.96R/\xi )$, where $\omega $ is the isotropic trap
frequency and the additional numerical factor $1.96$ in the logarithm is the
next correction to the logarithmic accuracy (see, for example, Ref.~\cite
{Rubi94}). With the JILA parameters $R\approx 22\ \mu $m and $\xi \approx
0.67\ \mu $m, this expression yields $|\omega _a|/2\pi =1.58\pm 0.16$ Hz,
where the uncertainty reflects the omission of corrections of relative order
$(\xi /R)\ln (R/\xi )\approx 0.1$. For comparison, the experimental value $%
1.8\pm 0.1$ Hz for the precession frequency is somewhat larger, but the
theoretical prediction is sensitive to the number $N$ of atoms in the
condensate and, as seen in Eq.~(\ref{prec3}), to the radial displacement of
the vortex~\cite{Fede00a,Lund00}.

The situation is very different for an elongated cigar-shape condensate with
$R_z\gg R_\perp$, when the solutions for the precessing normal-mode
amplitudes grow exponentially with $|z|$. In contrast to the two-dimensional
case, such solutions are now possible because the condensate is bounded
along the $z$ axis. In the simplest case of an axisymmetric trap with $%
R_x=R_y=R_\perp$, the mode with no nodes has a frequency $%
\omega_a(\Omega)=-\Omega_m+\Omega$. Although this expression has the same
form as Eq.~(\ref{87}) for a disk-shape condensate, the physical behavior is
very different because the metastable angular velocity

\begin{equation}
\label{89}\Omega_m=\frac{\hbar}{2MR_\perp^2} \,\frac{R_z^2}{R_\perp^2}
\ln\left(\frac{R_\perp}{\xi}\right)\approx \frac{R_z^2}{5\,R_\perp^2}
\,\Omega_c
\end{equation}
becomes large for a highly elongated TF condensate. For the ENS geometry~%
\cite{Madi99,Chev00}, where $\omega_\perp/\omega_z\approx R_z/R_\perp\approx
14.4$, Eq.~(\ref{89}) is far too large to fit the observations and can even
exceed the limit of rotational mechanical stability $\Omega=\omega_\perp$
that occurs when the centrifugal force cancels the confining trap potential.

For a harmonic transverse external potential $\propto r_{\perp }^2$, the
method of matched asymptotic expansions is valid if the vortex displacement $%
r$ from the $z$ axis satisfies the condition $r\gtrsim \xi $ (in the
vicinity of the vortex core the trap potential is approximated as a linear
function). For a long cigar-shape condensate, the solution for the lowest
mode has the form: $r=r_0\cosh (z/\alpha )$, where $r_0$ is the vortex
displacement at $z=0$. The condition of small vortex displacement implies
that $r_0\cosh (1/\alpha )\ll R_{\perp }$, while the condition of small
vortex curvature $k\xi \ll 1$ implies that $r_0\xi \cosh (1/\alpha
)/R_z^2\alpha ^2\ll 1$. A combination of these conditions gives the
following restriction on the validity Eq.~(\ref{89}): $\exp (1/\alpha )\ll
2R_{\perp }/\xi $. For the ENS experiments, $1/\alpha \approx 200$ and $%
R_{\perp }/\xi \approx 21$, so that this condition fails.

As mentioned in Sec.~IV.D.4, the frequency for the onset of metastability $%
\Omega_m$ in Eq.~(\ref{89}) can be larger than the thermodynamic critical
angular velocity $\Omega_c$ in Eq.~(\ref{41}). This behavior is readily
understandable because $\Omega_c$ characterizes the energy of a straight
vortex along the symmetry axis [compare Eq.~(\ref{40})], whereas the most
unstable normal-mode amplitude explicitly involves the small-amplitude
distortion with no nodes. For a very elongated condensate, the resulting
vortex dynamics is particularly sensitive to the large curvature of the
condensate surface near the two ends of the symmetry axis (in contrast to
the small curvature for the flattened condensate).

Recent numerical studies~\cite{Garc00a,Fede00a} of the most negative
anomalous modes for a trap geometry corresponding to the ENS experiments~%
\cite{Madi99,Chev00} yield values of $\Omega _m$ that are significantly
smaller than the prediction given in Eq.~(\ref{89}). Reference~\cite{Garc00a}
mentions the possible failure of the TF picture in the transverse direction,
even though the conventional TF ratio $R_{\perp }/\xi $ is large, at least
near the plane $z=0$. 
As confirmation of the validity of the GP equation and the particular role
of the anomalous modes, the numerically determined~\cite{Fede00a} $\Omega
_m/2\pi \approx 0.73\nu _{\perp }\approx 124$ Hz agrees well with the ENS
value $\Omega _{{\rm obs}}/2\pi \approx 120$ Hz for the appearance of the
first vortex.

\begin{figure}
\bigskip
\centerline{\epsfxsize=0.40\textwidth\epsfysize=0.45\textwidth
\epsfbox{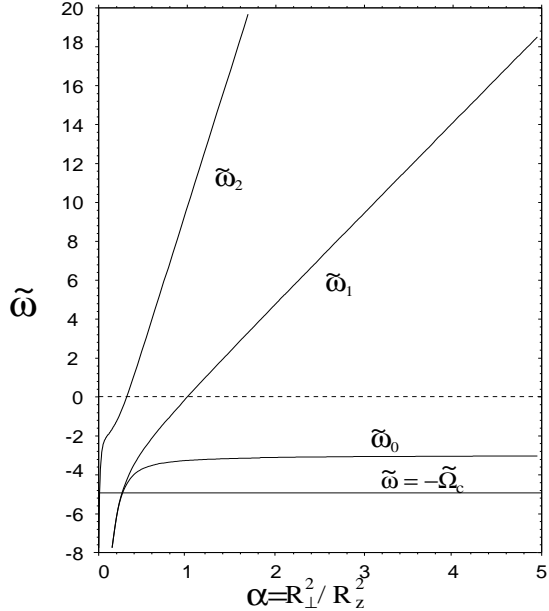}}

\vspace{0.3cm}

\caption{Dimensionless frequencies $\tilde\omega \equiv \tilde\omega(\Omega=0)$
for the first three normal modes of a vortex in an axisymmetric trap as a
function of the axial anisotropy $\alpha= R_\perp^2/R_z^2$.  The lower
horizontal line is the negative of the dimensionless thermodynamic critical
angular velocity $\tilde\Omega_c = 5$. Note that
$|\tilde\omega_0|>\tilde\Omega_c$ for $\alpha <0.26$.}
\label{fig11}
\end{figure}

For an axisymmetric trap ($\alpha=\beta$), we can seek normal-mode solutions
in the form $x(z)=y(z)$, leaving a single equation

\begin{equation}
\label{90}\left[ \tilde \omega (\tilde \Omega )-\tilde \Omega \right]
(1-z^2)x=-\frac d{dz}\left[ \alpha (1-z^2)\frac{dx}{dz}\right] -x
\end{equation}
that depends only on the Doppler-shifted frequency $\tilde \omega (\tilde
\Omega )-\tilde \Omega =\tilde \omega (0)$. The eigenfunctions are even or
odd functions of $z$ and can be classified by the number of times the vortex
crosses the $z$ axis (the number of nodes), $m=0,1,2,\cdots $. Figure~\ref
{fig11} shows the dimensionless frequency $\tilde \omega (0)$ as a function
of the trap anisotropy $\alpha =R_{\perp }^2/R_z^2$ for $m=0,1$, and $2$. In
agreement with the analytical results, a disk-shape trap ($\alpha \ge 1$)
has only a single mode with negative frequency $\tilde \omega _0$. For $%
\case{1}{3}<\alpha <1$, there are two such modes ($m=0$ and $m=1$) and
successively more negative-frequency modes appear for smaller $\alpha $. As
noted previously, the critical frequency $\tilde \Omega _m$ for
metastability is $|\tilde \omega _0|$, which is smaller than $\tilde \Omega
_c$ for disk-shape traps and for moderately elongated traps. Our numerical
analysis for the present TF limit predicts that $\tilde \Omega _m\ge \tilde
\Omega _c$ for $\alpha =R_{\perp }^2/R_z^2\le 0.26$, which is somewhat
larger than the value $0.2$ implied by the limiting expression in Eq.~(\ref
{89}).

As in the case of a two-dimensional condensate, the frequency of the
anomalous modes can become imaginary for an anisotropic trap with $R_x\neq
R_y$~\cite{Svid00a}. To demonstrate that result, let us consider Eqs.~(\ref
{81}) and (\ref{82}) for a trap close to axisymmetric with $|\alpha -\beta
|\ll \alpha $. The eigenfrequencies of the axisymmetric trap [with $\alpha
=\beta =\alpha _0=\frac 12(\alpha +\beta )$] are real and have the form $%
\tilde \omega _m(\tilde \Omega )=\tilde \omega _m+\tilde \Omega $, where $%
m=0,1,2,\cdots $ denotes the various modes. For an anomalous mode, the
frequency $\tilde \omega _m$ is negative, and the eigenfrequency $\tilde
\omega _m(\tilde \Omega )$ is equal to zero if the trap rotates with the
angular velocity $\tilde \Omega =|\tilde \omega _m|$. One can rewrite Eqs.~(%
\ref{81}) and (\ref{82}) as follows:
\begin{equation}
\label{i1}\tilde \omega (1-z^2)%
\pmatrix{x\cr y\cr}=\hat H_0
\pmatrix{x\cr y\cr}+\hat V
\pmatrix{x\cr y\cr},
\end{equation}
where
$$
\hat H_0=\left\{ -2-\alpha _0\partial _z[(1-z^2)\partial _z]+(1-z^2)|\tilde
\omega _m|\right\} 
\pmatrix{0&1\cr 1&0\cr},
$$

$$
\hat V=-\partial _z[(1-z^2)\partial _z]%
\pmatrix{0&\beta-\alpha_0\cr
\alpha-\alpha_0&0\cr} 
+(1-z^2)(\tilde \Omega -|\tilde \omega _m|)\pmatrix{0&1\cr 1&0\cr}. 
$$
Considering $\hat V$ as a perturbation, we obtain the following expression
for the normal-mode frequency in a nonaxisymmetric trap:

\begin{equation}
\label{i2}\tilde \omega =\pm \sqrt{\left( |\tilde \omega _m|-|\alpha -\beta
|I_m-\tilde \Omega \right) \left( |\tilde \omega _m|+|\alpha -\beta |I_m-%
\tilde \Omega \right) },
\end{equation}
where
\begin{equation}
\label{i3}I_m=\frac{\int_{-1}^1dz(1-z^2)\left( \partial _zx_m\right) ^2}{%
2\int_{-1}^1(1-z^2)x_m^2dz}>0
\end{equation}
and $x_m=x_m(z)$ describes shape of the $m$th vortex mode. As we increase
the trap rotation, the eigenfrequency is real for $\tilde \Omega <|\tilde
\omega _m|-I_m|\alpha -\beta |$ . Then, when $|\tilde \Omega -|\tilde \omega
_m||<I_m|\alpha -\beta |$, the frequency becomes imaginary. Finally, if $%
\tilde \Omega >|\tilde \omega _m|+I_m|\alpha -\beta | $, the frequency again
becomes real. For a given trap anisotropy (given $\alpha $ and $\beta $),
one or several normal modes of the vortex have negative frequency. Trap
rotation $\tilde \Omega $ shifts the frequencies in the positive direction.
When the frequency of a normal mode in the rotating frame approaches zero,
the frequency becomes imaginary until $|\tilde \omega _m+\tilde \Omega
|=I_m|\alpha -\beta |$. If we increase the trap rotation further, the
frequency (in the rotating frame) becomes positive.

For a disc-shape condensate (with $\alpha _0\gg 1$) there is only one
anomalous mode with $x_a=y_a=\varepsilon \left( 1+{z^2}/{2\alpha _0} \right)
$ and $\tilde \omega _a=-3- \frac{1}{5}\alpha _0^{-1}$. For a
nonaxisymmetric rotating trap, the frequency of this mode becomes imaginary
in the interval $|\tilde \Omega -|\tilde \omega _a||<\epsilon |\tilde \omega
_a|/(15\alpha _0)$, where $\epsilon =|R_x-R_y|/R_x$ is the trap anisotropy
in the transverse direction. Thus for a disk-shape condensate (with $\alpha
_0\gg 1$), the solution has an imaginary frequency in a relatively narrow
range of trap rotation.

For a cigar-shape condensate, several normal modes have negative
frequencies. In the limit $\alpha _0\ll 1$, the solution for the lowest
anomalous mode has the form $x_a=y_a=\varepsilon \cosh \left( z/\alpha
_0\right) $ and $\tilde \omega _a\approx -1/\alpha _0$. Consequently, the
frequency is imaginary if $|\tilde \Omega -| \tilde \omega _a||<\epsilon |%
\tilde \omega _a|$, namely in a relatively wide range of trap rotation. If
the transverse trap anisotropy is large enough, several different anomalous
normal modes can have imaginary frequencies in the same range of angular
velocities. In this case a vortex along the $z$ axis is stable (there are no
normal modes with imaginary frequencies) only if the trap rotates slightly
faster than the frequency of the lowest anomalous mode. This behavior could
be relevant to ENS experiments.

\subsubsection{Energy of a curved trapped vortex}

Consider a trap that contains a singly quantized vortex and rotates with
angular velocity $\Omega$ about the $z$ axis. At zero temperature, Eq.~(\ref
{76}) governs the dynamics of each element of the line

\begin{equation}
\label{91}{\bf V(r})=-\frac \hbar {2M}\left( \frac{\hat t\times {\bbox %
\nabla }V_{{\rm tr}}({\bf r})}{g|\Psi _{TF}|^2}+k\hat b\right) \ln \left(
\xi \,\sqrt{\frac 1{R_{\perp }^2}+\frac{k^2}8}\right) +\frac{2\,{\bbox\nabla
}V_{{\rm tr}}({\bf r})\times {\bbox\Omega }}{\Delta _{\perp }V_{{\rm tr}}(%
{\bf r})},
\end{equation}
where ${\bf r=(}x(z),y(z),z)$ determines the shape of the line.
Correspondingly, Eq.~(\ref{38}) serves as the energy functional

\begin{equation}
\label{92}E(\Psi )=\int dV\left( \frac{\hbar ^2}{2M}|\nabla \Psi |^2+V_{{\rm %
tr}}|\Psi |^2+\case{1}{2}g|\Psi |^4+\Psi ^{*}i\hbar \Omega \frac{\partial
\Psi }{\partial \phi }\right)
\end{equation}
in the rotating frame (for simplicity, we now use $E$ instead of $E^{\prime
} $). In Sec.~III.B, a physically motivated wave function served to evaluate
Eq.~(\ref{92}), yielding Eq.~(\ref{45}) for the energy of a straight vortex
displaced laterally from the trap axis. As noted previously, the assumption
of a straight vortex restricted the analysis to a disk-shape condensate.

To find the energy of a curved vortex, one can first find the condensate
wave function $\Psi $ and then substitute it into the functional (\ref{92}).
For a curved vortex line, however, this approach is complicated. Instead,
one can use Eq.~(\ref{91}) to find the vortex energy directly. As we know,
the stationary Gross-Pitaevskii equation can be obtained by varying the
energy functional (\ref{92}). The dynamical equation (\ref{91}) is, in fact,
the time dependent Gross-Pitaevskii equation, written in a way suitable to
describe the vortex motion. Consequently, if we formally put ${\bf V(r)=0}$
in Eq.~(\ref{91}) (namely omit the time derivatives), then the resulting
stationary equation must be an extremum of the energy functional $E_V$
associated with the presence of the vortex and considered as a functional of
the vortex shape $E_V=E_V\left( x(z),y(z)\right) $. An equivalent energy
functional has the form (in the TF limit):

\begin{equation}
\label{93}E_V\left( x(z),y(z)\right) =\frac{\pi \hbar ^2}M\int dz\,\left[
|\Psi _{TF}|^2\sqrt{1+(x^{\prime })^2+(y^{\prime })^2}\ln \left( \frac{
R_{\perp }}\xi \right) -\frac{2M}\hbar \frac{g|\Psi _{TF}|^4\Omega }{\Delta
_{\perp }V_{{\rm tr}}}\right] ,
\end{equation}
where the prime denotes the derivative with respect to $z$. Variation of
Eq.~(\ref{93}) with respect to $x(z)$ and $y(z)$ gives Eq.~(\ref{91}) with $%
{\bf V(r)=0}$, apart from terms of higher order $xx^{\prime }{}^2,xy^{\prime
}{}^2,\cdots $\ . Hence Eq.~(\ref{93}) provides an energy functional for the
small deformations of a vortex about a straight configuration along the $z$
axis (when the fourth-order terms in the displacement can be omitted) or for
arbitrary displacements of a straight vortex. Note that Eq.~(\ref{93})
involves only a one-dimensional line integral instead of the
three-dimensional expression in Eq.~(\ref{92}), which is a significant
simplification. In scaled dimensionless units $x\to R_xx$, {\it etc.}, this
energy functional becomes

\begin{equation}
\label{94}E_V\left( x(z),y(z)\right) =2\pi \mu R_z\xi ^2n(0)\int dz\,\left[
\left( 1-x^2-y^2-z^2\right) \sqrt{1+\alpha (x^{\prime })^2+\beta (y^{\prime
})^2}\ln \left( \frac{R_{\perp }}\xi \right) -\frac{2\mu \Omega
(1-x^2-y^2-z^2)^2}{\hbar (\omega _x^2+\omega _y^2)}\right] ,
\end{equation}
where $n(0)=\mu /g$ is the density at the center of the vortex-free
condensate, $\xi ^2=\hbar ^2/2M\mu $ and the integration is restricted to
the region $1-x^2-y^2-z^2\ge 0$. Using Eq.~(\ref{94}) one can obtain a
simple expression for the angular momentum of the condensate in the presence
of a curved vortex line:
\begin{equation}
L_z=-\frac{\partial E_V}{\partial \Omega }=\frac{15}8\hbar N\frac{R_xR_y}{%
R_x^2+R_y^2}\int dz(1-x^2-y^2-z^2)^2,
\end{equation}
where $N=8\pi R_xR_yR_zn(0)/15$ is the total number of particles in the
condensate.

The integration in Eq.~(\ref{94}) is particularly easy for a straight vortex
and readily reproduces Eq.~(\ref{45}). An expansion for small lateral
displacements yields Eqs.~(\ref{47}) and (\ref{48}) for $\Omega _m$ and $%
\Omega _c$ for a disk-shape TF condensate. In the more general case of
arbitrary small displacements, Eq.~(\ref{94}) can be expanded to second
order in the amplitudes $x$ and $y$ and their derivatives. Use of the
dynamical equations that lead to (\ref{81}) and (\ref{82}) gives the simple
expression

\begin{equation}
\label{95}E_V\left( x(z),y(z)\right) =\frac{8\pi }3\mu R_z\,\xi ^2n(0)\left[
\ln \left( \frac{R_{\perp }}\xi \right) -\frac 85\,\frac{\mu \Omega }{\hbar
\left( \omega _x^2+\omega _y^2\right) }\right] +\frac{15}8\hbar
N\int_{-1}^1dz\,(1-z^2)(x\dot y-y\dot x),
\end{equation}
The first term of Eq.~(\ref{95}) reproduces the value of $\Omega _c$ for a
general TF condensate, and the second term becomes a sum over all normal
modes of the form~(\ref{80})

\begin{equation}
\label{96}E_V\left( x(z),y(z)\right) =\frac{8\pi }3\mu R_z\,\xi ^2n(0)\left[
\ln \left( \frac{R_{\perp }}\xi \right) -\frac 85\,\frac{\mu \Omega }{\hbar
\left( \omega _x^2+\omega _y^2\right) }\right] +\frac{15}8N\sum_n\hbar
\omega _n(\Omega )\int_{-1}^1dz\,(1-z^2)x_n(z)y_n(z),
\end{equation}
where the orthogonality condition Eq.~(\ref{85}) eliminates the cross terms
between different normal modes. If any of the normal modes is anomalous
(namely with negative frequency), then the system is unstable with respect
to excitation of those modes. This analysis confirms the interpretation of $%
\Omega _m$ as the applied rotation frequency at which the frequency of the
last anomalous mode vanishes in the rotating frame. At this applied $\Omega $
the location of the vortex line along the $z$ axis becomes a local minimum
of energy. Note that this conclusion is wholly equivalent to that in Eq.~(%
\ref{56}) based on the Bogoliubov quasiparticles.

One should note that for a cigar-shape condensate with $R_z\gtrsim 2 R_\perp$%
, there is an interval of angular velocity of trap rotation when $\Omega
_c<\Omega<\Omega _m$. In this interval, the frequency of (at least) the
lowest vortex mode remains negative, but penetration of a vortex into the
condensate is energetically favorable. Under such a condition, the vortex
line can lower its energy by undergoing a finite-amplitude deformation, and
the ground state of the system corresponds to a curved vortex line displaced
from the trap axis (see also~\cite{Garc00a}).

\subsubsection{Precession and tilting of a straight vortex line in a nearly
spherical TF condensate}

The preceding discussion of vortex dynamics in a three-dimensional confined
condensate has focused on the small-amplitude displacements from
equilibrium. In the special case of a spherical trap, however, the presence
of a zero-frequency precessing mode (Sec.~V.D.2) allows a more general
analysis of the nonlinear dynamics, which is directly relevant to recent
JILA experiments on the evolution of an initially straight vortex in a
nearly spherical TF condensate~\cite{Halj00}. In practice, the trap deviates
slightly from spherical with $R_x\neq R_y\neq R_z$.

For a spherical condensate, a motionless straight singly quantized vortex
through the center of trap satisfies the general Eq.~(\ref{76}) for the
velocity of a vortex line because the axis of the vortex $\hat t$ lies along
$\bbox{\nabla}V_{{\rm tr}}$. Let

\begin{equation}
x=\gamma _xs,\quad y=\gamma _ys,\quad z=\gamma _zs
\end{equation}
specify the axis of the vortex line, where $s$ is the arc length measured
from the trap center and ($\gamma _x$, $\gamma _y$, $\gamma _z$) are the
direction cosines relative to the principal axes of the anisotropic trap.
For small anisotropy, the vortex remains approximately straight, but the
direction cosines become time dependent. To first order in the anisotropy,
the curvature $k$ can be omitted in Eq.~(\ref{76}) and $|\Psi _{TF}|^2$ can
be approximated by the TF density for a spherical vortex-free condensate
with TF radius $R$. Standard perturbation theory yields the nonlinear
dynamical equations

\begin{equation}
\label{98}\dot\gamma_x= \frac{5\hbar}{4\mu}\ln\left(\frac{R}{\xi}\right)
\left(\omega_z^2-\omega_y^2\right)\gamma_y\gamma_z,
\end{equation}

\begin{equation}
\label{99}\dot\gamma_y= \frac{5\hbar}{4\mu}\ln\left(\frac{R}{\xi}\right)
\left(\omega_x^2-\omega_z^2\right)\gamma_z\gamma_x,
\end{equation}

\begin{equation}
\label{100}\dot\gamma_z= \frac{5\hbar}{4\mu}\ln\left(\frac{R}{\xi}\right)
\left(\omega_y^2-\omega_x^2\right)\gamma_x\gamma_y.
\end{equation}

This set of equations is familiar in classical mechanics as Euler's
equations for the torque-free motion of a rigid body~\cite
{Land60,Klep73,Fett00b}, where they describe the motion of the
angular-velocity vector as seen in the body-fixed frame. In the present
context, this set of three coupled nonlinear equations has two first
integrals

\begin{equation}
\label{101}\gamma_z^2+\gamma_y^2+\gamma_z^2=1,
\end{equation}
which verifies that the first-order anisotropy simply rotates the vortex
axis and

\begin{equation}
\label{102}\omega_x^2\gamma_z^2+\omega_y^2\gamma_y^2+\omega_z^2\gamma_z^2=%
{\rm const},
\end{equation}
which is the condition of energy conservation.

The simplest situation is an axisymmetric trap with $\omega_x=\omega_y=%
\omega_\perp$, in which case the vortex line precesses uniformly about the $%
z $ axis (the symmetry axis) at a fixed polar angle $\arccos \gamma_z(0)$ at
a frequency~\cite{Svid00a}

\begin{equation}
\label{103}\omega= \frac{5\hbar(\omega_z^2-\omega_\perp^2)}{4\mu}%
\,\gamma_z(0)\ln\left(\frac{1.96 R}{\xi}\right)= \frac{5\hbar}{2M}\left(%
\frac{1}{R_z^2}-\frac{1}{R_\perp^2}\right)\,\gamma_z(0) \ln\left(\frac{1.96 R%
}{\xi}\right),
\end{equation}
where the numerical factor 1.96 inside the logarithm is the same as that
discussed below Eq.~(\ref{88}). For positive (negative) $\omega$, the
precession is counter-clockwise (clockwise). Recent experiments
at JILA have observed two recurrences of such precessional motion in a
slightly flattened trap with $\omega_z-\omega_\perp\approx 0.1\omega_z$ and
a polar tipping angle of $45^\circ$ from the $z$ axis. In this case, Eq.~(%
\ref{103}) predicts $\omega/2\pi\approx 0.33\pm 0.03$ Hz, in an agreement
with the observed value $0.25\pm 0.02$ Hz \cite{Halj00}.

\begin{figure}
\bigskip
\centerline{\epsfxsize=0.40\textwidth\epsfysize=0.45\textwidth
\epsfbox{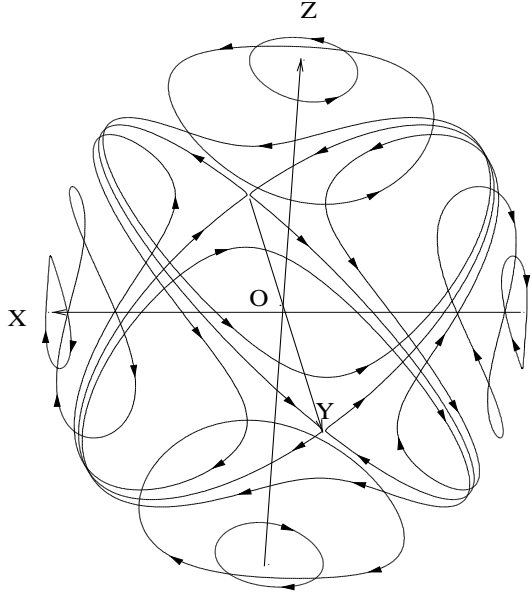}}

\vspace{0.3cm}

\caption{Typical trajectories of the end of a straight vortex line (that passes
through the condensate center) during its motion in a slightly nonspherical
trap with $R_x<R_y<R_z$.}
\label{fig12}
\end{figure}

More generally, for an anisotropic trap (with $\omega_x>\omega_y>\omega_z$),
the vortex executes closed trajectories (see Fig.~\ref{fig12}). For initial
positions close to the $x$ and $z$ axes (the smallest and largest TF radii),
the motion is ``stable," remaining nearby, but small-amplitude motion about
an initial position close to the $y$ axis (the intermediate TF radius)
yields imaginary frequencies. Thus such trajectories deviate far from the
initial neighborhood, even though they eventually return (this periodic
behavior is familiar from the corresponding solutions of the Euler equations~%
\cite{Land60,Klep73,Fett00b}). Reference~\cite{Svid00a} gives explicit
solutions for the resulting dynamical motion of a nearly straight vortex in
a totally anisotropic trap.

\section{Effect of thermal quasiparticles, vortex lifetime and dissipation}

In previous sections we considered a Bose condensate within the Bogoliubov
approximation, which omits the effect of thermal quasiparticles. At finite
temperatures, however, these noncondensate atoms can modify the frequencies
of the vortex modes and dissipate energy.

\subsection{Bogoliubov and Hartree-Fock-Bogoliubov Theories}

Let us consider a condensate in thermal equilibrium at temperature $T$.
Within the Hartree-Fock Bogoliubov (HFB) theory, the condensate wave
function $\Psi $ satisfies the following generalized Gross-Pitaevskii
equation (in a frame rotating with the angular velocity $\Omega \hat z$)\cite
{Grif96}:
\begin{equation}
\label{b1}\left( -\frac{\hbar ^2}{2M}\nabla ^2+V_{{\rm tr}}+g|\Psi
|^2+2g\rho ({\bf r})-\mu (\Omega )+i\hbar \Omega \partial _\phi \right) \Psi
+g\Delta ({\bf r})\Psi ^{*}=0,
\end{equation}
where $\phi $ is the azimuthal angle in cylindrical polar coordinates, $\rho
({\bf r})$ is the density of the noncondensed gas and $\Delta ({\bf r})$ is
the anomalous average of two Bose field operators describing the
noncondensate (as in Sec.~IV.A, $\hat \psi =\Psi +\hat \phi $ is the quantum
field operator, with $\Delta =\langle \hat \phi \hat \phi \rangle $ and $%
\rho =\langle \hat \phi ^{\dagger }\hat \phi \rangle $). The collective
excitation energies $E$ of the system are the eigenvalues of the generalized
Bogoliubov equations for the coupled amplitudes $u({\bf r})$ and $v({\bf r})$
$$
\left( -\frac{\hbar ^2}{2M}\nabla ^2+V_{{\rm tr}}+2g|\Psi |^2+2g\rho ({\bf r}%
)-\mu (\Omega )\right) 
\pmatrix{u\cr v}
$$
\begin{equation}
\label{b2}+
\pmatrix{i\hbar \Omega \partial _\phi & -g\left[ \Delta (
{\bf r})+\Psi {}^2\right] \cr -g\left[ \Delta ^{*}({\bf r})+\Psi
^{*}{}^2\right] & -i\hbar \Omega \partial _\phi }
\pmatrix{ u \cr v}
=E
\pmatrix{u \cr-v }.
\end{equation}
Eq. (\ref{b2}) is valid at least for temperatures much less than the
chemical potential $\mu $ when resonant contributions (the so-called
Szepfalusy-Kondor processes) to the self-energies are not substantial \cite
{Fedi98}. In addition, we have self-consistency relations for the
noncondensate density $\rho ({\bf r})$

\begin{equation}
\label{b3}\rho ({\bf r})=\sum_n\left[ \frac{|u_n({\bf r})|^2+|v_n({\bf r})|^2%
}{\exp (E_n/k_BT)-1}+|v_n({\bf r})|^2\right] ,
\end{equation}
and for the anomalous average $\Delta ({\bf r})$

\begin{equation}
\label{b4}\Delta ({\bf r})=-\sum_n\left[ \frac{2u_n({\bf r})v_n^{*}({\bf r})%
}{\exp (E_n/k_BT)-1}+u_n({\bf r})v_n^{*}({\bf r})\right] ,
\end{equation}
where $n$ denotes quantum numbers specifying the excited states with
energies $E_n$ ($n=0,1,2,\cdots $ ). The eigenfunctions $u_n({\bf r})$ and $%
v_m({\bf r})$ satisfy the normalization condition:
\begin{equation}
\label{b5}\int \left[ u_n^{*}({\bf r})u_m({\bf r})-v_n^{*}({\bf r})v_m({\bf r%
})\right] d{\bf r}=\delta _{nm}.
\end{equation}

Equations (\ref{b1})-(\ref{b5}) constitute a complete set of the
self-consistent equations for the HFB theory. Within this theory, the
quasiparticle eigenvalues $E_n$ in Eqs.~(\ref{b2})-(\ref{b4}) must be
positive because the condensate is defined to have zero energy. Thus a
negative eigenvalue means a failure of the self-consistency and the
associated thermal equilibrium of the system. If $\rho ({\bf r})$ and $%
\Delta ({\bf r})$ are set to zero, we recover the Bogoliubov theory. If we
set only $\Delta ({\bf r})=0$, we obtain the Popov approximation. For a
vortex-free condensate in the low-temperature limit, the Popov and
Bogoliubov theories give identical excitation spectra~\cite{Isos97}. The
excitation spectrum in the HFB theory has an unphysical gap because it does
not treat all condensate-condensate interactions consistently~\cite{Grif96}.
Gapless modifications of the HFB theory, the so-called G1 and G2
approximations, are discussed in \cite{Hutc98,Prou98}. Normally, the
zero-temperature limit of the Popov, G1 and G2 theories should be the
Bogoliubov theory (which does not take into account noncondensate atoms).
For a non rotating condensate with a vortex, however, this is not the case
because vortex is unstable.

Within the Bogoliubov theory, an isolated vortex in a nonrotating harmonic
trap has at least one normal mode with negative energy. Let us apply the HFB
theory for a condensate with a vortex. To find a self-consistent solution
for the lowest eigenvalue at low temperatures, one can use a perturbation
method analogous to those developed in Ref.~\cite{Svid98a}. We consider a
condensate in an axisymmetric trap that rotates with an angular velocity $%
\Omega $ around the $z$ axis. We assume that the condensate contains a
singly quantized vortex along the $z$ axis. For simplicity we consider a
disk-shape condensate, so one can omit vortex curvature in investigating the
lowest normal mode. The condensate wave function has the form $\Psi
=e^{i\phi }|\Psi |$, with $\Delta =e^{2i\phi }|\Delta |$, and we can rewrite
the generalized Bogoliubov equations as:
\begin{equation}
\label{b6}\hat H_0
\pmatrix{u \cr v }
+\hat V
\pmatrix{u \cr v }
=E
\pmatrix{u \cr -v },
\end{equation}
where 
\begin{equation}
\label{b7}\hat H_0=\left( -\frac{\hbar ^2}{2M}\nabla ^2+\frac 12M\omega
_z^2z^2+2g|\Psi _0|^2-\mu (\Omega )\right) 
\pmatrix{1 & 0 \cr 0 & 1 }+
\pmatrix{i\hbar \Omega \partial _\phi & -g\Psi _0{}^2 \cr -g\Psi
_0^{*}{}^2 & -i\hbar \Omega \partial _\phi },
\end{equation}
and $\hat V$ includes the remaining part of Eq.~(\ref{b2}). Here, $\Psi _0$
is the wave function for an unbounded condensate in the $xy$ plane with the
same chemical potential; its excitations obey the equation
\begin{equation}
\label{b9}\hat H_0
\pmatrix{ u_0 \cr v_0 }
=E
\pmatrix{u_0 \cr -v_0 }.
\end{equation}
Equation (\ref{b9}) has an exact pair of solutions (see Ref. \cite{Svid98a})
with positive norm 
and energy
\begin{equation}
\label{b11}E_0=\hbar \Omega .
\end{equation}


Let us now make the following assumption: $E_0\ll k_BT\ll E_1,E_2,\cdots $,
where $E_0$ is the energy of the lowest normal mode, which can depend on $T$%
. Then the term with $n=0$ gives the main contribution in the sum in Eqs.~(%
\ref{b3}), (\ref{b4}), and we obtain:
\begin{equation}
\label{b13}\rho ({\bf r})\approx \frac{k_BT}{E_0}\left[ |u_0({\bf r}%
)|^2+|v_0({\bf r})|^2\right] 
\end{equation}
\begin{equation}
\label{b14}\Delta ({\bf r})\approx -\frac{2k_BT}{E_0}u_0({\bf r})v_0^{*}(%
{\bf r})
\end{equation}
For a singly quantized vortex one can derive the expression

\begin{equation}
\label{b15}\rho (r_{\perp }=0,z)\approx \frac{1.44\mu k_BT}{E_0I^2g\xi ^2}%
\left( 1-\frac{z^2}{R_z^2}\right) ,
\end{equation}
where $I^2\approx 16\sqrt{2}\pi \mu ^{3/2}/3g\omega _z\sqrt{M}$ is a
normalization integral, and
\begin{equation}
\label{b16}|\Delta (r_{\perp }=0,z)|\approx 0.
\end{equation}

In first-order perturbation theory, the lowest energy eigenvalue $E_0$ is
defined by the equation:

\begin{equation}
\label{b19}E_0=\hbar \Omega +E_a\left( \frac{\gamma k_BT}{E_0}-1\right) ,
\end{equation}
where $E_a=(3\hbar ^2\omega _{\perp }^2/4\mu )\ln \left( R_{\perp }/\xi
\right) $ and $\gamma =0.077R_{\perp }^4/N\xi ^4$ are positive with $N=\frac
8{15}\pi n(0)R_zR_{\perp }^2$ the total number of particles in the
condensate, and $\mu $ can be taken as the chemical potential for a
nonrotating trap. Eq.~(\ref{b19}) has two solutions, one with positive
energy and one with negative energy that reproduces the previous anomalous
mode with $E_0=\hbar \Omega -E_a$ as $T\rightarrow 0$. The negative solution
can be formally omitted, satisfying the requirement of self-consistency.
The positive solution has the form:
\begin{equation}
\label{b20}E_0=\frac 12\left[ \sqrt{(E_a-\hbar \Omega )^2+4E_a\gamma k_BT}%
-(E_a-\hbar \Omega )\right] .
\end{equation}
For nonrotating trap ($\Omega =0$), we find

\begin{equation}
\label{b21}E_0=\frac{E_a}2\left[ \sqrt{1+\frac{4\gamma k_BT}{E_a}}-1\right] .
\end{equation}
If $T\rightarrow 0$, we obtain $E_0\approx \gamma k_BT$, so that $E_0$ is
proportional to $T$ in the low-temperature limit. In fact our method
generalizes the Beliaev theory \cite{Beli58} for the vortex state. Recently
Pitaevskii and Stringari actually generalized the Beliaev approach (in the
density-phase representation) for the trapped Thomas-Fermi condensate \cite
{Pita98}.

Virtanen, Simula and~Salomaa made numerical calculations of vortex normal
modes at finite $T$ within the Popov, G1 and G2 approximations and
demonstrated that for a singly quantized vortex there is a self-consistent
solution with only positive frequencies in the limit $
T\rightarrow 0$ \cite{Virt00}. Their lowest energy solution corresponds to
our Eq. (\ref{b21}). The vortex mode (\ref{b21}) arises from the presence of
quasiparticles (an external pinning potential~can also result in such motion
\cite{Isos99}). At low temperatures, the quasiparticles are mostly localized
in the vortex-core region and provide an extra repulsive potential [the term
$2g\rho ({\bf r})$ in Eq.~(\ref{b2})] that affects the elementary
excitations. At $T=0$, the residual localized noncondensate fraction arises
from the interaction between particles; this result follows from Eq.~(\ref
{b15}) if we take $E_0\propto T$ at low temperatures. The additional
potential has a peak at the vortex core and the vortex line precesses around
the quasiparticle potential center with a positive excitation energy.

However, this does not mean that quasiparticles stabilize the vortex in a
trap. The physics of the problem is the following. At any moment during the
vortex motion, quasiparticles fill the vortex core (the relaxation time of
quasiparticles is much less than the period of the vortex precession). The
vortex line participates in two motions: first, the vortex precesses around
the trap center with the frequency $\hbar \omega _a=-E_a<0$ ($\Omega =0$).
The trap potential is responsible for this unstable mode. The quasiparticles
are localized in the vortex core and move together with the vortex; their
presence simply slightly changes the chemical potential and slightly
decreases the normal mode frequency. In second motion, the
vortex line moves around the center of mass of the quasiparticles in a locally
uniform condensate (in $xy$ plane). The amplitude of this motion is less than
$\xi $ and the frequency can be found from Eq. (\ref{b21}) in the limit $%
R_x,R_y\rightarrow \infty $ or $E_a\rightarrow 0$:

\begin{equation}
\label{tm}\hbar \omega _T=\sqrt{\gamma E_ak_BT}=0.37\sqrt{\frac{\mu k_BT}{%
n_0R_z\xi ^2}\ln \left( \frac{R_{\perp }}\xi \right) },
\end{equation}
where $n_0$ is the density of the vortex free condensate at the vortex
location (in the plane $z=0$). For JILA parameters $\gamma \approx 0.3$, $%
E_a\approx 1.58$Hz, then for $T=0.8T_c$ we obtain $\omega _T\approx 13.6$ Hz.
If this mode is thermally excited, its amplitude is given by
\begin{equation}
A=\xi \left( \frac{6a}{R_z}\right) ^{1/2}\left( \frac{k_BT}{\hbar \omega _T}%
\right) ^{1/2},
\end{equation}
where $a$ is the scattering length. For parameters of JILA experiments $%
A=0.16\xi $. Taking into account $\omega _T\propto \sqrt{T}$ we obtain the
following temperature dependence $A\propto T^{1/4}$. It is interesting to
note that the thermal mode (\ref{tm}) exists only in 3D condensate; in the
limit $R_z=\infty $ both the mode frequency and the amplitude go to zero.

Recent measurements of the lowest vortex modes in the JILA experiments are
in a good quantitative agreement with solutions of the time-dependent
Gross-Pitaevskii equation~\cite{Ande00,Fede00a,Svid00a}. The JILA
experiments measure, in fact, not only the absolute value, but also the sign
of the lowest vortex mode. The negative value of the anomalous-mode
frequency means that the vortex precesses in the same direction as the
superfluid flow around the vortex core, which is seen in the experiments. An
experimental observation of the thermal mode (\ref{tm}) could be next
challenging problem of future investigations.

\subsection{Dissipation and Vortex Lifetimes}

It is valuable to consider dissipation and its role in the vortex lifetime.
In a nonrotating trap, the ground state of the system is a vortex-free
condensate, so that a condensate with a vortex necessarily constitutes an
excited state. In the absence of dissipation, however, the vortex line moves
along trajectories of constant energy, remaining inside the condensate. The
condensate with a vortex will be unstable only if there is a mechanism to
transfer the system to the lower-energy vortex-free state~\cite{Pu99}. The
dissipative dynamics of a straight vortex due to its interaction with the
thermal cloud in a trapped Bose-condensed gas was discussed by Fedichev and
Shlyapnikov~\cite{Fedi99a}. If the vortex line moves with respect to the
normal component, scattering of elementary excitations by the vortex
produces a friction force, like that in superfluid $^4$He (see Ch.~3 of Ref.~%
\cite{Donn91}). Such a mechanism can transfer energy and momentum to the
thermal cloud. The friction force ${\bf F}$ can be decomposed into
longitudinal and transverse components: ${\bf F}=-D{\bf u}-D^{\prime }({\bf u%
}\times \hat n),$ where ${\bf u}$ is the velocity of the vortex line with
respect to the normal component, $D$ and $D^{\prime }$ are the longitudinal
and transverse friction coefficients, respectively, and $\hat n$ is a local
tangent vector to the vortex line. The transverse friction coefficient is
independent of the scattering amplitude and is given by the universal
expression $D^{\prime }=\hbar \rho _n/M$, where $\rho _n$ is the local mass
density of the normal component \cite{Soni97}. The longitudinal friction
coefficient depends on the scattering process. In the limit $k_BT\gg \mu $,
one can treat the elementary excitations as single particles, with the
result that $\rho _n\approx 0.1M^{5/2}T^{3/2}/\hbar ^3$ and the longitudinal
friction coefficient is proportional to the temperature: $D\approx \hbar
n\,(na^3)^{1/2}T/\mu $, where $n=|\Psi |^2$ is the superfluid density for
the vortex-free condensate and $a$ is the $s$-wave scattering length~\cite
{Fedi99a}.

In the presence of dissipation, the vortex line moves toward a (local)
minimum of the energy. In a nonrotating condensate, an off-center vortex
precesses around the trap center and is expected to spiral out to the
condensate boundary due to the dissipation. Once the vortex reaches the
boundary, it presumably decays by emitting phonons and single-particle
excitations. The radial motion of the vortex is governed by the longitudinal
friction coefficient: $v_r\approx Du/\hbar n\approx (na^3)^{1/2}Tu/\mu \ll u$%
, where $u$ is the precessional speed. Using this expression, one can
estimate the characteristic lifetime of the vortex state~\cite{Fedi99a}. At
present, no dissipation of the moving vortex has been observed in the JILA
experiments~\cite{Ande00}. A characteristic decay time for the dissipative
mechanism of Fedichev and Shlyapnikov in the JILA conditions is
significantly larger than the life-time of the condensate. The temperature
and density are too small to see the dissipation.

Another factor that can influence the vortex lifetime is the possibility
that a moving vortex can emit phonons. It is known that a moving vortex in
an infinite compressible fluid emits phonons, leading to a slow loss of
energy~\cite{Ovch98}. Recently, ~Lundh and~Ao~\cite{Lund00} studied the
radiation of sound from a moving vortex in an infinite, uniform system. A
homogeneous two-dimensional superfluid described by a nonlinear
Schr\"odinger equation is equivalent to (2+1)-dimensional electrodynamics,
with vortices playing the role of charges and sound corresponding to
electromagnetic radiation~\cite{Arov97,Ambe80}. Thus, a vortex moving on a
circular trajectory in an infinite superfluid radiates sound waves, which
are analogous to the cyclotron radiation of an electrical charge moving
along a circular orbit. The power radiated by a vortex with unit length
executing circular motion with frequency $\omega $ at a radius $r_0$ is
given by the following Poynting vector~\cite{Lund00}:
\begin{equation}
\label{b22}P=\frac{\pi Q^2\omega ^3r_0^2}{4c_s^2},
\end{equation}
where $Q=-\hbar \sqrt{2\pi n/M}$ is the ``vortex charge,'' $n$ is the
uniform superfluid density, and $c_s=\sqrt{\mu /M}$ is the velocity of sound.

In a nonuniform system, such as a two-dimensional or a disk-shape
axisymmetric trapped condensate, an off-center vortex performs a circular
motion around the symmetry axis. If such motion excites sound waves
(radiates energy), the vortex will move outward toward regions of lower
potential energy, until it eventually escapes from the cloud. In a trapped
condensate, however, the excitations all remain confined within the
condensate, and no phonon radiation is expected. In particular, the
wavelength $\lambda $ of sound that would be emitted exceeds the size $R$ of
the condensate. Indeed, $\lambda \sim 2\pi c_s/\omega $ and the precession
frequency of the straight vortex is of the order of $\omega \sim \hbar \ln
(R/\xi )/MR^2$; as a result, $\lambda /R\sim (R/\xi) \ln (R/\xi )\gg 1$, and
the ``cyclotron'' radiation is prohibited.

Finally, let us discuss how vortex generation affects the dissipation in
superfluids. One classic manifestation of superfluidity is that objects
traveling below a critical velocity propagate through a superfluid without
dissipation. According to the Landau criterion~\cite{Land41}, which relies
on the use of Galilean invariance, the critical velocity is $v_L=\min
[E(p)/p]$, where $E(p)$ is the energy of an elementary excitation with
momentum $p$. For a homogeneous Bose condensate, the Bogoliubov spectrum
implies a Landau critical velocity equal to the speed of sound $v_L=c_s$.
The Landau critical velocity can usually be observed only by moving
microscopic particles through the superfluid. Such motion of microscopic
impurities through a trapped gaseous Bose condensate was studied recently in~%
\cite{Chik00}. As the impurities traverse the condensate, they dissipate
energy by colliding with the stationary condensate and radiating phonons.
When the impurity velocity was reduced below the speed of sound, however,
the collision probability decreased dramatically, providing evidence for
superfluidity in the condensate.

If a {\em macroscopic} object moves through the condensate, dissipation can
occur due to turbulence and vortex formation in the superfluid, even if the
object's velocity is much lower than the Landau critical velocity. Recently,
dissipation in a Bose-Einstein condensed gas was studied by moving a
blue-detuned laser beam through the condensate~\cite{Rama99,Onof00}. The
laser beam repels atoms from its focus and creates a moving macroscopic
``hole'' in the condensate. The observed heating of the system agrees with
the prediction of dissipation when the flow field becomes locally
supersonic. Numerical simulations of the nonlinear Schr\"odinger equation
were used to study the flow field around an object moving through a
homogeneous condensate~\cite{Fris92,Huep97,Jack98a,Nore00,Jack00}. When the
object moves faster than a critical velocity $v_c$, these studies show that
the superfluid flow becomes unstable against the formation of quantized
vortex lines, which gives rise to a new dissipative regime. Pairs of
vortices with opposite circulation are generated at opposite sides of the
object. The rate of the energy transfer to the condensate by the moving
object increases significantly above this critical velocity for vortex
formation. The heating rate can be expressed as $dE/dt=E_{{\rm pair}}f_s$,
where $E_{{\rm pair}}$ is the energy of a vortex pair and $f_s$ is the
shedding frequency. The rate of vortex-pair shedding $f_s$ is proportional
to $v-v_c$ and thus larger when the speed of sound is lower.

Other simulations of the GP equation have demonstrated that
vortex-antivortex pairs or vortex half-rings can be generated by superflow
around a stationary obstacle~\cite{Jack98a,Wini99,Cara99,Fris92} or through
a small aperture~\cite{Burk94}. One might expect similar excitations in a
rotating condensate. In addition, vortex half-rings can be nucleated at the
condensate surface when the local tangential velocity exceeds a critical
value.

\section{Vortex states in mixtures and Spinor condensates}

\label{mixt}

The advent of multicomponent BECs \cite{Myat97,Stam98,Sten98} has provided
many new possibilities for quantum-mechanical state engineering. Since there
is no intrinsic difficulty in loading and cooling more than one alkali
element in the same trap, interpenetrating superfluids can now be realized
experimentally. Binary mixtures of condensates can consist of different
alkalis, or different isotopes, or different hyperfine states of the same
alkali atom. Such binary mixtures of Bose condensates have a great variety
of ground states and vortex structures that are experimentally accessible by
varying the relative particle numbers of different alkalis~\cite{Ho96}. In
particular, one can move continuously from regimes of interpenetrating
superfluids to those with separated phases. Many alkali binary mixtures
contain a coexistence region, which is the analog of $^3$He-$^4$He
interpenetrating superfluids in ultralow-temperature physics~\cite{Whea70}.

\subsection{Basic Phenomena}

Most experiments on Bose-Einstein condensation of atomic gases of $^{87} $Rb%
\cite{Ande95}, $^7$Li\cite{Brad95}, and $^{23}$Na\cite{Davi95} have used
magnetic traps to condense atoms with a hyperfine spin $F=2$ (or $F=1$).
Such a condensate of spin-$F$ bosons constitutes a spinor field

\begin{equation}
\label{m1}\langle \hat \psi _m({\bf r},t)\rangle =\zeta _m({\bf r},t)\Psi (%
{\bf r},t),
\end{equation}
where $\hat \psi _m$ is the field operator, $m$ labels $F_z$ (where $-F\leq
m\leq F$), $\Psi $ is a scalar, and $\zeta _m$ is a normalized spinor. In
magnetic traps, the spins of the alkali atoms are frozen and maximally
aligned with the local magnetic field ${\bf B}$ \cite{Ho96}. As a result, $%
\zeta $ is given by the eigenvalue equation $\hat {{\bf B}}\cdot {\bf F}%
\zeta =F\zeta $, where ${\bf F}$ is the hyperfine spin operator and $\hat {%
{\bf B}}$ is a unit vector along ${\bf B}$. The dynamics of $\langle \hat
\psi _m\rangle $ is therefore completely specified by the scalar field $\Psi
$, as in $^4$He. Thus, even though the alkali atoms carry a spin, they
behave in magnetic traps like scalar particles. In contrast to the scalar
field, however, the spinor field in Eq.~(\ref{m1}) possesses a local
spin-gauge symmetry: a local gauge change $\exp [i\chi ({\bf r},t)]$ of $%
\langle \hat \psi _m\rangle $ can be undone by a local spin rotation $\exp
[-i(\chi /F)\hat {{\bf B}}({\bf r},t)\cdot {\bf F}]$. Because of this
symmetry, the effective Hamiltonian of the scalar field $\Psi $ is not that
of $^4$He, but that of a neutral superfluid in a velocity field ${\bf u}_s$.
The velocity (or gauge field) ${\bf u}_s$ is a direct reflection of the
spin-gauge symmetry and it is given by

\begin{equation}
\label{m2}{\bf u}_s=-\frac{i\hbar }M\zeta ^{\dagger}{\bbox \nabla }\zeta .
\end{equation}
The velocity ${\bf u}_s$ can be calculated from the vorticity ${\bf \Omega }%
_s$ of ${\bf u}_s$, which satisfies the Mermin-Ho relation\cite
{Merm76,Voll90},

\begin{equation}
\label{m3}{\bf \Omega }_s=\frac 12{\bbox \nabla }\times {\bf u}_s=\left(
\frac \hbar {2M}\right) \epsilon _{\alpha \beta \gamma }\hat B_\alpha {\bbox %
\nabla } \hat B_\beta \times {\bbox \nabla }\hat B_\gamma .
\end{equation}
Equation (\ref{m3}) shows that the spatial variations of ${\bf B}$ necessary
to produce the trapping potential will inevitably generate a nonvanishing
superfluid velocity ${\bf u}_s=(2\hbar/M)\left( 1-B_z/B\right) {\bbox \nabla
}\,[\arctan\,(B_y/B_x)]$~\cite{Ho96}. If ${\bf B}_0=B_0\hat z$ is the
magnetic field at the center of an axisymmetric harmonic trap and $\omega _0$
is the maximum trap frequency, then the spin-gauge effect generates the
following constant effective ``rotation'' $\Omega _s$ around the $\hat z$
axis~\cite{Ho96}:

\begin{equation}
\label{m4}\,\frac{{\bf \Omega }_s}{\omega _0}\sim -\hat z\,\frac{\hbar
\omega _0}{\mu _BB_0},
\end{equation}
where $\mu _B$ is the Bohr magneton. The superfluid velocity ${\bf u}_s$
splits the degeneracy of the harmonic energy levels, breaks the inversion
symmetry of the vortex-nucleation angular velocity $\Omega _{c}$, and can
produce vortex ground states in the absence of external rotation if $\Omega
_s>\Omega _{c}$~\cite{Ho96}. In current experiments, the spin-gauge effect
is small; for example, if $\omega _0=10$ Hz and $B_0=1$ G, we obtain $\Omega
_s/\omega _0\sim 10^{-5}$. In oblate traps with $\omega _z\gg \omega _{\perp
}$, however, the spin-gauge effect can be significant ($\Omega _s$ could be
comparable with $\omega _{\perp }$ for large enough values of $\omega _z$).

Recently, the MIT group has succeeded in trapping a $^{23}$Na Bose
condensate by purely optical means~\cite{Stam98,Sten98}. In contrast to a
magnetic trap, the spins of the alkali atoms in such an optical trap are
essentially free, so that the spinor nature of the alkali Bose condensate
can be fully realized. Specifically, $^{23}$Na atoms possess a hyperfine
spin, with $F=1$ in the lower multiplet. All three possible projections of
the hyperfine spin can be optically trapped simultaneously. Thus the
condensate is described by a spin-1 spinor. The internal vortex structure of
a trapped spin-1 BEC was investigated in Ref.~\cite{Yip99}. Such vortices
and their stability were also discussed in~\cite{Ho98,Ohmi98}. In an optical
trap, the ground state of spin-1 bosons such as $^{23}$Na, $^{39}$K, and $%
^{87}$Rb can be either ferromagnetic or ``polar,'' depending on the
scattering lengths in different angular momentum channels~\cite{Ho98}. The
ferromagnetic state also has coreless (or skyrmion) vortices, like textures
found in superfluid $^3$He-A. Because of the wide range of hyperfine spins
of different alkalis, the optical trap has provided great opportunities to
study different spin textures in dilute quantum gases of atoms with large
spins. This is a fruitful subject for future experiments.

Although most of the theoretical effort has concentrated on
single-condensate systems, the first experimental realization of BEC
vortices was achieved with a two-species $^{87}$Rb condensate~\cite{Matt99},
following the proposal of Ref.~\cite{Will99}. Several other proposals have
been made for the dynamical production of a vortex using the internal
structure of atoms \cite{Marz97,Bold98,Dum98,Ruos99}. The spin-exchange
scattering rate is suppressed for $^{87}$Rb, which makes possible the study
of magnetically trapped multicomponent condensates of these atoms. The two
species correspond to two different hyperfine energy levels of $^{87}$Rb,
denoted $|1\rangle $ and $|2\rangle $; they are separated by the
ground-state hyperfine splitting. Since the scattering lengths are
different, both states are not equivalent. Typically, the $|1\rangle \equiv
|F=1,m=-1\rangle $ state is trapped and cooled to the condensation point.
Once the atoms in $|1\rangle $ have formed the condensate ground state, a
two-photon microwave field is applied, inducing transitions between the $%
|1\rangle $ state and the $|2\rangle \equiv |F=2,m=1\rangle $ state~\cite
{Matt99}. As a result, the atoms cycle coherently between the two hyperfine
levels with an effective Rabi frequency $\Omega _{{\rm eff}}$~\cite{Will00}.
Two parameters characterize the coupling: the detuning and the power. The
detuning $\delta $ denotes the mismatch of the frequency of the coupling
electromagnetic field to the frequency difference between the two internal
atomic states. The power is characterized by the Rabi frequency $\Omega $;
it is the rate at which population would oscillate between the two states if
$\delta $ were zero. When $\delta $ is larger than $\Omega $, the population
oscillations occur at the effective Rabi frequency $\Omega _{{\rm eff}}=%
\sqrt{\Omega ^2+\delta ^2}$, which obviously exceeds $\Omega $.

In principle, both states could be cooled simultaneously, so that the
condensate forms in a mixture of states. In practice, however, the typical
lifetime of atoms in the $|2\rangle $ state is about 1 s due to inelastic
spin-exchange collisions, which makes it very difficult to achieve runaway
evaporation for this state. In contrast, atoms in the $|1\rangle $ state
have a much longer lifetime of about 75 s~\cite{Matt99}. The advantage of
using the $|F=1,m=-1\rangle $ and $|F=2,m=1\rangle $ states is that their
magnetic moments are nearly the same, so that they can be simultaneously
confined in identical and fully overlapping magnetic trap potentials. Unlike
the more familiar single-component superfluids [see a discussion after Eq.~(%
\ref{m7})], where the topological constraints make it difficult to implant a
vortex within an existing condensate in a controlled manner, the coupled
two-component condensate has a different order parameter and hence different
topological constraints. Indeed, the coupled two-component system allows the
direct creation of a $\left| 2\right\rangle $ (or $\left| 1\right\rangle $)
state wave function having a wide variety of shapes out of a {$\left|
1\right\rangle $} (or $\left| 2\right\rangle $) ground-state wave function~%
\cite{Will99}.

\begin{figure}
\bigskip
\centerline{\epsfxsize=0.45\textwidth\epsfysize=0.23\textwidth
\epsfbox{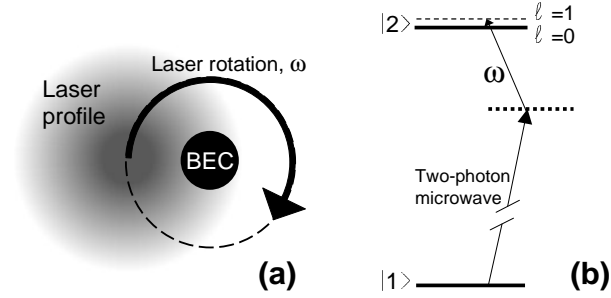}}

\vspace{0.3cm}

\caption{(a) A basic schematic illustration of the technique used to create a
vortex.
An off-resonant laser provides a rotating force on the
atoms across the condensate as a microwave drive of detuning
$\delta$ is applied. (b) A level diagram showing the microwave transition to
very near the $\left | 2 \right>$ state, and the modulation due to
the laser rotation frequency that couples only to the angular momentum $l=1$
state when $\omega \approx \delta $. }
(Taken from Ref.~\cite{Matt99})
\label{mf1}
\end{figure}

For example, to form a vortex in the two-component system, one should impose
a perturbation $\hat H_1$ that couples the ground state of the system to the
vortex state (namely, the matrix element of the perturbation operator
between these two states must be nonzero). The time-dependent GP equation
describing the driven, two-component condensate is~\cite{Will99}

\begin{equation}
\label{m0}i\hbar \frac \partial {\partial t} \pmatrix{ \Psi _1 \cr
\Psi _2}=
\pmatrix{\hat H_0+U_{11}|\Psi _1|^2+U_{12}|\Psi _2|^2+\hat H_1+\hbar
\delta /2 &
\hbar \Omega /2 \cr
\hbar \Omega /2 & \hat H_0+U_{21}|\Psi _1|^2+U_{22}|\Psi _2|^2-\hat
H_1-\hbar \delta /2} \pmatrix{\Psi _1 \cr
\Psi _2},
\end{equation}
where $\hat H_0=-(\hbar ^2\nabla ^2/2M)+\frac 12 M\omega _0^2(r_\perp^2+z^2)$
for a spherical trap, $M$ is the atomic mass, $\omega _0$ is the trap
frequency, $U_{ij}=4\pi \hbar ^2a_{ij}/M$, with $a_{ij}$ the $s$-wave
scattering lengths for binary collisions between constituents $i$ and $j$.
Williams and Holland considered the perturbation $\hat H_1$ in the following
form~\cite{Will99}:

\begin{equation}
\label{m5}\hat H_1=\kappa [f({\bf r})\cos (\omega t)+g({\bf r})\sin (\omega
t)],
\end{equation}
where $\kappa $ is a coupling coefficient and $f({\bf r})$ and $g({\bf r})$
are prefactors that depend on ${\bf r}$. The explicit form of $\hat H_1$
determines the symmetry of the quantum state being prepared, so that general
$f$ and $g$ can serve to prepare a macroscopic quantum state of arbitrary
symmetry. To create a vortex state with one unit of angular momentum, one
can take $\kappa =M\omega _0^2\rho _0$, $f({\bf r})=x$ and $g({\bf r})=y$ in
Cartesian coordinates. This form of perturbation effectively confines the
two hyperfine states in separate axially symmetric harmonic-oscillator
potentials with the same trap frequency $\omega _0$. The trap centers are
spatially offset in the $xy$ plane by a distance $\rho _0$ (from the center)
and rotate about the symmetry axis at an angular velocity $\omega $. To
achieve this configuration experimentally, Ref.~\cite{Matt99} shone a laser
beam into the trap along the $\hat z$ axis so that the cloud sits in the
middle of the Gaussian beam waist where the gradient of the beam intensity
is approximately linear (see Fig.~\ref{mf1}a). This arrangement produces a
constant force on the atoms. If the frequency of the laser beam is tuned
between the two hyperfine states, the optical dipole force acts in opposite
directions for each state, displacing the trap centers for each state. When
the beam rotates around the condensate at the angular velocity $\omega $, we
obtain the desired result.

To create a vortex, the angular velocity $\omega $ should be close to the
value at which a resonant transfer of population from the nonrotating
condensate into the vortex state takes place. Consider the frame co-rotating
with the trap centers at an angular frequency $\omega $. In this frame, the
energy of the vortex with one unit of angular momentum is shifted by $\hbar
\omega $ relative to its value in the laboratory frame. When this energy
shift compensates for both the energy mismatch $\hbar \delta $ of the
internal coupling field and the small chemical potential difference between
the vortex and the nonrotating condensate, resonant transfer of population
takes place (see Fig.~\ref{mf1}b). It is obvious that if we change the sign
of detuning $\delta $ while keeping the trap rotation fixed, a vortex will
be created with opposite circulation. Vortices with opposite circulations
experience opposite energy shifts in transforming to the rotating frame and
therefore require opposite signs of detuning in order to achieve the
resonant coupling.

In practice, $\omega \gg \omega _0$ and $\delta \gg \Omega $. The first
inequality allows the vortex to be generated rapidly. The main problem with
a slow drive (when $\omega \approx \omega _0$) is that the time scale for
coupling to the vortex state is very long, on the order of seconds in a trap
with $\omega _0=10$ Hz. The weak-coupling limit, given by the second
inequality, allows the resonance condition $\omega \approx \delta $ to
select energetically the desired state with high fidelity.

\begin{figure}
\bigskip
\centerline{\epsfxsize=0.45\textwidth\epsfysize=0.45\textwidth
\epsfbox{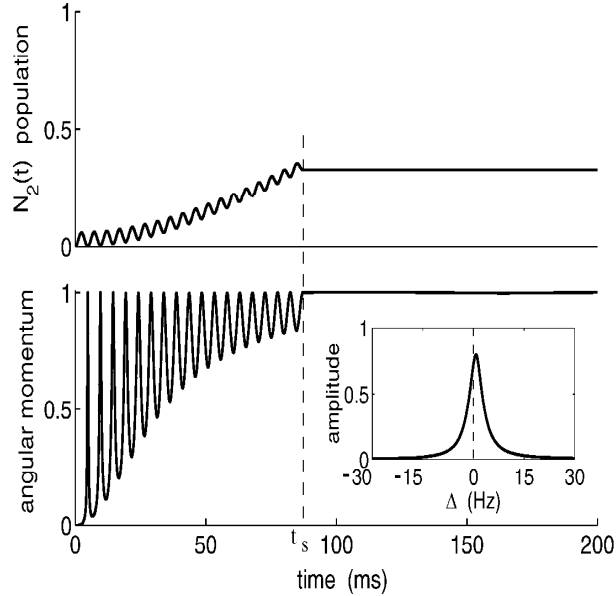}}

\vspace{0.3cm}

\caption{Dynamical evolution that can create a vortex. The top graph shows the
fractional population of atoms in the $|2\rangle$ internal state. The bottom
graph shows the angular momentum of the $|2\rangle$ state, in units of
Planck's constant
$\hbar$. The inset shows the amplitude of population transfer to the
vortex as a function of the trap rotation frequency $\omega$, with
$\Delta=\Omega_{\rm{eff}}-\omega$. The various parameters used in
the calculation are: $\omega_0=10$ Hz, $\delta=200$ Hz, $\omega=205.4$ Hz,
$N=8\times 10^5$ atoms, $M$ is the mass of the ${}^{87}$Rb atom, for
simulations the values of scattering lengths are taken to be
$a_{11}=a_{22}=a_{12}=5.5$ nm, and for $t<t_s$, $\Omega=50$ Hz and $\rho_0=1.7
\mu$m. Reprinted by permission from Nature {\bf 401}, 568, (1999), \copyright
1999 Macmillan Magazines Ltd.}
\label{mf2}
\end{figure}

Figure~\ref{mf2} shows the results of a numerical integration of Eq.~(\ref
{m0}) in two dimensions ($\omega _z=0$), with the condensate initially in
the nonrotating ground state and in the internal state $|1\rangle $~\cite
{Will99}. The coupling drive is turned on at time $t=0$, and is turned off
at time $t=t_s$ by setting both $\Omega $ and $\rho _0$ to zero. The top and
the bottom graphs show the fractional population and the angular momentum
per atom of the $|2\rangle $ state as a function of time.

The small-amplitude rapid oscillations on the top graph correspond to the
cycling between internal levels at the effective Rabi frequency $\Omega _{%
{\rm {eff}}}$. The gradual rise of this line reflects coupling from the
ground state to the vortex mode caused by the drive $\hat H_1$ in Eq.~(\ref
{m0}). Once during each Rabi cycle, the angular momentum approaches unity
(bottom graph), and, at that time, the $|2\rangle $ state wave function
approaches a pure vortex mode. By turning off the coupling at a precise time
$t=t_s$ on a given Rabi cycle, the $|2\rangle $ state can be prepared to
have unit angular momentum. The maximum possible population transfer to the
vortex state using this scheme obeys a Lorentzian response curve as $\omega $
is varied near $\Omega _{{\rm eff}}$, exhibiting a narrow resonance. This
situation is shown in the inset of Fig.~\ref{mf2}, where $\Delta =\Omega _{%
{\rm {eff}}}-\omega $.

In an experiment, it is possible put the initial condensate into either the $%
\left| 1\right\rangle $ or $\left| 2\right\rangle $ state, and then make a
vortex in the $\left| 2\right\rangle $ or $\left| 1\right\rangle $ state,
respectively. The evolution of the vortex can be watched over time scales
from milliseconds to seconds. In Ref.~\cite{Matt99}, the vortex was found to
be stable in only one of the two possible configurations corresponding to
the vortex in the $|1\rangle $ state, which is the one with the larger
scattering length (with the $|2\rangle $ state in the core). The other
possibility (the vortex in the $|2\rangle $ state, which is the one with the
lowest self-interaction coefficient) produces an instability.

\subsection{Stability Theory}

We use the following notation for the states: $(1,0)$ for the state with the
vortex in $|1\rangle $ and $(0,1)$ for the state with the vortex in $%
|2\rangle $. In the JILA experiment~\cite{Matt99}, the number of particles
is the same for each component ($N_1=N_2=N$) but, in general, one could
consider any ratio between the populations of the different levels. The
scattering lengths for binary collisions depend on the internal hyperfine
level of the atom. For $^{87}$Rb the values of scattering lengths are nearly
degenerate and in the proportion $a_{11}:a_{12}:a_{22}=1.00:0.97:0.94$~\cite
{Hall98}. Because of the relation $U_{11}>U_{12}>U_{22}$, the experiment is
performed in a regime in which the first component separates from the second
one. Consequently, a favored configuration has the first component spread
over the largest part of the space. Numerical simulations show that in the
equal population case, $N_1=N_2=N$, and for arbitrary nonlinearities, the
stationary states $(1,0)$ is stable while the other state $(0,1)$ is
unstable.

The origin of the instability of the state $(0,1)$ is purely dynamical~\cite
{Ripo00} and can be understood within the framework of mean-field theories
for the double-condensate system without dissipation. Actually, the
instability mechanism does not lead to expulsion of the vortex from the
condensate, but to periodic transfer of the phase singularity from one
species to the other. To study the vortex stability, one can start from a
pair of coupled Gross Pitaevskii equations for the condensate wave functions
of each species

\begin{equation}
\label{m6}i\hbar \frac \partial {\partial t}\Psi _1=\left[ -\frac{\hbar
^2\nabla ^2}{2M}+V_1+U_{11}|\Psi _1|^2+U_{12}|\Psi _2|^2\right] \Psi _1,
\end{equation}

\begin{equation}
\label{m7}i\hbar \frac \partial {\partial t}\Psi _2=\left[ -\frac{\hbar
^2\nabla ^2}{2M}+V_2+U_{21}|\Psi _1|^2+U_{22}|\Psi _2|^2\right] \Psi _2,
\end{equation}
where $V_1$ and $V_2$ are trap potentials for the condensate components.
These equations are a particular case of Eq. (\ref{m0}) when the drive is
turned off ($\hat H_1$, $\Omega$, $\delta =0$). Equations (\ref{m6}) and (%
\ref{m7}) conserve the number of particles in each hyperfine level. However,
the angular momentum of each component is no longer a conserved quantity,
and the topological charge of each species can change through the time
evolution. Instead, what is conserved is the total angular momentum of the
system

\begin{equation}
\label{m8}L_z=i\hbar \int d^3r{\Psi }_1^{*}\partial _\phi \Psi _1+i\hbar
\int d^3r{\Psi }_2^{*}\partial _\phi \Psi _2.
\end{equation}
As in the JILA experiments, we assume that both potentials are spherically
symmetric and have the form $V_1({\bf r})=V_2({\bf r})=\frac 12M\omega
_0^2(r_{\perp }^2+z^2)$. For stationary configurations in which each
component has a well-defined value of the angular momentum, the time and
angular dependence are factored out
\begin{equation}
\label{m9}\Psi _i(r_{\perp },z,\phi )=e^{-i\mu _it/\hbar }e^{iq_i\phi }\psi
_i(r_{\perp },z),
\end{equation}
with $i=1,2$. We focus on three particular configurations, which are the
lowest energy states with vorticity $(q_1,q_2)=(0,0),(1,0),(0,1)$. They
correspond to the ground state of the double condensate, and to the single
vortex states for the $|1\rangle $ and $|2\rangle $ species, respectively.

Linear stability analysis of the three states gives the following results
\cite{Ripo00}. For the $(0,0)$ state, the frequencies of all normal modes
are positive, as expected for the ground state of the system. Among the
normal modes of the $(1,0)$ family, there is a negative eigenvalue, which
means that there is a path in the configuration space along which the energy
decreases (this is just the analog of the anomalous mode in the
one-component system with a vortex). This path belongs to a perturbation
that takes the vortex out of the condensate. As in the case of a
single-component condensate, however, the lifetime of the vortex state is
only limited by the presence of dissipation (without dissipation, the
configuration is dynamically stable). Finally, in the $(0,1)$ family, there
are normal modes with complex frequencies. The shape of the unstable modes
is similar to the energy-decreasing modes of the $(1,0)$ family --- that is,
they are perturbations that push the vortex out of both clouds. The
imaginary part of the eigenvalues implies that vortices with unit charge in $%
|2\rangle $ are unstable under a generic perturbation of the initial data,
whereas those in $|1\rangle $ can be long-lived. This conclusion is
consistent with the JILA experiments, where a vortex in the $|2\rangle $
species was found to be unstable~\cite{Matt99}.

Numerical simulations of the vortex behavior for large perturbations show
that the linearly stable state $(1,0)$ is robust and survives under a wide
range of perturbations, suffering at most a precession of the vortex core
plus changes of the shapes of both components \cite{Ripo00}. This behavior
arises in both two- and three-dimensional simulations. In contrast, the
unstable configuration $(0,1)$ develops a recurrent dynamics. In the first
stage, the first component and the vortex oscillate synchronously (the hole
in $|2\rangle $ pins the peak of $|1\rangle $). These oscillations grow in
amplitude, and the vortex spirals out. Finally the first component develops
a tail and later a hole which traps the second component. The hole is a
vortex that has been transferred from $|2\rangle $ to $|1\rangle $. Though
not completely periodic, this mechanism exhibits some recurrence, and the
vortex eventually returns to $|2\rangle $. The preceding behavior persists
even for strong perturbations in a two-dimensional condensate. However, for
large perturbations of a three-dimensional condensate, the dynamics may lead
to a turbulent behavior \cite{Ripo00}.

\begin{figure}
\bigskip
\centerline{\epsfxsize=0.22\textwidth\epsfysize=0.45\textwidth
\epsfbox{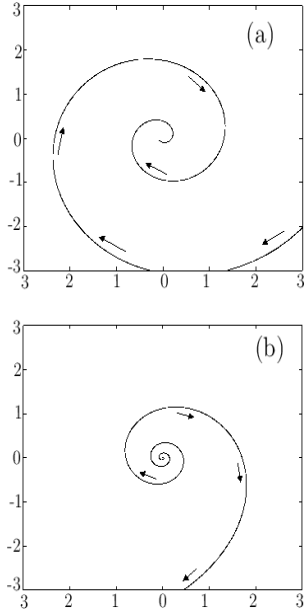}}

\vspace{1cm}

\caption{Evolution of the position of the phase singularity in the $xy$ plane.
Lengths are given in units of the trap characteristic length
$d=\sqrt{\hbar/M\omega_0}$.  (a) Phase singularity in $|1\rangle $,
(b) phase
singularity in $|2\rangle $.}
(Taken from Ref.~\cite{Garc00}).
\label{mf3}
\end{figure}

In Fig.~\ref{mf3}, it is shown how a small initial perturbation makes the
phase singularity in $|2\rangle $ spiral out of the system while a phase
singularity appears in $|1\rangle $ and occupies the center of the atomic
cloud. This dynamics is recurrent.

The preceding results are valid for the equal population case, $N_1=N_2$.
For any ratio of the populations $N_1/N_2$ and any values of the nonlinear
coefficients $U_{ij}$, the stability conditions are the following \cite
{Garc00}: The configuration $(1,0)$ is stable if
\begin{equation}
\label{m10}\left( \sqrt{\frac{N_1}{N_2}}-1\right) ^2>1-\frac{a_{11}}{a_{12}}%
.
\end{equation}
For $^{87}$Rb, the inequality (\ref{m10}) is always satisfied, which proves
that the configuration with a vortex in $|1\rangle $ is always linearly
stable, as found in ~\cite{Matt99}. Note that the stability properties do
not depend on the total number of particles but only on the ratio between
the populations.

The stability condition of the configuration $(0,1)$ is
\begin{equation}
\label{m11}\left( \sqrt{\frac{N_2}{N_1}}-1\right) ^2>1-\frac{a_{22}}{a_{21}}
\end{equation}
This inequality fails for a certain range of $N_1/N_2$. For the case of $%
^{87}$Rb the unstable range is $N_1/N_2\in [0.73,1.49]$, which means that
certain choices of the population imbalance allow stabilization of the
vortex in $|2\rangle $. These results predict the possibility of stable
vortex states for various multiple-condensate systems~\cite{Garc00}.

Energetic considerations show that the extra degree of freedom associated
with the second component allows a more intricate structure for the
free-energy surface. As a result, in a two-component system, it is possible
to achieve a local minimum in the free energy at the center of the trap~\cite
{McGe00}. The presence of such a minimum implies the existence of a region
of energetic stability where the vortex cannot escape and might generate a
persistent current.

\section{Conclusions and outlook}

In this paper, we have provided an introductory description of vortices in
trapped Bose condensed gases. The main conclusion of our analysis is that
the vortex dynamics in such systems is well described by the time-dependent
Gross-Pitaevskii equation (at least for low temperatures). The nonuniform
nature of the condensate results in the appearance of anomalous vortex
mode(s) with negative frequency and positive norm. Trap rotation shifts the
normal-mode frequencies and can stabilize the vortex state. To date,
experimental measurements of vortex dynamics and other properties of vortex
states are in a good quantitative agreement with theoretical predictions
based on solutions of the GP equation. Deviations from the mean-field
predictions could arise when the gas parameter $\bar n|a|^3$ is not very
small (semiclassical corrections to the mean-field approximation were
calculated in~\cite{Ande99} ) or from ``mesoscopic'' effects associated with
the finite systems. However, there is no experimental evidence for these
effects so far.

We have been able to cover only part of the existing literature on vortices
in trapped condensates. Among important issues that we have not discussed
are: different methods of vortices generation and detection, kinetics of
vortex nucleation and decay, vortices in BECs with attractive interactions
and in Fermi condensates, other defects in BECs (solitons, instantons,
vortex solitons, skyrmions, wave-function and spin monopoles).

In the case of superfluid helium, vortex nucleation is associated with
pinning of vortex lines at the walls of the container. Trapped condensates
have no rough surfaces, and the nucleation process of quantized vorticity
has a different origin~\cite{Mart00,Garc00a,Dalf00}. An important question
in vortex nucleation is the role of the thermal component and transverse
anisotropy of magnetic traps \cite{Reca01,Madi01}.

The literature of the past few years contains many different proposals for
the creation of vortices in trapped BECs, although we considered only a few
of them in this review. To illustrate the diversity of different methods,
let us cite some other schemes. An experimental setup for vortex creation by
Berry's phase induced Bose-Einstein condensation is proposed by Olshanii and
Naraschewski~\cite{Olsh98}. A related vortex-production scheme employing the
Aharonov-Casher effect is discussed by Petrosyan and You~\cite{Petr99}.
Other proposals suggest the creation of the vortex state by opto-mechanical
stirring \cite{Bold98}; by a rotating force \cite{Marz98a}; by an adiabatic
population transfer of a condensate from the ground to the excited
Bose-condensed state via a Raman transition induced by laser light \cite
{Marz97,Dum98,Will99}; the accidental generation of vortices in a quench
\cite{Angl99,Drum99} or in self-interference measurements \cite{Ruos99}.

A possible way to create rotating states from a trapped ground-state BEC by
using light-induced forces is proposed by Marzling and Zhang~\cite{Marz98}.
They show that the dipole potential induced by four traveling-wave laser
beams with an appropriate configuration in space, phase and frequency can be
used to realize such a system. Vortex states can be trapped in an
evaporative cooling process if the evaporation length is less than the size
of the thermally excited state \cite{Mars99,Drum99}. In order to nucleate
vortices, the trapped gas can be rotated at temperatures above the BEC
transition. Recently, it has been suggested that vorticity could be
imprinted by imaging a BEC through an absorption plate \cite{Dobr99}. The
method consists of passing a far-off-resonant laser pulse through an
absorption plate with an azimuthally dependent absorption coefficient,
imaging the laser beam onto a BEC, and thus creating the corresponding
nondissipative Stark-shift potential and condensate phase shift. A vortex
ring may be formed by translating one condensate through another one \cite
{Jack99} (this process is analogous to ring nucleation by moving ions in
superfluid $^4$He \cite{Rayf64}), or by three-dimensional soliton decay \cite
{Jone86,Joss95}. Recently the JILA group generated vortex rings by the decay of
dark solitons through the snake instability \cite{Ande00b}.

Many different proposals for the detection of vortices in BECs have been
mentioned in literature. Some of them are used in current experiments. The
spatial size of the vortex core in the TF regime is too small to be
observed; for visualizing the vortex state, it was suggested to switch off
the trap and let the cloud expand ballistically~\cite{Lund98}. After
expansion, the size of the vortex core is magnified by approximately the
same factor as the size of expanding condensate~\cite{Cast99,Dalf99a}, so
that the core becomes observable. Also the vortex state can be detected by
the splitting of the collective condensate modes in axisymmetric traps \cite
{Sinh97,Svid98,Zamb98} or by looking at the phase slip in the interference
fringes produced by two expanding condensates \cite{Bold98,Cast99}. Dobrek
{\it et al.} \cite{Dobr99} proposed an interference method to detect
vortices by coherently pushing part of the condensate with optically induced
Bragg scattering. A detection scheme that reveals the existence of vortex
states in a cylindrically symmetric trap is discussed by Goldstein, Wright
and Meystre~\cite{Gold98}. This scheme relies on the measurement of the
second-order correlation function of the Schr\"odinger field and yields
directly the topological charge of the vortex state.

Also one can detect the vortex state by observing the off-resonance
absorption image of the rotational cloud \cite{Marz97}. For a vortex state
one should expect a bright ``hole'' in the image which accounts for the
vortex core in the density distribution. Another possibility is to observe
the Doppler frequency shift due to the quantized circular motion of the
atoms \cite{Marz97}, or by scattering fast atoms in a pure momentum state
off a trapped atomic cloud \cite{Kukl99}.

Another question that has recently attracted significant theoretical and
experimental interest is the dynamics and stability of dark solitons and
vortex solitons in trapped condensates. Solitary waves (kinks) have been
studied in many physical contexts \cite{Raja87} and exist in different
physical, chemical and biological systems \cite{Kern89}. Recent theoretical
studies discuss the dynamics and stability of dark solitons \cite
{Zhan94,Rein97,Morg97,Jack98b,Marg99} (the range of parameters where the
solitons are dynamically stable has been determined in \cite{Mury99,Fede00}%
, while the theory of dissipative dynamics of a kink at finite temperature
condensates has been developed in \cite{Fedi99b}), as well as suggestions
for their creation~ \cite{Dum98,Scot98,Dobr99}. Recently dark
solitons inside a condensate were generated by a phase-imprinting method
\cite{Burg99,Dens00}. Unlike vortices, dark solitons are not topologically
stable. At finite temperature, they exhibit thermodynamic and dynamic
(small-amplitude) instabilities. The interaction of the soliton with the
thermal cloud causes dissipation that accelerates the soliton. There is an
interesting analogy between solitons and relativistic particles, in which
the soliton velocity and speed of sound correspond to the particle velocity
and speed of light in vacuum \cite{Fedi99b}. However, the kinematic mass of
the soliton decreases when its velocity increases. This behavior is opposite
to the case of relativistic particle, where the kinematic mass increases
with velocity, and an infinite force is required to accelerate the particle
beyond the velocity of light. In contrast to the particle, the soliton can
reach the velocity of sound. An interesting problem is to create a soliton
and a vortex simultaneously (this object is known as vortex soliton). The
vortex soliton has a topological charge and therefore could be stable.

Another challenging perspective for future experiments is the creation of
vortex-like states in optically confined BECs. By relaxing the condition of
spin polarization imposed by magnetic trapping, this new method of
confinement permits the study of diverse textures that can be formed by the
spinor order parameter, like those in superfluid $^3$He-A \cite{Ho98}. Also
optical traps allow strong variation of the scattering length via Feshbach
resonances, which provides new possibilities for manipulating the condensate
states.

Among other challenging problems, one should mention measurements of vortex
normal modes at higher temperatures, which could establish the connection
between the Bogoliubov approximation and self-consistent mean field
theories. Also, it would be interesting to observe vortex dissipation and
damping of vortex normal modes.

\acknowledgments

We are grateful to B.~Anderson, E.~Cornell, J. Dalibard, D.~Feder,
M.~Holland, M. Linn, G. Shlyapnikov and S.~Stringari for valuable
correspondence and discussions. This work benefited from our participation
in recent workshops at the Lorentz Center, Leiden, The Netherlands and at
the European Centre for Theoretical Studies in Nuclear Physics and Related
Areas, Trento, Italy; we thank H.~Stoof and S.~Stringari for organizing
these workshops and for their hospitality. This research was supported in
part by the National Science Foundation, Grant No. DMR 99-71518, and by
Stanford University (A.A.S.).

\end{document}